%% file: aanda.tex
\documentclass{aa}  

\usepackage{graphicx}
\usepackage{txfonts}
\usepackage[colorlinks=true,linkcolor=bluea,allcolors=blue]{hyperref}
\usepackage{multirow}
\usepackage{multicol}
\usepackage{placeins}
\usepackage[dvipsnames]{xcolor}

\newcommand{\kms}{$\rm km~s^{-1}~$}  
\newcommand{\ergs}{$\rm erg~s^{-1}~$}    
\newcommand{\lbol}{$\rm L_{bol}$}               
\newcommand{\msun}{$\rm M_{\odot}$}              
\newcommand{\msunyr}{$\rm M_{\odot}~yr^{-1}$} 

\newcommand{\apr}{$\rm \sim $} 

\newcommand{\hb}{$\rm H\beta$~}          
\newcommand{\ha}{$\rm H\alpha$~}         
\newcommand{\pa}{$\rm Pa\alpha$}         
\newcommand{\brg}{$\rm Br\gamma$}         
\newcommand{\hii}{$\rm H_2$~}                
\newcommand{\sivi}{$\rm [Si~VI]$}                

\newcommand{\loglbol}{$\rm log(L_{bol}[erg~s^{-1}])$~}                

\begin{document} 

   \title{QSOFEED: Investigating warm molecular, low- and high-ionization atomic gas in six type-2 quasars with GTC/EMIR}



   \author{P. H. Cezar
          \inst{1,2}
          \and M. Coloma Puga \inst{3,4,5}
          \and
          C. Ramos Almeida \inst{1,2} \and J. A. Acosta-Pulido \inst{1,2} \and G. Speranza\inst{6} \and L. R. Holden\inst{7} \and C. N. Tadhunter\inst{8} \and M. V. Zanchettin\inst{9} \and A. Audibert\inst{1,2}
          }

   \institute{Instituto de Astrofísica de Canarias, Calle Vía Láctea, s/n, 38205 La Laguna, Tenerife, Spain\\ 
   \email{pedrocezar.astro@gmail.com}
         \and Departamento de Astrofísica, Universidad de La Laguna, 38206 La Laguna, Tenerife, Spain 
        \and  Dipartimento di Fisica, Università degli Studi di Torino, Via Pietro Giuria 1, 10125 (Torino), Italy 
        \and  INAF – Osservatorio Astrofisico di Torino, Via Osservatorio 20, I-10025 Pino Torinese, Italy 
        \and European Southern Observatory (ESO), Alonso de Córdova 3107, Casilla 19, Santiago 19001, Chile
        \and Instituto de Física Fundamental, CSIC, Calle Serrano 123, 28006 Madrid, Spain
        \and Centre for Astrophysics Research, University of Hertfordshire, Hatfield, AL10 9AB, United Kingdom.
        \and School of Mathematical and Physical Science, University of Sheffield, Sheffield S3 7RH, UK. 
        \and INAF - Osservatorio Astrofisico di Arcetri, Largo E. Fermi 5, I-50125 Firenze, Italy 
        }
        
   \date{Received 22 October 2025 / Accepted 29 January 2026}

 
  \abstract{

  We present long-slit near-infrared spectroscopic observations of six nearby (z$\sim$0.1) radio-quiet type-2 quasars (QSO2s) from the Quasar Feedback (QSOFEED) sample. They have bolometric luminosities of $\rm 10^{45-46}~erg~s^{-1}$ and stellar masses of $\rm 10^{10.6-11.3}~M_{\odot}$. The observations were obtained with the instrument Espectrógrafo Multiobjeto Infra-Rojo (EMIR) on the 10.4 m Gran Telescopio Canarias.
  The nuclear K-band spectra (central $\sim$1-3 kpc of the QSO2s) reveal signatures of high-velocity outflows in either the \pa~or \brg~lines, depending on the redshift, and in the [Si VI] lines. The broadest kinematic components have full width at half maximum (FWHM) of $\sim$1200-2500 km $\rm s^{-1}$. From the near-infrared hydrogen recombination lines we derived ionized outflow masses of $\rm  M_{Hion}\rm \sim0.08-20\times 10^{6}~M_{\odot}$, mass outflow rates of $\rm \dot{M}_{Hion}\rm \sim0.03-6~M_{\odot}~yr^{-1}$, and kinetic powers of $\rm \dot{E}_{Hion}\rm \sim 10^{37.8-40.8}~erg~s^{-1}$. These ionized gas outflow masses and mass outflow rates have median values that are 5.9 and 5.8 times larger, respectively, than those derived from the [Si VI] line. 
  Our study provides evidence, at least for these six QSO2s, that the near-infrared recombination lines and [Si VI] are tracing
  the same outflow (i.e., they have similar kinematics and radii), but they carry different amounts of mass. We detected warm molecular lines in the six QSO2s, from which we measured total (nuclear) gas masses from 1.1 (0.7) to 32 (13) $\rm \times~10^3~M_{\odot}$, similar to other QSO2s with warm \hii measurements reported in the literature, but we did not find any molecular outflow associated with them. Comparing with other five QSO2s with \hii measurements reported in the literature, 
   we find that the four QSO2s with detected H$_2$ outflows have total (nuclear) \hii masses 2.2 (2.7) times larger, on average, than the seven QSO2s without detected \hii outflows.} 

 \titlerunning{Investigating warm molecular, low- and high-ionization atomic gas in six type-2 quasars}
    \authorrunning{P. H. Cezar et al.}

   \maketitle

%

\section{Introduction}\label{section:Intro}

Accreting supermassive black holes (SMBHs) in the nuclei of galaxies can  undergo several accretion events \citep{Hickox+14,Schawinski+15} 
 during the life-cycle of galaxies, producing 
luminous activity episodes known as Active Galactic Nuclei (AGN). The energy released in these recurrent 
AGN events can have 
significant effects on the host galaxy, the so-called AGN feedback. This feedback is capable of 
disturbing the gas kinematics (e.g., driving gas outflows), 
producing changes in gas excitation and metallicity, and ultimately having 
an impact on the formation of stars (see \citealt{Harrison+24} for a recent review). Although AGN-driven outflows might lead to 
star formation quenching when they are powerful enough (negative feedback), simulations \citep{Mercedes-Feliz+23} and observational works have also found evidence of 
positive feedback (i.e., AGN inducing star formation; \citealt{Cresci+15,Cresci+15b,Santoro+16,Maiolino+17,Gallagher+19,Bessiere+22}). AGN feedback has become a key component for analytical, semi-analytical models, and simulations \citep{Harrison+18} to reproduce the observed Universe \citep{DiMatteo+05,Dave+19,Su+19,Zinger+20,Zubovas+23}.



To infer the  impact of AGN feedback on galaxies, it is essential to determine how energy couples with the multiphase gas. Several studies, mostly focusing on optical emission lines, reported the presence of ionized outflows \citep{VillarMartin+16,Fiore+17,Kakkad+20,Hervella+23,Speranza+24,Bertola+25}, which appear to be ubiquitous, at least for luminous quasars \citep{Harrison+14,Woo+16,Rupke+17,Bessiere+24}. However, the role of outflows in other gas phases, which may be more relevant in terms of mass and energy 
budget, is still largely unconstrained in comparison with the ionized gas phase \citep{Cicone+18}. Besides, there is no consensus yet on whether or not the outflows detected in different phases are different faces of the same phenomenon \citep{Harrison+24}. 

Coronal emission lines are important tracers of outflows in the highly-ionized phase (see \citealt{Ardila+25} for a recent review). They are transitions of highly-ionized species with ionization potential (IP) $\gtrsim$100 eV, mainly associated with AGN activity (but see \citealt{Hernandez+25}).  
Their usually high critical densities make them good outflow tracers closer to their launching region. 
Furthermore, \cite{Trindade+22} showed that coronal lines with IP>138 eV could be used to trace the X-ray emitting-gas, 
confirming that these ions can probe the highest-ionized component of outflows. Despite their relevance, coronal line outflow studies are mostly restricted to nearby low-luminosity AGN (LLAGN; \lbol$\rm <$10$^{45}$\ergs; 
\citealt{Muller+11,RodriguezArdila+17,May+18,Fonseca+23,Delaney+25}). Only a few studies targeting quasars are reported in the literature \citep{RamosAlmeida+17,RamosAlmeida+19,RamosAlmeida+25,Speranza+22,VillarMartin+23,Doan+25}.

The study of outflows in the cold molecular gas phase is of particular interest, since it is the fuel for 
star formation. Recent studies were able to study cold molecular outflows at submillimeter wavelengths, 
from nearby AGN and ultra-luminous infrared galaxies (ULIRGs, \citealt{Pereira-Santaella+18,Zanchettin+21,Zanchettin+23,Lamperti+22,DallAgnol+23}) to luminous quasars \citep{Feruglio+10,Cicone+14,Vayner+21,RamosAlmeida+22,Audibert+25}. The few multi-phase outflow studies 
of AGN suggest 
that although slower than the ionized outflows, cold molecular outflows carry the bulk of mass in the outflows \citep{Fiore+17,Fluetsch+19,Fluetsch+21,Garcia-Bernete+21,Zanchettin+23,Holden+24,Speranza+24}, at least in the local Universe. However, little is known about the warm molecular gas phase observed in the near-infrared (NIR, H$_2$ gas at temperatures (T) $\gtrsim$1000 K) and mid-infrared (MIR, H$_2$ gas at T$\approx$100-1000 K), that represents a small fraction of the total molecular gas reservoir \citep{Dale+05,Mazzalay+13,Emonts+17,CostaSouza+24,Zanchettin+25,Kakkad+25}. While in nearby low-luminosity AGN some works detected NIR warm molecular outflows \citep{Tadhunter+14,Bianchin+22,Riffel+23}, at higher luminosities, only six QSO2s have their \hii kinematics studied in the NIR \citep{Rupke+13b,RamosAlmeida+17,RamosAlmeida+19,Speranza+22,VillarMartin+23,Zanchettin+25} with four of them showing warm molecular outflows. Thanks to the James Webb Space Telescope (JWST), 
it is now possible to study the rich MIR spectra of AGN with unprecedented spectral resolution and sensitivity, 
including the warm molecular \hii lines that are observed at those wavelengths \citep{CostaSouza+24,Davies+24,Esparza-Arredondo+25,Dan+25,Marconcini+25,RamosAlmeida+25}. Recent JWST results show that just a few AGN with strong jet-ISM interactions have signatures of warm molecular outflows (e.g., \citealt{Bohn+24,CostaSouza+24,Riffel+25}).



QSO2s are excellent laboratories to study both their multi-phase outflows and host galaxies, thanks to their high-luminosity, which contributes to drive more powerful outflows \citep{Cicone+14,Fiore+17}, and to obscuration \citep{Ramos&Ricci+17}. Obscuration makes it possible to study the impact of outflows on e.g. the young stellar populations of galaxies when high angular resolution data are used (e.g., \citealt{Bessiere+22}).
Studying luminous quasars at NIR wavelengths is key for multiphase studies of gas, since emission lines from warm molecular, ionized, and highly-ionized gas are observed simultaneously. NIR data have 
the additional advantage of reduced dust attenuation, enabling a clearer view of innermost regions of the outflows \citep{RamosAlmeida+17,RamosAlmeida+19}. 

In this work we present K-band spectroscopic observations of six nearby QSO2s, from which we study the properties of the low and high-ionization and warm molecular gas. This study is organized as follows: in Sect. \ref{sample} we present the QSO2 sample studied in this work. Sect. \ref{sec:observations} details the observations and data reduction. In Sect. \ref{sec:Results} we analyze the nuclear spectra and study the gas kinematics, including outflow properties. Finally, in Sect. \ref{sec:discussion} and \ref{sec:conclusions} we discuss and summarize our findings. Throughout this work we assume the following cosmology: H$_{0}$ = 70.0 km s$^{-1}$ Mpc$^{-1}$, $\rm \Omega_M$ = 0.3, and $\rm \Omega_\Lambda$ = 0.7.

\section{Sample selection}
\label{sample}
\input{Tables/table_sample}
The Quasar Feedback (hereafter \href{https://research.iac.es/galeria/cristina.ramos.almeida/qsofeed/}{QSOFEED}) sample \citep{RamosAlmeida+22,Pierce+23,Bessiere+24} includes all the QSO2s in the catalogue of \cite{Reyes+08} with $ \rm L_{[OIII]}>10^{8.5}~L_{\odot}~(L_{bol}> 10^{45.6}~erg~s^{-1})$ and redshifts $z < 0.14$, comprising 48 galaxies. These criteria guarantee that we selected the most luminous QSO2s in this optically-selected catalogue, but at a suitable distance to spatially resolve the multi-phase gas outflows \citep{RamosAlmeida+22,Speranza+24}, the dust distribution \citep{RamosAlmeida+26}, and the stellar populations \citep{Bessiere+22}.  

\begin{figure}
    \centering
    \includegraphics[width=0.3\linewidth]{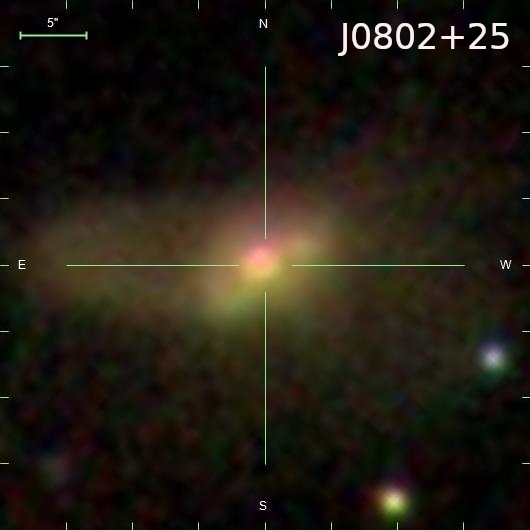}
    \includegraphics[width=0.3\linewidth]{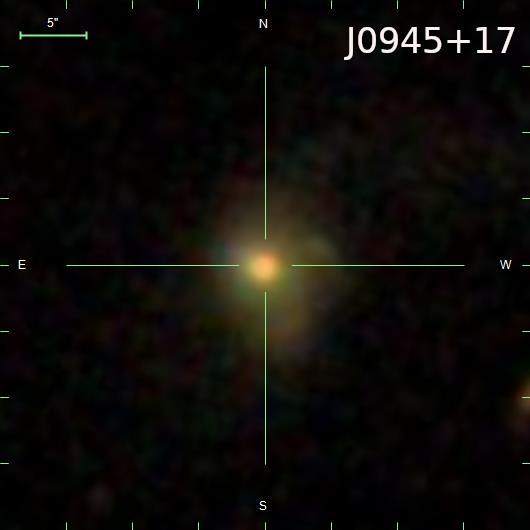}
    \includegraphics[width=0.3\linewidth]{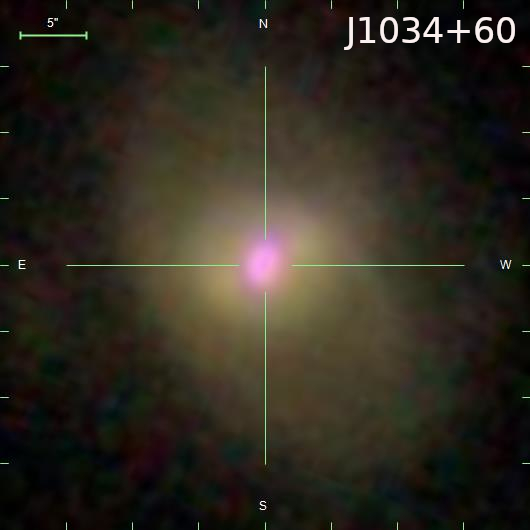} \\[1ex]  
    \includegraphics[width=0.3\linewidth]{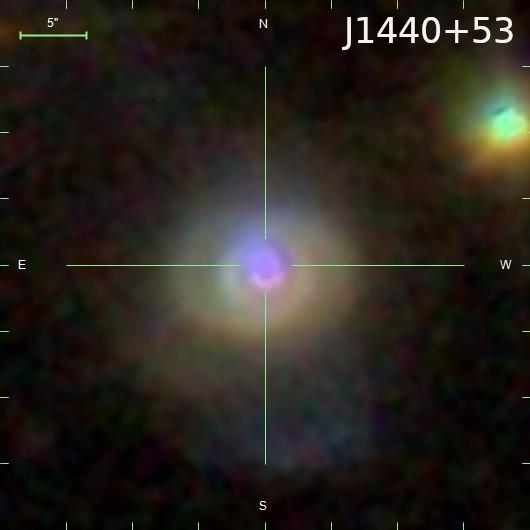}
    \includegraphics[width=0.3\linewidth]{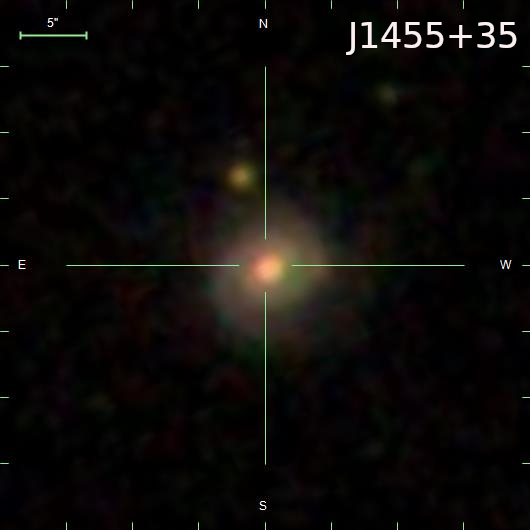}
    \includegraphics[width=0.3\linewidth]{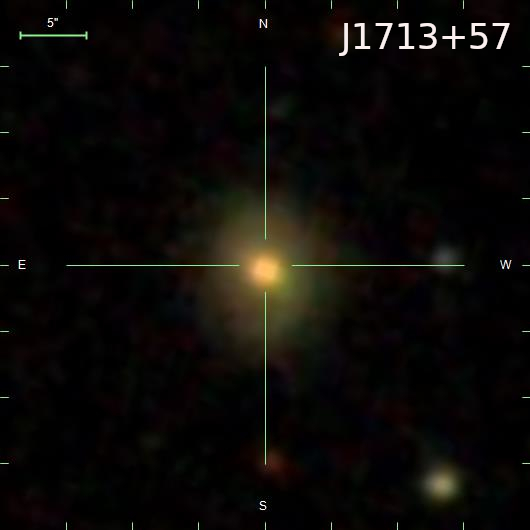}
    \caption{Sloan Digital Sky Survey (SDSS) gri images of the QSO2s. North is up and east to the left. Horizontal bars in the top left corner of each image indicate the physical size of 5\arcsec. Images size is 40\arcsec$\times$40\arcsec.} 
    \label{fig:1}
\end{figure}

Forty QSO2s from the QSOFEED sample have been observed  in the K-band with the Espectrógrafo Multiobjeto Infra-Rojo (EMIR) at the 10.4m Gran Telescopio Canarias (GTC). From those we selected six QSO2s showing both recombination (either \pa~or \brg) and \hii emission lines, allowing a multiphase gas characterization (see Table \ref{tab:sample}). 
Composite gri images of the QSO2s are shown in Figure \ref{fig:1}. Four of the QSO2s are in the post-coalescence phase of a galaxy merger, J1440+53 is in the pre-coalescence phase, and J1713+57 is a seemingly undisturbed galaxy \citep{Pierce+23}.

\section{GTC / EMIR Observations and data reduction}
\label{sec:observations}

The six QSO2s were observed with the NIR multi-slit spectrograph EMIR \citep{Garzon+06,Garzon+14,Garzon+22} installed at the Naysmith-A focal station of the GTC at the Roque de los Muchachos Observatory, in La Palma. EMIR has a $\rm 2048\times2048$ Teledyne HAWAII-2 HgCdTe NIR-optimized chip with a pixel size of 0.2\arcsec. The observations were carried out in service mode between June 2018 and August 2019, as part of proposals GTC77-18A, GTC62-18B, and GTCMULTIPLE2G-19A (PI: Ramos Almeida). Observing conditions were either clear or spectroscopic. Previous observations taken as part of these programs were published in \cite{RamosAlmeida+19} and \cite{Speranza+22}. All spectra were taken with the K-grism, with a nominal dispersion 1.71 $\rm \mathring{A}$/pix and resolving power $\rm\lambda/\delta \lambda$ = 4000. The long slit used has a width of 0.8$\arcsec$. The instrumental width estimated from ArHg, Ne, and Xe arc lamps was $\sim$6.0 $\mathring{A}$ ($\sim$82 km~s$^{-1}$). EMIR has a Configurable Slit Unit (CSU) that allows the user to configure and position slits over the  4'$\rm\times$6.67' spectroscopic field of view (FOV). Therefore, depending on the redshift of our targets, we positioned the 0.8\arcsec~slit either in the center of the FOV or shifted to the left/right to ensure that the wavelength coverage included the emission lines of interest. Before each QSO2 observation, J-band acquisition images were taken for slit positioning and seeing-determination, by measuring the FWHM of several stars in the FOV. The spectra were then taken following a nodding pattern ABBA, with an offset of 30\arcsec, for improved sky subtraction. Standard star K-band spectroscopic observations followed each QSO2 (with an exposure time of 4 x 60 s = 240 s), for both flux calibration and telluric absorption correction. Table \ref{tab:obs} summarizes the main details of the observations.     

The data reduction was done as follows. First, two-dimensional frames were flat fielded, using blue and red flats with individual exposures of 1.5 s and 6 s, to improve the correction in the blue and red part of the spectra. Bad pixels were masked using the IRAF \citep{Tody+86,Tody+93} task {\it ccdmask}. The wavelength calibration was done using HgAr, Ne, and Xe arc lamps spectra, and the IRAF tasks {\it identify} and  {\it reidentify}. Consecutive pairs of AB and BA frames were subtracted to remove the sky background, and then combined with the IRAF package {\it lirisdr} to obtain the two-dimensional spectra. 
We then extracted one-dimensional nuclear spectra of each QSO2, centred at the peak of the continuum emission and using the apertures listed on the last column of Table \ref{tab:obs} using the IRAF package {\it apextract}. These apertures were chosen based on the seeing FWHM measured from stars present in the J-band acquisition images of the QSO2s, which range between 0.8 and 1.4\arcsec. At the redshift of the QSO2s, this corresponds to the central 1.0-3.2 kpc of the galaxies. For J1713+57, the nuclear spectrum is the combination of the two one-dimensional spectra obtained from the data of two different nights (see Table \ref{tab:obs}).
Flux calibration was performed by fitting the continuum of the standard star observed just after each QSO2, and comparing them to their known integrated magnitude in the K-band, making use of IRAF tasks {\it standard, sensfunc}, and {\it calibrate}. Telluric absorption correction was done using the standard star spectrum and the IRAF task {\it telluric}, although for some of the targets 
the sky transmission was modeled instead using the ESO tool SKYCALC \citep{Noll+12,Jones+13,Moehler+14}. The nuclear spectra of the six QSO2s are shown in Fig. \ref{fig:fig2_allspectra}.

\input{Tables/table_observations}

\section{Methodology and results}
\label{sec:Results}

\subsection{Nuclear spectra}
\label{sec:nuclear_spectra}

\begin{figure*}
    \centering
    \includegraphics[width=0.65\linewidth]{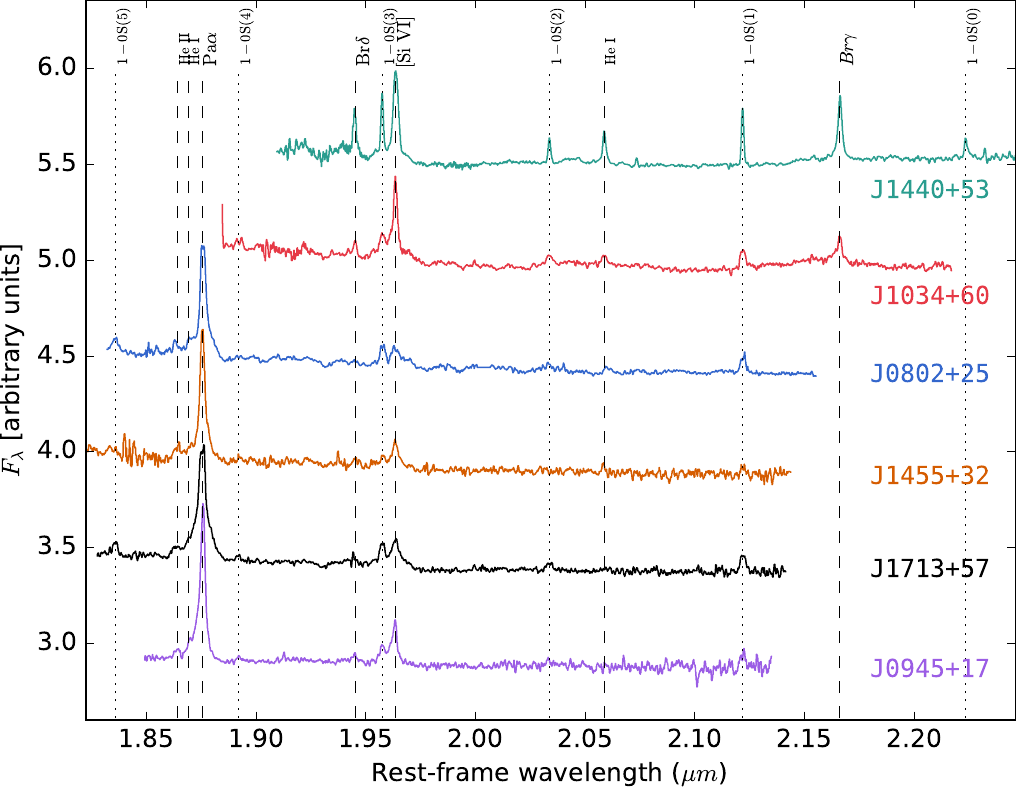}
    \caption{EMIR K-band nuclear spectra of the QSO2s showing the emission lines detected. The spectra were scaled in the Y-axis with the purpose of better visualization and smoothed using a 10 pixels boxcar. 
    The spectra of the QSO2s are displayed from top to bottom in order of increasing redshift. 
    Black dashed vertical lines indicate the position of the atomic lines and dotted lines the \hii lines.}
    \label{fig:fig2_allspectra}
\end{figure*}

The nuclear spectra of the QSO2s show emission lines tracing different gas phases, from low- to high-ionized gas and warm molecular gas, as shown in Fig. \ref{fig:fig2_allspectra}. As can be seen from Fig. \ref{fig:fig2_allspectra}, the spectra of the two QSO2s with the lowest redshifts, J1440+53 and J1034+60 (z = 0.04 and 0.05 respectively) do not include the Pa$\alpha$ line but Br$\gamma$, and in the case of J1440+53, also \hii 1-0S(0). The other four QSO2s, which have redshifts between 0.08 and 0.13, include \pa. 
Helium lines, such as He II$\lambda1.8637 \mu m$, He I$\rm \lambda1.8691 \mu m$, and He I$\rm \lambda2.0587 \mu m$ are also detected for some sources. Furthermore, the coronal emission line of $\rm [Si~VI]\lambda1.963\mu m$ with ionization potential $\rm \chi_e$ = 167 eV \citep{Oliva+90,RamosAlmeida+06,Mazzalay+13} is ubiquitous in our sample and it is the second most prominent emission line in the spectra of the QSO2s, after \pa. All QSO2s present H$_2$ emission lines in their spectra, which trace molecular gas at $\rm T>1000~K$. We detect \hii lines from 1-0S(1) to S(4) in all the QSO2s, also S(0) in J1440+53, and S(5) in J0802+25, J1455+35, and J1713+57. 
 


\begin{figure*}
    \centering
    \includegraphics[width=0.4\linewidth]{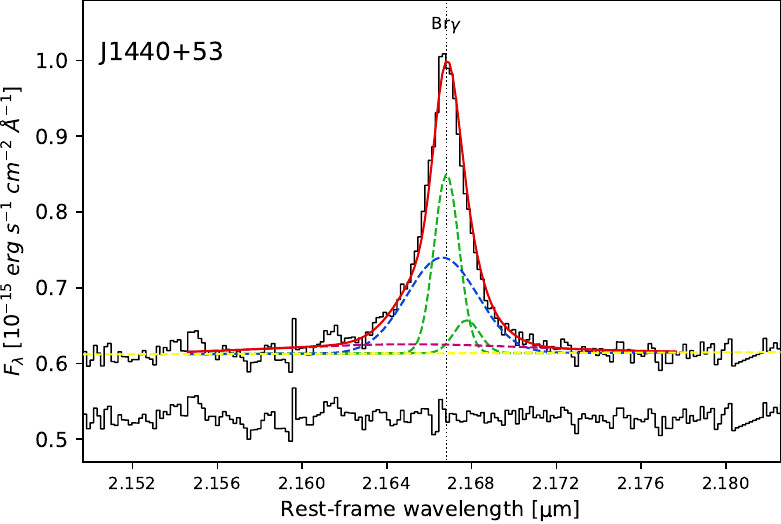}
    \includegraphics[width=0.4\linewidth]{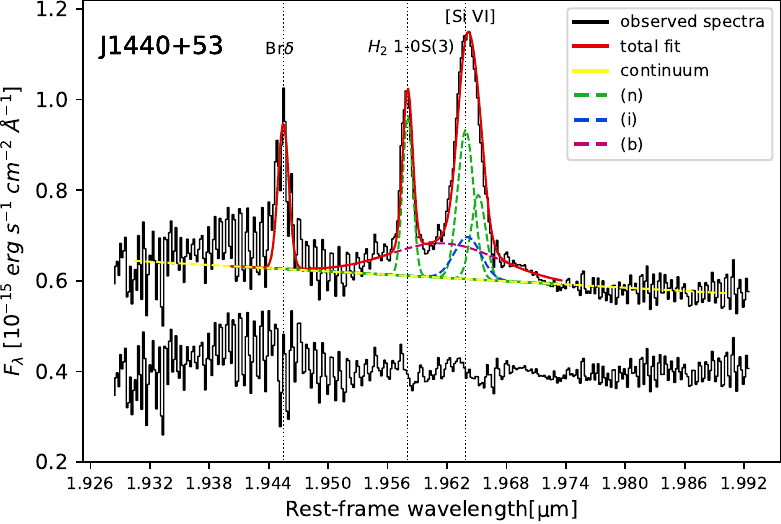}
    \caption{Examples of emission line fits. The rest-frame spectrum of J1440+53 and residuals are shown in black and the fitted model in red. Narrow (n), intermediate (i), and broad (b) components are shown in green, blue, and magenta. 
    The spectra were smoothed using a 4 pixels boxcar and residuals were scaled up from zero to reduce blank space. Vertical dotted lines correspond to the peak of the narrow component fitted to each line.} 
    \label{fig:3}
\end{figure*}

To quantify the ionized and warm molecular gas kinematics and fluxes we fitted the emission line profiles using Gaussians, as shown in Fig. \ref{fig:3} for J1440+53. We fitted a first-order polynomial function to two continuum bands blue- and red-wards of the emission lines. Then we modelled the emission line profiles with single or multiple Gaussian components using an in-house developed code \citep{Speranza+22,Speranza+24,Musiimenta+24} that makes use of the \texttt{Astropy} python library \citep{Astropy+22}. The fits are shown in Figs. \ref{fig:3} and \ref{fig:fit_J0802}-\ref{fig:fit_J1713}. Tables \ref{tab:f_J0802+25}-\ref{tab:f_J1713+57} list the FWHM, velocity shift, flux, and flux fraction of each of the Gaussian components fitted to the emission lines. The velocity shift is relative to the central wavelength of the narrow component fitted to the Pa$\alpha$/Br$\gamma$ profiles. When two narrow components are fitted 
instead, the velocity shifts are relative to the amplitude weighted value of the central wavelengths of the two narrow components. The flux uncertainties are the quadratic sum of the fit uncertainty obtained from Monte Carlo simulations and the flux calibration error estimated using the standard stars (average flux calibration error of 26\%). In addition, since the emission lines were measured in the rest frame, a multiplicative factor of (1+z) was applied to the fluxes. 



Since some of the NIR emission lines present complex kinematics, are blended (e.g., [Si VI] and H$_2$ 1-0S(3), and/or they have relatively low signal-to-noise), we performed optical emission line fits using the available optical Sloan Digital Sky Survey (SDSS, \citealt{York+00,Abazajian+09}) spectra of the QSO2s to use them as reference for the NIR fits, following \cite{RamosAlmeida+19} and \cite{Speranza+22}. 
The criteria to decide the number of Gaussian components needed to fit the optical lines was 1) an improvement of $\ge$10\% in the reduced $\rm \chi^2$, following \cite{Bessiere+24} and \cite{Speranza+24}, and 2) line fluxes representing $\ge$10\% of the total emission line flux. We then fitted the NIR lines using the same number of Gaussian components as in the optical, and the optical kinematics as initial values. \hb was often used as a reference for \pa, \brg, He I, and He II since they must follow similar kinematics. For Br$\delta$, either Pa$\alpha$/Br$\gamma$ or \hb were considered, and [OIII]$\rm\lambda$5007$\rm\mathring{A}$ was used as reference for [Si VI]. For fitting \hii 1-0S(3), that is blended with [Si VI], the fit of \hii 1-0S(1) was taken as reference. Exceptions to the previous are described in Tables \ref{tab:f_J0802+25}-\ref{tab:f_J1713+57}.

\input{Tables/Table_kinematics}

For the warm molecular lines that are not blended, no initial parameters were considered. In all cases, only one Gaussian component of FWHM$\sim$120-460 \kms was enough to reproduce the \hii line profiles. The velocity shifts that we measured for these components, relative to the narrow component(s) fitted to either \pa~or \brg, go from -130 to 60 km~s$^{-1}$. 
In the case of the atomic lines (i.e., either \pa~or \brg, and \sivi), we fitted narrow components of FWHM $<$ 600 \kms~(see Table \ref{tab:kinematic}). 
Two narrow components instead of one were needed to reproduce the profiles of J1034+60, J1440+53, and J1713+57. 
For all QSO2s but J1034+60, we also fitted intermediate components. These components are $\ge$160 \kms broader than the narrowest component fitted to a given QSO2, and narrower than the broadest component (see Table \ref{tab:kinematic}). 
Finally, broad components (b) with FWHM $\rm\ge$ 1200 \kms~and showing blueshifts of up to -600 \kms~were fitted to all the QSO2s. The parameters derived from the individual fits are shown in Tables \ref{tab:f_J0802+25}-\ref{tab:f_J1713+57}. 


In this work we assume that the narrow components, which exhibit velocities typical of galactic rotation, are tracing gas within the Narrow Line Region (NLR). Since the intermediate and broad components exhibit velocities and widths that cannot be explained by rotation alone, and considering that these QSO2s have ionized outflows detected in [O III]\footnote{QSO2s with [OIII] outflows are those having a W80 larger than the W80 measured for the stellar component \citep{Bessiere+24}.} \citep{Bessiere+24}, we consider them to be associated with outflowing gas. 
It is important to stress that these broad components are not related to the broad line region (BLR) of the AGN since this analysis is restricted to a sample of obscured (type-2) AGN, and the intermediate and broad components are detected in both the permitted and forbidden emission lines. The fraction of flux in the outflow components (i.e., adding the flux of the intermediate and broad components, when both are present) is larger than that in the narrow component, with the exception of J0802+25. This is common in the case of QSO2s \citep{RamosAlmeida+17,RamosAlmeida+19,Speranza+22,Hervella+23,VillarMartin+23,Zanchettin+25}. 



\subsection{Outflow properties}
\label{sec:outflow_extent}

From the analysis described in Section \ref{sec:nuclear_spectra} and following our criteria to identify outflows, we find ionized outflows in all six QSO2s, both in the recombination lines and [Si VI]. Outflow components are not found in the \hii emission lines. The following sections are focused on estimating the corresponding outflow properties, to infer their energetic output, whether there is a connection between the outflows found at different ionization levels, and their potential impact on the host galaxies. As previously mentioned, components (i) and (b) are considered outflow, and the derived outflow properties take both components into account. To estimate the outflow masses ($\rm M_{out}$), mass outflow rates ($\rm \dot{M}_{out}$), and kinetic powers ($\rm \dot{E}_{kin}$), we need the fluxes and kinematics measured from the nuclear spectra and reported in Tables \ref{tab:f_J0802+25}-\ref{tab:f_J1713+57}, electron densities ($\rm n_e$), and outflow extents ($\rm r_{out}$).  

\subsubsection{Electron densities}
\label{sec:electron_density}

Electron density ($\rm n_e$) and temperature ($\rm T_e$) are important indicators of the physical conditions of the ionized gas, and the former is essential to determine its mass. Moreover, $\rm n_e$ is one of the largest sources of uncertainty for ionized gas masses, potentially leading to 
orders of magnitude uncertainties on measured values \citep{Harrison+18,Rose+18,Davies+20,Holden+23a,Holden+23b,Speranza+24}. Here we estimate $\rm n_e$ using two distinct methods.     

The first method uses the $\rm n_e$-sensitive [SII]$\lambda\lambda$6716,6731$\rm \mathring{A}$ doublet in combination with the $\rm T_e$-sensitive lines of [OIII]$\lambda$4363$\rm\mathring{A}$ and [OIII]$\lambda$5007$\rm\mathring{A}$ \citep{Osterbrock+06}. This method (hereafter [SII] method) is sensitive to electron densities in the range $\rm 10^{2} \le log~n_e \le 10^{3.5}~cm^{-3}$ \citep{Rose+18}. The emission line fluxes of the [SII] and [OIII] lines were measured from the stellar continuum-subtracted SDSS spectra of the QSO2s from \cite{Bessiere+24}, adopting one Gaussian component to reproduce each emission line. Since [OIII]$\lambda$4363$\rm \mathring{A}$ is faint and blended with the much brighter H$\gamma$ line, we fixed its FWHM to the one obtained for [OIII]$\lambda$5007$\mathring{A}$. We then used the emission line ratios as inputs to the python package \texttt{Pyneb} (version 1.1.19; \citealt{Luridiana+15}), obtaining $\rm n_e$ and $\rm T_e$ using the function {\it getCrossTemDen} and the errors from the shaded regions of the diagnostic diagrams generated in {\it Diagnostic}, shown in Fig. \ref{fig:A5}. We obtain 2.6 $\le \rm{log~n_e~(cm^{-3})}\le$ 3.1 and $\rm 12000~K\le T_e\le~17000~K$ (see Table \ref{tab:density}). We note that the line fluxes used to infer $\rm n_e$ and $\rm T_e$ were not corrected for extinction. On average, the impact of applying the extinction correction to the fluxes to derive $\rm T_e$ is $\rm\sim1000-3000~K$.  For $\rm n_e$ the difference is much smaller ($\rm\sim15~cm^{-3}$) than the measurement errors, since $\rm n_e$ is only weakly dependent on $\rm T_e$, as shown in the diagnostic diagrams shown in Fig. \ref{fig:A5}.


The second method consists of using two line ratios that involve trans-auroral lines, namely 
TR([OII]) = F(3726+3729) / F(7319+7331) and TR([SII]) = F(4068+4076)/F(6716+6731) (hereafter TR method). The [SII] and [OII] line ratios are then compared to a grid of photoionization models generated with \texttt{Cloudy} \citep{Ferland+13}, following the procedure described in \cite{Holt+11}, which allows the simultaneous determination of reddening and $\rm n_e$. This method is sensitive up to much higher $\rm n_e$ than the [SII] method, in the range 2.0 $\le \rm{log~n_e~(cm^{-3})}\le$ 6.5 \citep{Rose+18}. More details can be found in Sec. 2.4 of \cite{Bessiere+24}, from where the corresponding values of $\rm n_e$ were taken. This second method results in electron densities of 3.0 $\le \rm{log~n_e~(cm^{-3})}\le$ 4.1 and reddening values of $\rm 0.2~mag\le E(B-V)\le 0.4~mag$ (see Table \ref{tab:density}). Therefore, except in the case of J1034+60, for which we get $\rm n_e$ values consistent within the errors using both methods, the densities measured from the TR method are significantly higher ($\rm n_e^{TR} \sim$6-15$\rm \times n_e^{[SII]}$) than the ones calculated from the [SII] method.

\input{Tables/Table_ne}



\subsubsection{Outflow extent}
\label{sec:outflow_ext}


To measure the spatial extent of the outflows, we followed the methodology described in \cite{Rose+18} and \cite{RamosAlmeida+19}. We built spatial profiles of the blue and red wings of Pa$\alpha$/Br$\gamma$ and [Si VI] (blue and red shaded areas in the upper panels of Figs. \ref{fig:out_ext_J0802}-\ref{fig:out_ext_J1713}), avoiding the region covered by the fitted narrow component(s) and any other adjacent emission lines. Then we produced an average continuum spatial profile from spatial slices blueward and redward of the broad emission line, and subtracted it from the blue and red broad line profiles. We finally averaged these continuum-subtracted blue and red wing spatial profiles and fitted the resulting profile with a Gaussian to measure the outflow radial size (FWHM$\rm_{out}$). We also calculate the FWHM of the spatial profile of the narrow component (FWHM$\rm_{nar}$) in the same way (green shaded areas in the upper panels of Figs. \ref{fig:out_ext_J0802}-\ref{fig:out_ext_J1713}). In order to inspect the regions where the broad wings, narrow component, and continuum profiles were extracted, we produce continuum-subtracted maps of the emission lines, shown in the middle panels of Figs. \ref{fig:out_ext_J0802}-\ref{fig:out_ext_J1713}.

Finally, to investigate whether the outflows are spatially resolved we determined the seeing FWHM (FWHM$\rm _{seeing}$) from the K-band spectrum of the corresponding standard star of each QSO2, using $\lambda$>21000 $\rm \mathring{A}$ (to avoid telluric absorption) to extract the continuum profile and fit it with a Gaussian. 
We consider an outflow resolved if its continuum-subtracted profile FWHM is:
\begin{equation}
\rm     FWHM_{out} > FWHM_{seeing} + 3\delta_{seeing},
    \label{eq1}
\end{equation}
where $\rm FWHM_{seeing}$ and $\rm \delta_ {seeing}$ are the seeing FWHM and its corresponding error derived from the standard star spatial profile. If we find the outflow to be unresolved, we adopt FWHM$_{seeing}$ as an upper limit for the outflow extent (r$\rm _{out}$). If it is resolved, then we compute the size as: 

\begin{equation}
\rm    r_{out} = \sqrt{\rm FWHM_{out}^2-FWHM_{seeing}^2}\mathpunct{\kern0.2em.}
\label{eq2}
\end{equation}

Table \ref{tab:extent} shows the seeing FWHM derived from the K-band spectra of the standard stars (FWHM$\rm _{seeing}$), the FWHM of the outflow (FWHM$\rm _{out}$) and narrow (FWHM$\rm _{nar}$) components, and the seeing-deconvolved values (r$\rm _{out}$ and r$\rm _{nar}$). The outflow errors were estimated by adding in quadrature the error propagation of Equation \ref{eq2} and the standard deviation of five computations of $\rm FWHM_{out}$,  slightly varying the green, red, blue, and continuum regions shown in Figs. \ref{fig:out_ext_J0802}-\ref{fig:out_ext_J1713}. From Pa$\rm \alpha$ and Br$\rm \gamma$ we measured outflow extents ranging from $\rm 0.3~kpc\le r_{out} \le 2.1~kpc$. In [Si VI] the outflow is not resolved for J1455+35, J0802+25, and J1440+53. For the other three QSO2s, the [Si VI] outflows have extents of $\rm 1.0~kpc\le r_{out} \le 2.7~kpc$. The outflow extent found for J0945+17 in Pa$\alpha$ is 2.13 kpc, which is smaller than 3.37 kpc reported by \cite{Speranza+22} based on integral field data from Gemini/NIFS. This is because the latter authors measured the outflow extent from 2D outflow flux maps, which are sensitive to fainter structures and probe different position angles, unlike our long-slit data. In fact, \cite{Speranza+22} reported an outflow $\rm PA\sim125^{\circ}$, where its extent is maximum, and our slit PA is $\rm 25^{\circ}$.
Finally, for the narrow component we measure extents which are larger or comparable to the outflow extents within the errors (see Table \ref{tab:extent}).

 

\input{Tables/Table_extent_23Jun25}

\subsubsection{Outflow energetics} 
\label{sec:outflow_energy}

For computing the ionized gas mass in the outflow from the recombination lines, we used either the Pa$\alpha$ or Br$\gamma$ extinction corrected fluxes of their intermediate and broad components to calculate their luminosities ($\rm L = 4\pi D_L^2 F^{corr}$) and then convert them to H$\beta$ luminosity, assuming Case B recombination and T$_{\rm e}$ = $10^{4}$ K ($\rm L_{H\beta} = L_{Pa\alpha}/0.352$ and  $\rm L_{H\beta} = L_{Br\gamma}/0.0281$; \citealt{Osterbrock+06}). We performed the extinction correction using the E(B-V) values reported in Table \ref{tab:density}, and assumed $\rm R_V = 3.1$ and the extinction law from \cite{Cardelli+89}. Then, following Equation 1 from \cite{Rose+18}:  
\begin{equation}
\rm  M_{Hion} = \frac{L_{H\beta} m_p}{\alpha^{eff}_{H\beta}h\nu_{H\beta}n_e} \raisebox{0.5ex}{,}
\label{eq3}
\end{equation}
where $\rm m_p = 8.41 \,{\rm \times}\, 10^{-58} M_{\odot}$, $\rm \alpha^{eff}_{H\beta} = 3.03 \,{\rm \times}\, 10^{-14} cm^{3} \, s^{-1}$, $\rm \nu_{H\beta} = 6.167 \,{\rm \times}\, 10^{14} s^{-1}$, with $\rm h = 6.626 \,{\rm \times}\, 10^{-34} Js$, we calculated the total ionized gas mass in the outflow.




For estimating the outflow mass of the coronal [Si VI] emission line, we can describe its luminosity as: 
\begin{equation}
\rm L_{[Si~VI]} = \int_V f \,n_e \,n(Si^{5+}) \,j_{[Si~VI]}(n_e,T_e) \,dV,   
\end{equation}
where f is the filling factor, $\rm n(Si^{5+})$ the density of $\rm Si^{5+}$, and $\rm j_{[Si~VI]}$ its emissivity that is a function of $\rm n_e$ and $\rm T_e$. The $\rm n(Si^{5+})$ can be also defined as:
\begin{equation}
\rm n(Si^{5+}) = \left[\frac{n(Si^{5+})}{n(Si)}\right]\left[\frac{n(Si)}{n(H)}\right]\left[\frac{n(H)}{n_e}\right]n_e.      
\end{equation}

Following \cite{Carniani+15} and \cite{Belli+24}, we assume that the $\rm Si^{5+}$ is dominant over neutral $\rm n(Si^{5+})/n(Si)$ = 1. As in \cite{Carniani+15} we can assume $\rm n(H)/n_e = (1.2)^{-1}$, as $\rm n_e \approx n(H) + 2\times n(He) = n(H) + 2 \times 0.1\times n(H) = 1.2\times n(H)$ when we consider 10\% of helium atoms. Then, using Equation 3 in \cite{Carniani+15}, we can write:
\begin{equation}
\rm L_{[Si~VI]} = (1.2)^{-1} \, j_{[Si~VI]}(n_e,T_e) \,\langle n_e^2 \rangle \, \left(\frac{n(Si)}{n(H)}\right)_{\odot} 10^{[Si/H]-[Si/H]_{\odot}} \, f \,V,  
\label{3}
\end{equation}
and based on Equation 4 of the same paper:
\begin{equation}
\rm M_{[Si~VI]} \approx 1.06 \, m_p \,\langle n_e \rangle \, f \, V.   
\label{4}
\end{equation}

Connecting Equations \ref{3} and \ref{4} using $\rm f \, V$, assuming that $\rm \langle n_e^2 \rangle = \langle n_e \rangle^2$, we obtain:
\begin{equation}
\rm  M_{[Si~VI]} \approx \frac{1.4 \, L_{[Si~VI]} \, m_p}{n_e \, j_{[Si~VI]} \, \left(\frac{n(Si)}{n(H)}\right)_{\odot} 10^{[Si/H]-[Si/H]_{\odot}}}\raisebox{0.5ex}{.}
\label{5}
\end{equation}

Using Chianti IDL 10.1 \citep{Dere+97,DelZanna+20} we obtain the emissivities (dividing EMISS\_CALC output by $\rm n_e$) for the mean $\rm [SII]$ and TR methods electron densities found in the sample: $\rm j_{[Si~VI]}(T_e = 10^4~K, n_e = 752~cm^{-3}) = 2.4772~\times~10^{-21}~erg~s^{-1} \,cm^{3}$ and $\rm j_{[Si~VI]}(T_e = 10^4~K, n_e = 5741~cm^{-3}) = 2.4828~\times~10^{-21}~erg \,s^{-1} \,cm^{3}$. Since there is no significant variation in the derived emissivities, we adopted the value of $\rm j_{[Si~VI]} =$ 2.48$\rm ~\times~10^{-21}~erg \,s^{-1} \,cm^{3}$. Furthermore, we adopted the solar Si abundance\footnote{Since Si is a refractory element, the assumed Solar abundances imply that a significant fraction of ISM dust has been destroyed, perhaps as a consequence of the outflow acceleration process (\citealt{McKaig+24}, \textcolor{blue}{Holden \& Tadhunter in prep.}).} from \cite{Asplund+21} as log($\rm \epsilon_{Si}$) = 7.51 and considering that $\rm log(\epsilon_{Si}) = log\left(\left[\frac{n(Si)}{n(H)}\right]\right) + 12$, we find $\rm \left[\frac{n(Si)}{n(H)}\right]$ = 3.2359$\rm ~\times~10^{-5}$. Assuming the values of emissivity and solar Si abundance described above, and $\rm m_p = 8.41 \times 10^{-58} \, M_{\odot}$ 
, we obtain the [Si VI] mass as:
\begin{equation}
\rm M_{[Si~VI]} = 1.415\, {\rm \times} \,10^{-32} \, \frac{L_{[Si~VI]}}{n_e} \, M_{\odot}.
\label{6}
\end{equation}

The corresponding outflow masses calculated from Pa$\alpha$/Br$\gamma$ and [Si VI] are reported in Table \ref{tab:outfenergetics}. We find ionized gas masses of $\rm M_{Hion} = 0.08-20 \times 10^{6}~M_{\odot}$ from the recombination lines and of $\rm M_{[Si~VI]} = 0.02-2 \times 10^{6}~M_{\odot}$ from [Si~VI] when assuming TR-method electron densities. When assuming [SII]-method electrons densities we obtain ionized gas masses of $\rm M_{Hion} = 0.5-33 \times 10^{6}~M_{\odot}$ from the recombination lines and of $\rm M_{[Si~VI]} = 0.1-4 \times 10^{6}~M_{\odot}$ from [Si~VI]. Then, to compute the mass outflow rates, we assumed spherical geometry for the outflow, following \cite{Fiore+17}:  
\begin{equation}
\rm \dot{M}_{out} = 3 \, v_{out} \, \frac{M_{out}}{r_{out}},
\end{equation}
\noindent
and we follow the kinetic power definition from \cite{Hervella+23}:
\begin{equation}
\rm \dot{E}_{kin} = \frac{\dot{M}_{out}}{2} \times v_{out}^2.
\end{equation}

From these properties, we have also calculated the coupling efficiency as $\rm \xi = \dot{E}_{kin}/L_{Bol}$ and the mass loading factor $\rm \eta = \dot{M}_{out}/SFR$. Since the measured velocity shifts ($\rm v_s$) are projected velocities, besides using $\rm v_{out} = v_s$ to infer the outflow properties, we also computed their upper limits
adopting the maximum outflow velocity $\rm v_{max} = |v_{s}| + 2\sigma$ (i.e., to account for projection effects, we measured the velocity in the wing of the line, under the assumption that this corresponds to the velocity of the gas moving directly towards us; \citealt{Rupke+13b,Fiore+17,Speranza+22}), where $\rm \sigma \sim FWHM/2.355$. Table \ref{tab:outflowprop} shows the observed outflow properties (e.g., $\rm n_e$, $\rm r_{out}$, $\rm v_{s}$, $\rm v_{max}$) directly measured from the spectra, while Table \ref{tab:outfenergetics} summarises the derived outflow properties (e.g., $\rm M_{out}$, $\rm \dot{M}_{out}$, $\rm \dot{E}_{kin}$, $\rm \xi$, $\rm \eta$) using the electron densities estimated with the two methods here considered. The properties have been calculated separately for each outflow component (i.e., b and i, if present) and then added. Assuming the TR-method electron densities we estimate mass outflow rates of $\rm \dot{M}_{Hion} \sim 0.03-6 ~M_{\odot}~yr^{-1}$ from the recombination lines and of $\rm \dot{M}_{ [Si ~VI]} \sim 0.004-1 ~M_{\odot}~yr^{-1}$ from [Si VI]. Their kinetic powers are $\rm \dot{E}_{Hion} \sim 10^{37.8-40.8}~erg~s^{-1}$ for recombination lines and $\rm \dot{E}_{[Si~VI]} \sim 10^{36.6-40.5}~erg~s^{-1}$ for [Si VI]. If instead we assume [SII]-method electron densities the recombination lines mass outflow rates are $\rm \dot{M}_{Hion} \sim  0.2-10~M_{\odot}~yr^{-1}$ and from [Si VI] of $\rm \dot{M}_{[Si~VI]} \sim 0.02-2~M_{\odot}~yr^{-1}$. The corresponding kinetic powers are $\rm \dot{E}_{Hion} \sim 10^{38.6-41.6}~erg~s^{-1}$ for recombination lines and $\rm \dot{E}_{[Si~VI]} \sim 10^{37.4-41.3}~erg~s^{-1}$ for [Si VI].   





\section{Discussion}
\label{sec:discussion}

Here we discuss the results of our study of the ionized and warm molecular gas properties of six nearby QSO2s. In Sections \ref{disc:ionized} and \ref{disc:coronal} we focus on the ionized gas outflow properties derived using TR-based electron densities instead of the [SII]-based ones, as the findings of recent studies indicate that ionized outflows exhibit higher electron densities than non-outflowing gas \citep{Speranza+22,Holden+23a}. Since the TR lines are sensitive to higher densities than [SII], they are more appropriate for calculating
the properties of ionized outflows (see \citealt{Holden+25}). In addition, the electron densities derived from the mid-infrared coronal lines of [NeV] detected in JWST observations of five QSO2s in the QSOFEED sample are in good agreement with those obtained from the TR-method and higher than those from the [SII]-method \citep{RamosAlmeida+25}. 


\subsection{Energetics of the ionized outflows in QSO2s}\label{disc:ionized} 

\input{Tables/Table_comparison}

Here we have characterized the ionized gas outflows of the QSO2s by means of their Pa$\alpha$ or Br$\gamma$ emission lines, which are less affected by obscuration than the optical emission lines. 
To investigate whether these NIR emission lines provide different outflow properties than those obtained from optical lines (e.g., H$\beta$ and [OIII]), in Table \ref{tab:NIROPCOMP} we compare them with those measured for the same six QSO2s and for the whole QSOFEED sample of 48 QSO2s, but using optical SDSS spectra. For these QSO2s,  
\cite{Bessiere+24} measured the ionized outflow properties using a non-parametric analysis of the [OIII]$\lambda$5007 \AA~emission line, considering TR-based densities, and assuming an outflow extent of $r_{\rm out}$=0.62 kpc. The [OIII] masses were multiplied by three in order to estimate the total ionized gas masses ($\rm M_{Hion} \sim 3\times M_{[OIII]}$; \citealt{Fiore+17}). 

Focusing first on the ionized outflow properties of the six QSO2s studied here, we measure higher outflow masses from the NIR recombination lines, with median $\rm log(M_{Hion}[M_{\odot}])$ = 5.81, than \citet{Bessiere+24}, who reported median $\rm log(M_{Hion}[M_{\odot}])$ = 4.91 for the same targets (i.e., 7.9 
times lower than our median measurement; see Table \ref{tab:NIROPCOMP}). The largest difference that we found is between the NIR and optical outflow masses measured for J1034+60 ($\rm log( M_{Hion}[M_{\odot}])$ = 7.30 here versus $\rm log(M_{Hion}[M_{\odot}])$ = 5.59 in \citealt{Bessiere+24}). This is due to the large integrated flux of the broad component fitted to Br$\gamma$ (78\% of the total flux in the line; see Table \ref{tab:f_J1034+60} and Fig. \ref{fig:fit_J1034}). Part of these differences can be accounted for by the different fitting techniques employed in the two works, i.e., parametric versus non-parametric. \citet{Hervella+23} reported parametric-based median 
$\rm M_{\rm Hion}$ values 
3.4 times 
 higher than the non-parametric ones measured for the same targets. This happens because the parametric analysis considers the integrated flux of the broad and intermediate components fitted to the line profiles, while non-parametric methods just use the integrated flux of the high-velocity wings of the lines. This, together with the relatively large flux calibration uncertainty of the NIR fluxes (see Section \ref{sec:nuclear_spectra}), explains part of the  difference between the optical and NIR masses of the same targets. Finally, it is also possible that the NIR data, less affected by extinction than optical spectra and probing a smaller physical region (seeing limited versus 3\arcsec~SDSS fiber size), allow us to peer deeper in the outflow regions, recovering part of the flux that remains undetected in the optical \citep{RamosAlmeida+17}. 

The NIR and optical mass outflow rates ($\rm \dot{M}_{Hion}$) are consistent within the errors, having median values of 0.6 and 0.35 M$\rm_{\odot}~yr^{-1}$ respectively. This happens because the higher NIR outflow masses 
are compensated by 1) the lower NIR outflow velocities that we measure from the parametric method, compared to those derived from the non-parametric analysis (see \citealt{Hervella+23} for a comparison of the outflow mass rates derived from different methods) and 2) by the larger outflow radii that we used as compared to the fix value of 0.62 kpc used by \citet{Bessiere+24}\footnote{Consequently, the corresponding mass loading factors ($\rm \eta_{Hion} =  \dot{M}_{Hion} $/SFR) are also similar between them (median values of 0.022 and 0.013 respectively).}. 

The lower NIR outflow velocities that we measured from our parametric fits also result in lower kinetic energies ($\rm \dot{E}_{Hion}$) and coupling efficiencies ($\rm \xi_{Hion}$) than those derived from optical data by \citet{Bessiere+24} (median values of $\rm log(\dot{E}_{Hion}[erg~s^{-1}])$ = 40.2 and $\rm \xi_{Hion}$ = 0.0004\% in the NIR and 41.1 and 0.003\% in the optical), as $\rm \dot{E}_{Hion}\propto v_{\rm out}^3$. \citet{Hervella+23} reported even larger differences between the kinetic energies derived from the parametric and non-parametric analysis of the same targets. 

Our results suggest that, at least for the QSO2s studied here, there are no significant variations between the NIR and optical outflow mass rates and kinetic energies if we account for the different methodologies used to characterize the physical outflow properties. However, the outflow masses that we measure in the NIR are 7.9 times higher than the optical ones, considering median values. This is partly due to the use of a parametric method to fit the NIR lines and to the flux calibration uncertainty of the NIR spectra, and possibly to the higher angular resolution and lower reddening of the NIR data as well, which allow us to probe deeper outflow regions. 
The outflow properties of the six QSO2s studied here are representative of the whole QSOFEED sample looking at the median values and ranges of the different outflow properties reported in Table \ref{tab:NIROPCOMP}.



We can also compare our outflow energetics with those obtained from the parametric analysis of the [OIII] line of 18 QSO2s at z = 0.3-0.41 observed with the Gemini South Telescope \citep{Hervella+23}. The latter authors also assumed $\rm M_{Hion} \sim 3~\times~M_{[OIII]}$. This sample of QSO2s have \loglbol = 44.9-46.7, with a median value of 45.5, coinciding with the median bolometric luminosity of our subset of six QSO2s. Their outflow extent assumption, of 1 kpc, agrees well with our measurements (median $\rm  r_{out}$=1.1~kpc). Since, they assumed $\rm n_e = 200~cm^{-3}$, we multiplied their outflow masses by that value of $\rm n_e$, and divided them by the median value of the QSOFEED sample measured with the TR-based method, which is $\rm n_e = 2570~cm^{-3}$ \citep{Bessiere+24}. We did this to make their ionized outflow properties comparable to ours. 
The outflow masses measured for the two samples have medians of log($\rm M_{Hion}[M_{\odot}]$) = 5.81 and 5.35 at z = 0.1 and z = 0.3-0.41, respectively (2.9 times higher in the case of the low-z QSO2s). The outflow mass rates and kinetic powers are also lower in the case of the QSO2s at z = 0.3-0.41, with medians of 0.14 \msunyr and $\rm log(\dot{E}_{Hion}[erg~s^{-1}])$ = 39.2, 
but they span similar ranges: $\rm log(\dot{E}_{Hion}[erg~s^{-1}])$ \apr 37-41~at z=0.3-0.41 and \apr 38-41 ~at z=0.1. 
Overall, this comparison suggests that there is no significant evolution in the outflow properties of QSO2s 
from redshift z = 0.1 and z = 0.3-0.41, albeit the samples are small.

Finally, we compared our outflow energetics with those of QSO2s at higher redshifts based on data of the JWST Near Infrared Spectrograph (NIRSpec).      
 \cite{Bertola+25} and \cite{Perna+25} reported outflow properties from \ha observations of four QSO2s at z$\rm \sim$2.05-3.51 and log L$_{\rm bol}$ = 45.2-46.2 (see Table \ref{tab:NIROPCOMP}). 
 Both studies used fixed electron density values and used different definitions of outflow velocity. In order to compare to our results, we corrected their outflow properties values assuming $\rm n_e = 2570~cm^{-3}$ and $\rm v_{out} = v_s$ (see Table \ref{tab:NIROPCOMP}). 
 The median values of log ${\rm M_{Hion}}$ and $\rm \dot{M}_{Hion}$ of the high-redshift QSO2s are 31 and 10 times higher than those of the six low-redshift QSO2s studied here,  
 but the median kinetic power is only a factor 1.3 higher (see Table \ref{tab:NIROPCOMP}). 
 It is also noteworthy that two of the high-z QSO2s have $\rm log(M_{Hion}[\rm M_{\odot}])$ = 5.8-6.4 and $\rm \dot{M}_{Hion}$ = 0.2-0.7 \msunyr, which are similar to the properties of the local QSO2s. Despite the small number of targets involved in this comparison, the outflow masses and mass outflow rates of QSO2s at z$\sim$2-3.5 are larger than those of local QSO2s, but their kinetic powers are comparable.

\subsubsection{Scaling relations}

\begin{figure}
    \centering
    \includegraphics[width=0.48\textwidth]{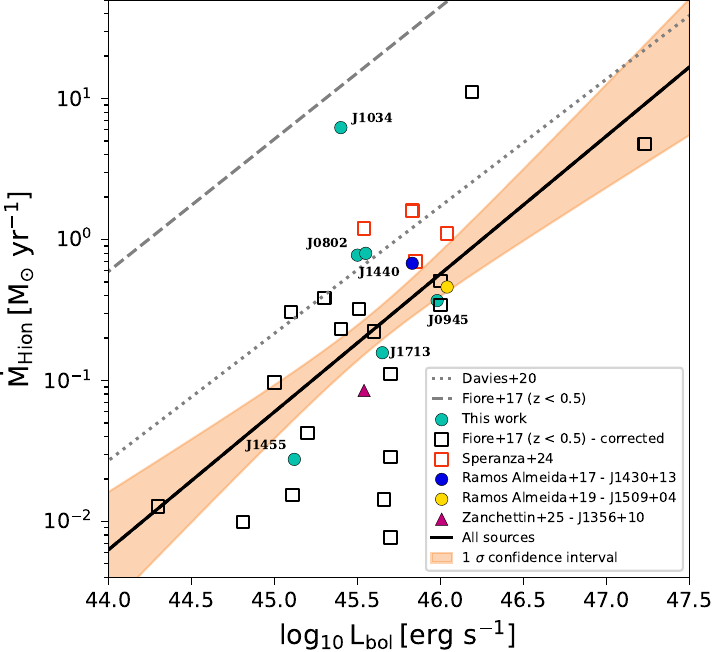}
    \caption{$\rm \dot{M}_{Hion}$ versus $\rm L_{bol}$. Turquoise points are our ionized mass outflow rates ($ \rm \dot{M}_{Hion}$), calculated from either Pa$\alpha$ or Br$\gamma$ and using TR-based electron densities. Data of other QSO2s are shown with different colors and symbols. Open squares are mass outflow rates from \citet{Fiore+17}, divided by 130 (see text). 
    The grey dashed and dotted lines are the scaling relations from \cite{Fiore+17} and \cite{Davies+20}, respectively, 
    and the black solid line is the linear fit to all the data points included in the plot. The orange shaded area represents the 1$\sigma$ confidence interval of the fit.} 
    \label{fig:8}
\end{figure}

In Fig. \ref{fig:8} we show
the relation between mass outflow rate and bolometric luminosity. Our data points correspond to the mass outflow rates derived from the hydrogen recombination lines (either Pa$\alpha$ or Br$\gamma$), using $\rm v_{out} = v_s$ and TR-based electron densities. In Fig. \ref{fig:8} we also include the mass outflow rates of AGN and ULIRGs at z<0.5 from \citet{Fiore+17}. These were calculated assuming n$\rm _e$=200 $\rm cm^{-3}$, $\rm v_{out} = v_{max}$, $\rm r_{out}$=1 kpc, and using either H$\alpha$, H$\beta$, or [OIII] (assuming $\rm  M_{Hion} \simeq M_{H\beta} \simeq 3~\times~M_{[OIII]}$). Therefore, we divided the \citet{Fiore+17} mass outflow rates by 130 (2.1 dex) to account for the differences in how their outflow velocity was measured relative to ours, and to avoid assuming an electron density based on [SII] measurements, which are sensitive to  $\rm n_e\sim100-3000~cm^{-3}$ (we refer the reader to \citealt{Speranza+24} and \citealt{Holden+25} for a discussion on the influence of electron density on the mass outflow rates). This value of 130 comes from first converting from $\rm v_{out} = v_{max}$ to $\rm v_{out} = v_s$, using the ratio of the median values reported in \citet{Hervella+23}, of $\rm \sim$ 10 (1.0 dex). We then converted from $\rm n_e = 200~cm^{-3}$ to $\rm n_e = 2570~cm^{-3}$, dividing the mass rates by $\rm \sim13$ (1.1 dex)\footnote{To do this we are assuming that the median TR-electron density of the QSOFEED sample is representative of the sample of ULIRGs and AGN in \citet{Fiore+17}, which might not be the case.}. 
After doing this, the values from \citet{Fiore+17} are similar to ours, showing the strong influence of methodology and assumptions on the outflow mass rate determination \citep{Hervella+23,Harrison+24}. We also included the mass outflow rates of another four QSO2s from the QSOFEED sample \citep{Speranza+24}. They were calculated from [OIII] observations obtained with the integral field unit of GTC/MEGARA, using a non-parametric analysis, and TR-based electron densities. Probably because of the non-parametric analysis 
and the better sensitivity and spatial coverage of the optical IFU data, their mass outflow rates are among the highest in Fig. \ref{fig:8}. 

We also calculated the mass outflow rates of another two QSOFEED QSO2s that were observed in the NIR using VLT/SINFONI observations: J1430+13 and J1356+10. For J1356+10 we used the Pa$\alpha$ fluxes and velocity shifts reported in \citet{Zanchettin+25}, and the TR-based electron densities and A$_V$ values reported in \citet{RamosAlmeida+25}. 
Since there are no outflow sizes reported for the Pa$\alpha$ outflows in \citet{Zanchettin+25}, we took the smaller [OIII] outflow radius reported by \citet{Speranza+24} for J1356+10. For J1430+13 we used the Pa$\alpha$ fluxes, velocity shifts, and outflow sizes from \cite{RamosAlmeida+17}, adopting the TR-based electron density and extinction from \cite{Bessiere+24}.    
Finally, we added the Pa$\alpha$ outflow mass rate reported by \citet{RamosAlmeida+19} for the QSO2 J1509+04, which was observed with GTC/EMIR and whose outflow properties were calculated using the same methodology as here. 
Fig. \ref{fig:8} shows that the QSO2s studied here occupy the same locus as the other QSO2s with NIR measurements. 
Considering all the data points shown in Fig. \ref{fig:8} with $\rm L_{bol}\sim 10^{45-46}~erg~s^{-1}$, the mass outflow rates are clustered between $\rm 0.01~M_{\odot}~yr^{-1}$ and $\rm 2~M_{\odot}~yr^{-1}$, with the exception of J1034+60. Finally, if we fit all the data points included in Fig. \ref{fig:8} we derive a scaling relation that is 2.6 times lower than that derived by \cite{Davies+20} from a sample of 291 type-II AGN with 43.0 $<$ \loglbol $<$ 48.0. The higher mass outflow rates reported in \cite{Davies+20} arise from their definition of outflow velocity, which is $\rm v_{out} \sim 3\times v_s$ \citep{Baron+19}.

\subsection{Coronal line outflows}\label{disc:coronal}

The QSO2s studied here 
 show [Si VI] outflows with similar kinematics and radii as those measured from the recombination lines. 
 The outflow extents reported for [Si VI] outflows range from tens to hundreds of parsecs in Seyfert galaxies \citep{Muller+11,RodriguezArdila+17,May+18} and $\sim$1 kpc in QSO2s \citep{RamosAlmeida+17,RamosAlmeida+19,Speranza+22,VillarMartin+23}. 
Despite the small statistics, recent results including ours 
suggest that coronal line outflows in QSO2s show similar extents as those detected in the recombination lines.





The outflow physical properties measured from [Si VI] when adopting the TR-based electron densities are 
$\rm \dot{M}_{[Si~VI]} \sim 0.004-1.2~M_{\odot}~yr^{-1}$ and $\rm \dot{E}_{[Si~VI]} \sim 10^{36.5-40.5}~erg~s^{-1}$ (see Table \ref{tab:outfenergetics}). When we consider the velocity of the outflow as $\rm v_{max}$ we find 
$\rm \dot{M}_{[Si~VI],max} \sim 0.08-9.3~M_{\odot}~yr^{-1}$ and $\rm \dot{E}_{[Si~VI],max} \sim 0.05-6\times 10^{42}~erg~s^{-1}$. Comparing our findings with the literature is not straightforward, since the few works that computed these quantities were restricted to LLAGN and Seyferts, and the method used to estimate those properties did not rely on the flux of the lines. \cite{Muller+11} analysed SINFONI data of seven nearby Seyfert galaxies, detecting [Si VI] outflows and construcing biconical outflow models for them. The mass outflow rates were inferred using their models maximum velocities and lateral surface area, assuming $\rm n_e$ = 5000 $\rm cm^{-3}$ and filling factor of f = 0.001. They found [Si VI] mass outflow rates of 4.0-9.0 M$\rm _{\odot}~yr^{-1}$ for five galaxies, and 120 M$\rm _{\odot}~yr^{-1}$ for NGC 2992. They reported kinetic powers in the range 0.06-5$\rm ~\times~10^{42}~erg~s^{-1}$. \cite{RodriguezArdila+17} studied the Seyfert galaxy NGC 1386 in the K-band with SINFONI and using their measured [Si VI] velocities and angular scales, $\rm n_e\sim$~930 cm$^{-3}$ (estimated from the [Ne V] ratio measured with Spitzer), and f=0.1, they measured a mass outflow rate of 11 M$\rm _{\odot}~yr^{-1}$ and kinetic power of 1.7$\rm ~\times~10^{41}~erg~s^{-1}$. \cite{May+18} studied the Seyfert ESO 428-G14, and, using similar assumptions as \cite{RodriguezArdila+17}, found mass outflow rates of 3-8 M$\rm _{\odot}~yr^{-1}$ and kinetic powers of 2-11$\rm ~\times~10^{40}~erg~s^{-1}$. 
However, we stress that none of the previously discussed works used the line fluxes to estimate the outflow mass. Here we provide an equation to work out the coronal outflow mass using the line flux.

\begin{figure*}[htbp]
\centering
\includegraphics[width=0.45\linewidth]{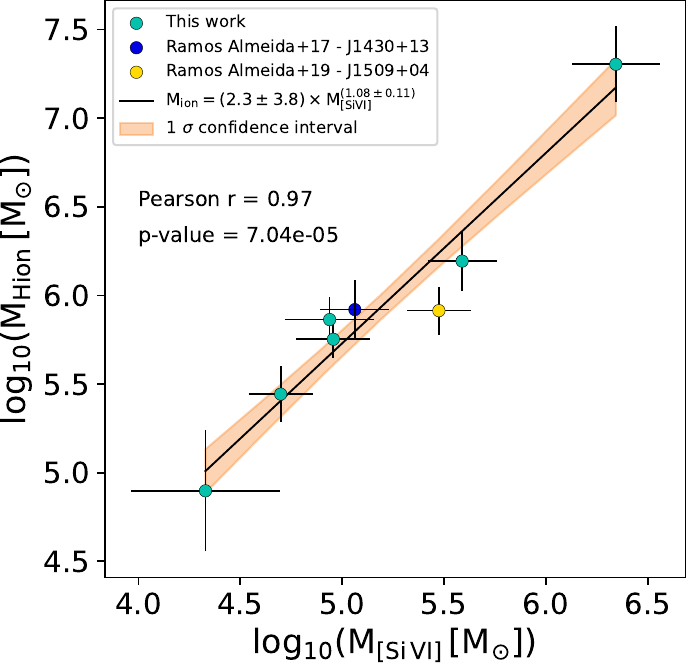}
\includegraphics[width=0.46\linewidth]{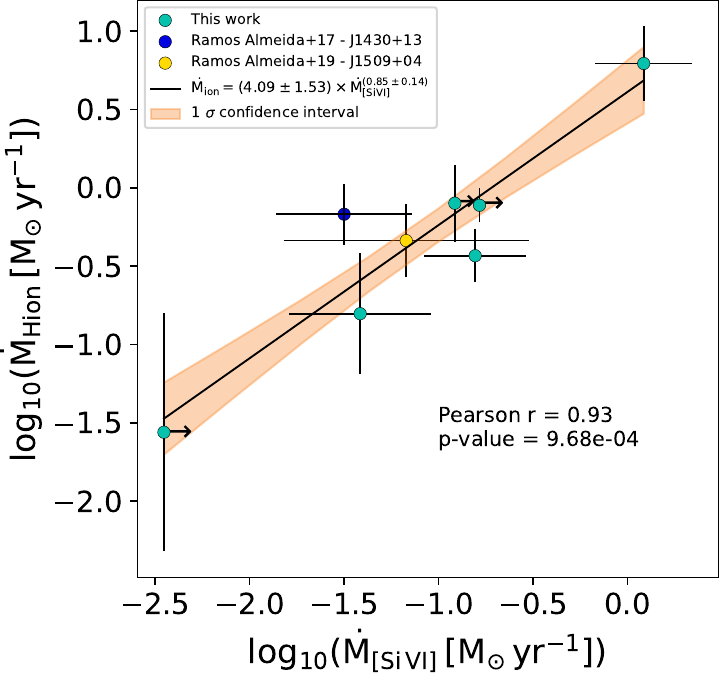}
\caption{Total ionized (Hion) versus [Si VI] gas outflow mass (left panel) and mass outflow rate (right panel). Turquoise circles are the QSO2s studied here, with outflow masses and mass rates calculated with TR electron densities, and blue and yellow circles are the QSO2s J1430+13 and J1509+04 from \cite{RamosAlmeida+17,RamosAlmeida+19}. Black solid lines are the corresponding fits to the data and the orange shaded area represents the 1$\sigma$ confidence interval of the fit. The Pearson correlation coefficient and the associated p-value are indicated in each panel.}
\label{fig:7}
\end{figure*}

As in \cite{Fiore+17}, where they found that the mass in the \hb outflows was 3 times those calculated using [OIII], 
here we attempted to estimate what fraction of the ionized gas mass in the outflow is carried by [Si VI].  
Fig. \ref{fig:7} shows the ionized and [Si VI] outflow mass and mass rates, 
including our six QSO2s and another two with [Si VI] outflow measurements \citep{RamosAlmeida+17,RamosAlmeida+19}. The ionized mass outflow rates from the recombination lines, and their corresponding outflow masses, are the same as in Fig. \ref{fig:8}. In the case of 
[Si VI], for J1430+13 we adopted the outflow flux, velocity shift, and radius reported in \cite{RamosAlmeida+17} and the TR-based electron density and reddening from \citet{Bessiere+24}. For J1509+34, we use all the measurements in \cite{RamosAlmeida+19} and assumed that the [Si VI] outflow radius is the same as that reported for \pa.
For the eight QSO2s, and considering the lower limits as values, $\rm M_{Hion}/M_{[Si~VI]}$=[3.0-9.2], with a median of 5.9, and 
$\rm \dot{M}_{Hion}/\dot{M}_{[Si~VI]}$=[2.4-21.5], with a median of 5.8. 
In Fig. \ref{fig:7} we also show the linear fits to the data, from which we found a stronger correlation for the outflow mass (Pearson r = 0.97, p-value = 7.0$\times 10^{-5}$). 
These fractions need to be confirmed in larger samples of QSO2s and AGN. Considering this, together with the similar kinematics and extents of the outflows found for [Si VI] and the NIR recombination lines, we conclude that the low- and high-ionization lines are tracing the same outflow events, and not different phenomena \citep{Cicone+18}.

\subsection{The warm molecular gas phase}
\input{Tables/Table_H2}

The six QSO2s studied here were selected because they all showed recombination and \hii emission lines in the NIR. However, we did not find any outflow associated with the \hii emission. 
The same was found for another five QSO2s in the QSOFEED sample using the rotational \hii lines detected in the nuclear JWST/MIRI spectra \citep{RamosAlmeida+25}. It is noteworthy that the five QSO2s observed with JWST/MIRI show massive outflows of cold molecular gas, detected in CO(2-1) using ALMA observations \citep{RamosAlmeida+22,Audibert+25}. Nevertheless, recent analysis of deep NIR integral field spectra from VLT/SINFONI of two of these QSO2s, J1430+13 and J1356+10, revealed their elusive warm molecular outflows. These outflows have $\rm r_{out}<$2 kpc and $\rm M_{out}\sim2.6$ and $\rm 1.5\times10^3 M_{\odot}$ \citep{Zanchettin+25}. This indicates that detecting warm molecular outflows is not easy, and indeed, there are only two other QSO2s with warm molecular outflows reported in the literature so far (F08572+3915 NW; \citealt{Rupke+13b} and J1509+04; \citealt{RamosAlmeida+19}). This last QSO2, J1509+04, has a NIR \hii outflow reported, but it appears to lack its MIR counterpart \citep{RamosAlmeida+25}, although further analysis of the MIRI IFU data is required.

Following \cite{Speranza+22}, in order to investigate what can be driving the presence or not of warm molecular outflows, we first estimated the \hii mass of the six QSO2s studied here using the relation from \citet{Mazzalay+13}:

\begin{equation}
  \rm  M_{H_2} \approx 5.0875 \times 10^{13} \left(\frac{D_L}{Mpc}\right)^2 \left(\frac{F_{H_2~1-0S(1)}^{corr}}{erg~s^{-1}cm^{-2}}\right),  
\end{equation}
where $\rm D_L$ is the luminosity distance and $\rm F_{H_2~1-0S(1)}^{corr}$ is the flux measured in $\rm H_2~1-0S(1)$ corrected by extinction (see section \ref{sec:outflow_energy} for more details). The H$_2$ masses are shown in Table \ref{tab:H2} together with measurements of other QSO2s 
reported in the literature.

The nuclear \hii masses of the six QSO2s (i.e., measured from the nuclear spectra shown in Fig. \ref{fig:1}) are in the range (0.7-12.9)$\rm ~\times~10^{3}~M_{\odot}$. They correspond to regions with sizes of 0.6-3.2 kpc (median of 1.3 kpc). In addition, we also computed \hii masses from the lines detected in spectra of the central 3'' of the QSO2s (total \hii masses hereafter, corresponding to scales of 2.1-6.1 kpc, with a median of 4.6 kpc), obtaining values of (1.1-31.9)$\rm ~\times~10^{3}~M_{\odot}$. 

Using the measurements shown in Table \ref{tab:H2}, we find that the mean nuclear and total \hii masses for QSO2s without \hii outflows are (7.7 $\pm$ 3.9) $\times$ 10$^3$ \msun~and (15.4 $\pm$ 9.2) $\times$ 10$^3$ \msun, while for the four QSO2s with \hii outflows they are (20.6 $\pm$ 18.9) $\times$ 10$^3$ \msun~ and (34.0 $\pm$ 15.4) $\times$ 10$^3$ \msun\footnote{We included the lower limit reported for F08572+3915NW by \cite{Rupke+13b} in the nuclear and total gas mass averages.}. 
Therefore, the mean nuclear (total) \hii mass in QSO2s with \hii outflows 2.7 (2.2) times larger than that of QSO2s without \hii outflows. 
If confirmed for a larger sample, the nuclear and total warm molecular mass can be relevant factors for driving and/or detecting outflows, but there are counterexamples. 
For example, J1347+12 (4C12.50) has a higher total \hii mass than 
J1356+10, but there is no warm molecular outflow reported for it \citep{VillarMartin+23}, despite the massive cold $\rm H_2$ molecular outflow detected in high angular resolution CO ALMA observations \citep{Holden+24}. Another example is J0945+17, for which \citet{Speranza+24} reported a total \hii mass of $\rm \sim 59\times 10^3~M_{\odot}$ measured in the central $\rm \sim$ 1'' x 1'' region using Gemini/NIFS data and no outflow detection. It is noteworthy that the discrepancy between the Gemini/NIFS and our GTC/EMIR measurements is likely due to the lower sensitivity of our data to fainter structures (see exposure times in Table \ref{tab:H2}) and to the fact that we are using long-slit observations along a certain PA, whilst Gemini/NIFS is an IFU.  

Other factors that could potentially favor the presence of warm molecular outflows are the radio luminosities \citep{Mullaney13,Zakamska14} and the AGN luminosities \citep{Fiore+17}, which are listed in Table \ref{tab:H2}. 
The 1.4 GHz luminosities of the QSO2s with warm molecular outflows have mean $\rm log(L_{1.4 GHz}[W~Hz^{-1}])$ = 23.6$\pm$0.7, 
similar to the sources with no outflow detection, 
$\rm log(L_{1.4 GHz}[W~Hz^{-1}])$ = 23.4$\pm$0.5, 
if we exclude 4C12.50.  
Regarding the bolometric luminosities, the average values for the QSO2s with and without H$_2$ outflows are 45.8$\pm$0.2 and 45.5$\pm$0.3. These average values are therefore consistent within the dispersion of the sample. 
Therefore, despite the small number of QSO2s observed in the NIR, we find that sources with higher radio and AGN luminosities are not more likely to have warm molecular outflows \citep{Speranza+22}, at least within the luminosity ranges here considered. However, that does not exclude the possibility that compact, low-power jets 
might be contributing to launch molecular outflows in some of the QSO2s \citep{Mukherjee+18,Girdhar+24,Audibert+23,Audibert+25}.

The widespread lack of warm molecular outflows 
could also be explained if the NIR \hii lines represent an intermediate and short phase in a post-shock cooling sequence (see \citealt{Holden+23a}). If the outflow is accelerated by either accretion-disc wind or jet-induced shocks, the shocked gas will cool from the highly-ionized to the cold-molecular phase. Assuming that the ionized gas phase is somewhat stable and that the cold molecular gas accumulates over time, the warm-molecular phase observed in the NIR could represent a short-lived transitional phase. This could explain both the lack of warm molecular outflows and the absence of correlation with \lbol.

Finally, we cannot rule out that deeper observations than those used here are required to detect \hii outflows (see Table \ref{tab:H2}). Recently, \cite{Zanchettin+25} detected and characterized the elusive warm molecular outflow of J1430+13 by using deeper NIR SINFONI observations (doubling the exposure time) than those used by \cite{RamosAlmeida+17}, and also in the QSO2 J1356+10. The on-source EMIR exposure times of the six QSO2s studied here but J1713+57 were 1920 seconds, which is shorter than the exposure times of three of the QSO2s with \hii outflows (see Table \ref{tab:H2}). 
However, there are also QSO2s with long exposure times and no molecular outflows detected, as J1713+57 and J1347+12, and also the Gemini/NIFS data of J0945+17 (4000 s; \citealt{Speranza+22}).
Finally, another factor that can be relevant for the detection of \hii outflows is slit orientation. If the slit does not follow the outflow PA, part of the outflow emission may be undetected. The four QSO2s with \hii outflows detected were observed with NIR IFUs, unlike the QSO2s without \hii outflows but J0945+17 \citep{Speranza+22}.


Therefore, based on a small sample of eleven QSO2s with NIR spectra, we find tentative evidence that warm H$_2$ outflows may be associated with larger H$_2$ gas masses. This apparent trend could reflect more efficient coupling between winds and/or jets and the ISM in systems hosting more massive H$_2$ reservoirs, or it may simply arise because outflows in such systems are intrinsically brighter and therefore easier to detect. Nevertheless, warm H$_2$ outflows are also observed in QSO2s with lower H$_2$ masses than some sources without detected outflows, indicating that a large H$_2$ reservoir is not a necessary condition for the presence of outflows. Larger samples of QSO2s with deep NIR observations are required to confirm or refute this tentative trend.

\section{Summary and conclusions}
\label{sec:conclusions}

In this paper we investigate the warm molecular, low and high-ionization emission lines of six QSO2s from the QSOFEED sample using K-band spectroscopic observations from GTC/EMIR. We analyzed their nuclear spectra, modelling their line 
emission to characterize the gas kinematics, which revealed low- and high-ionization gas outflows in all targets. None of the sources exhibits warm molecular outflows. We estimated the outflow properties from both the recombination and coronal lines. We summarize our main findings below.   

   \begin{itemize}
      \item The QSO2s show low and high-ionization gas outflows with similar kinematics. For all QSO2s, except J1034+60, besides a broad component of $\rm FWHM\sim 1200-2500~km~s^{-1}$, it was necessary to fit an intermediate component of $\rm FWHM\sim 500-1200~km~s^{-1}$. 
      \item The spatially resolved outflow extents measured from the recombination lines, of 0.3-2.1 kpc and from [Si VI], of 1.1-2.7 kpc, are similar.
      \item 
      From the NIR recombination lines and adopting trans-auroral electron densities we found outflow masses of $\rm M_{Hion}\rm \sim0.08-20\times 10^{6}~M_{\odot}$, mass rates of $\rm \dot{M}_{Hion}\rm \sim0.03-6~M_{\odot}~yr^{-1}$, and kinetic powers of $\rm  \dot{E}_{Hion}\rm \sim 10^{37.8-40.8}~erg~s^{-1}$. From [Si VI] we found $\rm M_{[Si~VI]}\rm \sim0.02-2\times 10^{6}~M_{\odot}$, $\rm  \dot{M}_{[Si~VI]}\rm \sim0.004-1~M_{\odot}~yr^{-1}$, and $\rm \dot{E}_{[Si~VI]}\rm \sim 10^{36.6-40.5}~erg~s^{-1}$.

      \item Considering the direct and physical properties of outflows measured for eight QSO2s, we find median outflow mass ratios of $\rm M_{Hion}/M_{[Si~VI]}\sim 5.9$ (spanning from $\rm M_{Hion}/M_{[Si~VI]} = 3.0-9.2$) and median outflow mass rate ratios of $\rm \dot{M}_{Hion}/\dot{M}_{[Si~VI]}\sim 5.8$ ($\rm \dot{M}_{Hion}/\dot{M}_{[Si~VI]} = 2.4-21.5$).
    
    From this we conclude that the recombination lines and [Si VI] trace the same outflows but they carry different amounts of mass. 

      \item 
      Despite the different methods used to measure the direct outflow properties from NIR and optical data of the same QSO2s, as well as the higher flux calibration uncertainty of the NIR data, the median outflow mass that we measure in the NIR is 7.9 times higher than the optical one. This likely indicates that the lower extinction and higher angular resolution of the NIR data might be allowing us to probe deeper and more obscured outflow regions.
      
      \item We find similar outflow mass rates and kinetic powers between those measured for local QSO2s ($\rm z~\sim~0.1$), and a sample of QSO2s at $\rm z~\sim~0.3-0.41$. This suggest no strong evolution of the ionized outflow properties of QSO2s in this redshift range. 
       
      \item We expanded the sample of QSO2s with detected warm molecular emission in the NIR to eleven targets (four with H$_2$ outflows and seven without). 
      We did not find any connection between the presence of \hii outflows and either radio or bolometric luminosity, but  QSO2s with \hii outflows have nuclear (total) \hii masses 2.7 (2.2) times larger, on average, than those without. 
    \end{itemize}

Our findings add information to multiphase studies of outflows in luminous quasars. 
In particular, this work connects low- and high-ionization gas outflows, 
providing an equation to calculate the coronal gas mass. 
Furthermore, this study also investigated the warm molecular gas, stressing the need to further investigate this elusive outflow gas phase in statistically significant quasar samples. 

\begin{acknowledgements}
     Based on observations made with the Gran Telescopio Canarias (GTC), installed in the Spanish Observatorio del Roque de los Muchachos of the Instituto de Astrofísica de Canarias (IAC), in the island of La Palma. This work is based on data obtained with the instrument EMIR, built by a Consortium led by the IAC. EMIR was funded by GRANTECAN and the National Plan of Astronomy and Astrophysics of the Spanish Government. PHC, CRA, JAP, and AA acknowledge support from project ``Tracking active galactic nuclei feedback from parsec to kiloparsec scales'', with reference PID2022-141105NB-I00. LRH acknowledges support from the UK Science and Technology Facilities Council (STFC) in the form of grant ST/Y001028/1. M.V.Z. acknowledges the support from project "VLT-MOONS" CRAM 1.05.03.07. AA also acknowledges funding from the European Union (WIDERA ExGal-Twin, GA 101158446).
     We thank the anonymous referee for valuable and constructive suggestions that helped to improve this manuscript.
\end{acknowledgements}

\bibliographystyle{aa}
\bibliography{References}

\begin{appendix}
\section{Results from the emission line fits}
\label{sec:ApA}

\begin{figure}[!htbp]
    \centering
    \includegraphics[width=0.8\linewidth]{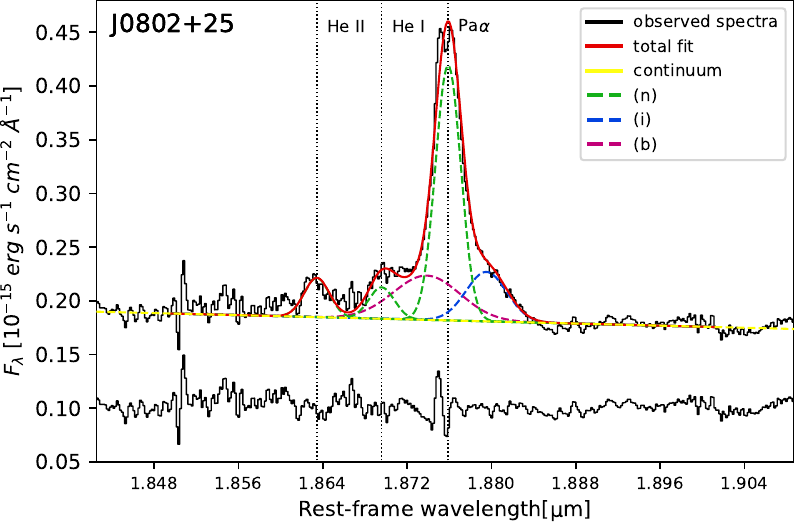}
    \includegraphics[width=0.8\linewidth]{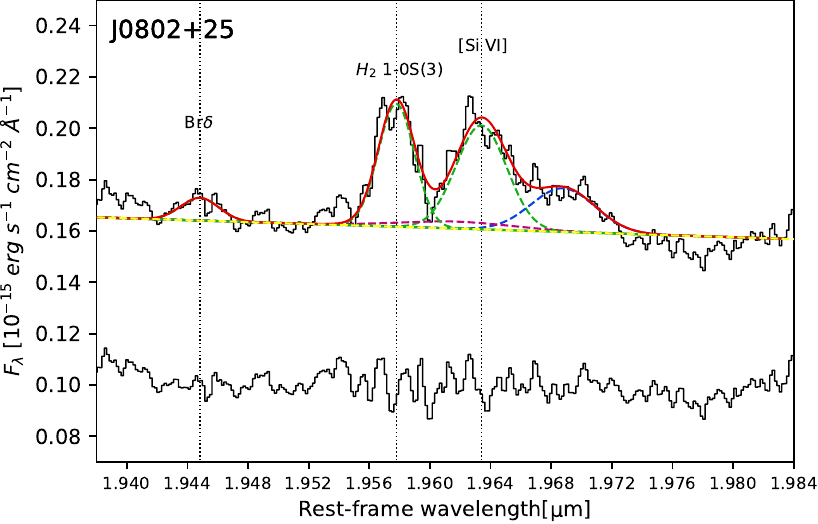}
    \includegraphics[width=0.8\linewidth]{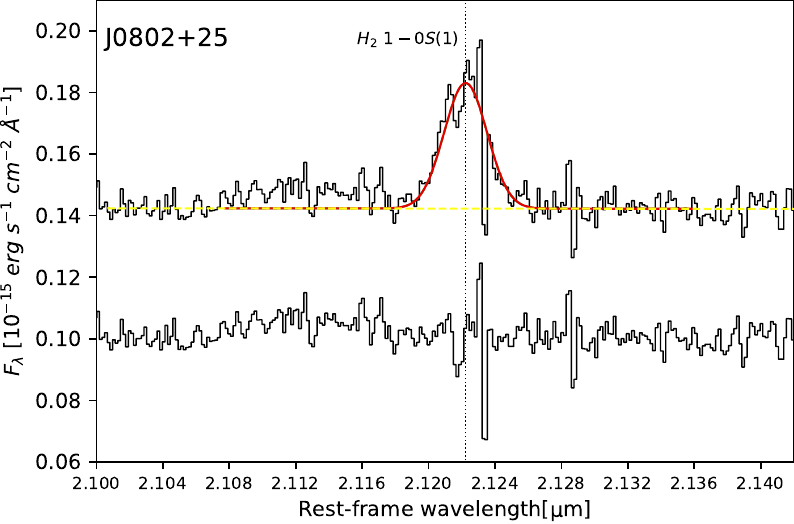}
    \caption{Same as Fig. \ref{fig:3}, but for J0802+25. The spectra were smoothed using a 4-6 pixels boxcar.} 
    \label{fig:fit_J0802} 
\end{figure}

\begin{figure}[!htbp]
    \centering
    \includegraphics[width=0.8\linewidth]{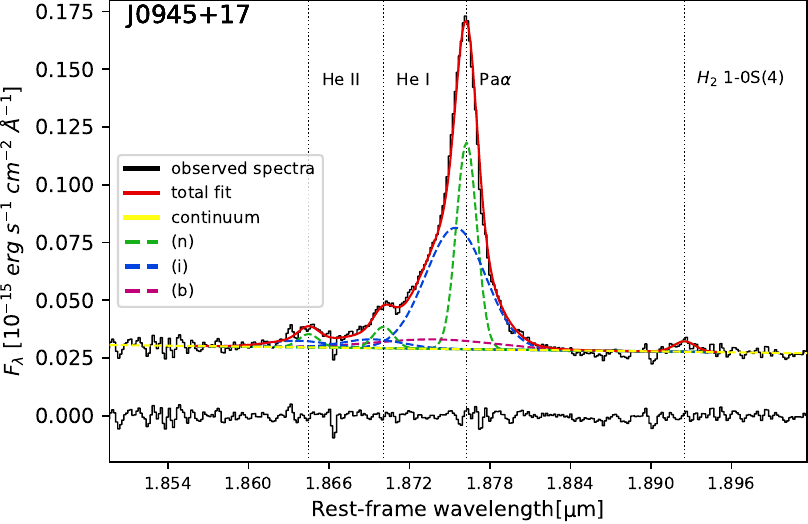}
    \includegraphics[width=0.8\linewidth]{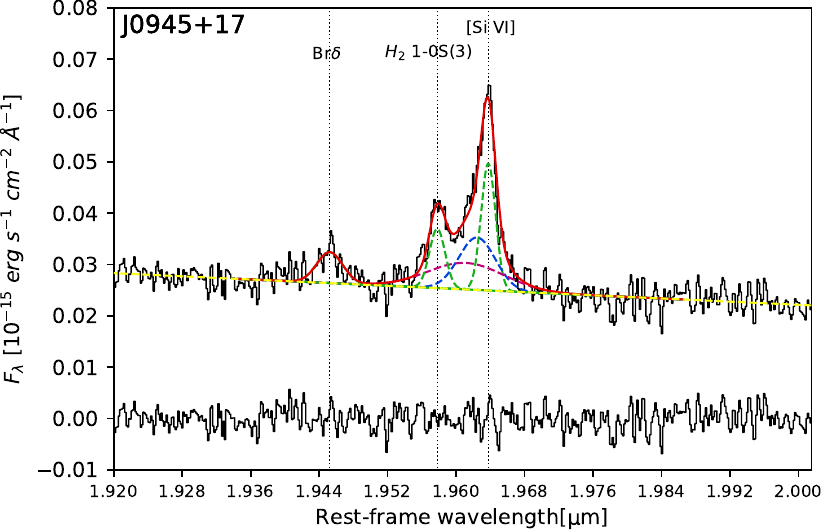}
    \includegraphics[width=0.8\linewidth]{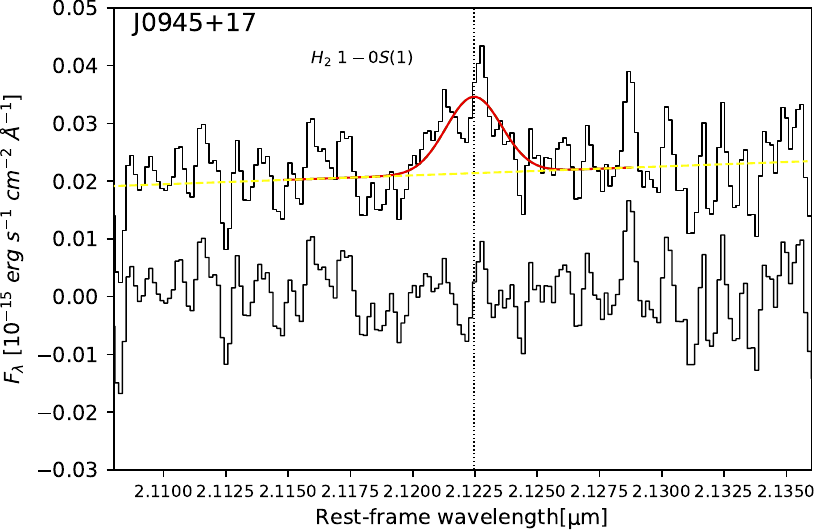}
    \caption{Same as Fig. \ref{fig:3}, but for J0945+17. The spectra were smoothed using a 5 pixels boxcar.}
    \label{fig:fit_J0945} 
\end{figure}

\begin{figure}[!htbp]
    \centering
    \includegraphics[width=0.8\linewidth]{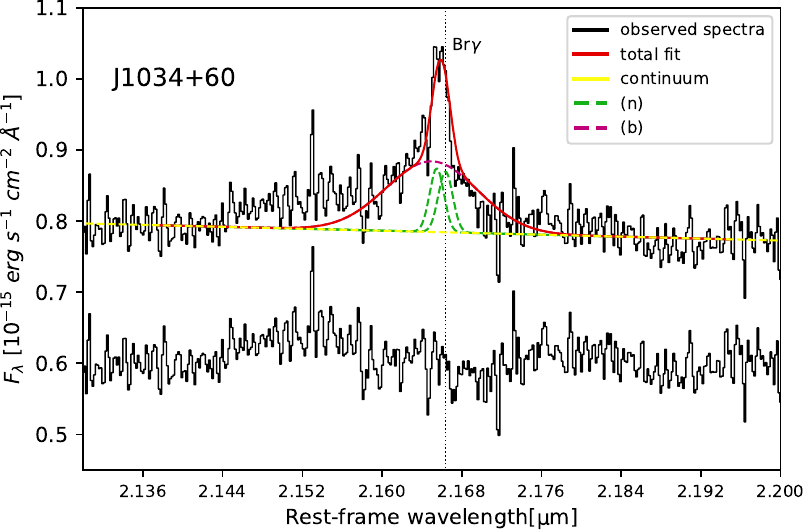}
    \includegraphics[width=0.8\linewidth]{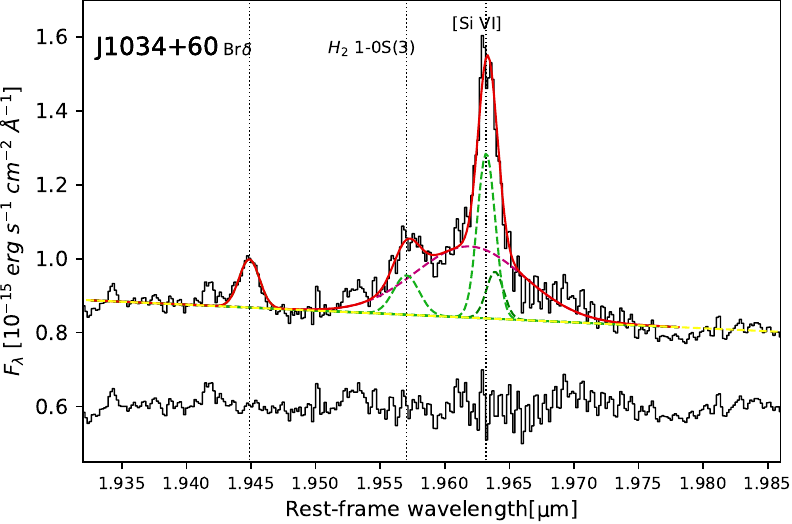}
    \includegraphics[width=0.8\linewidth]{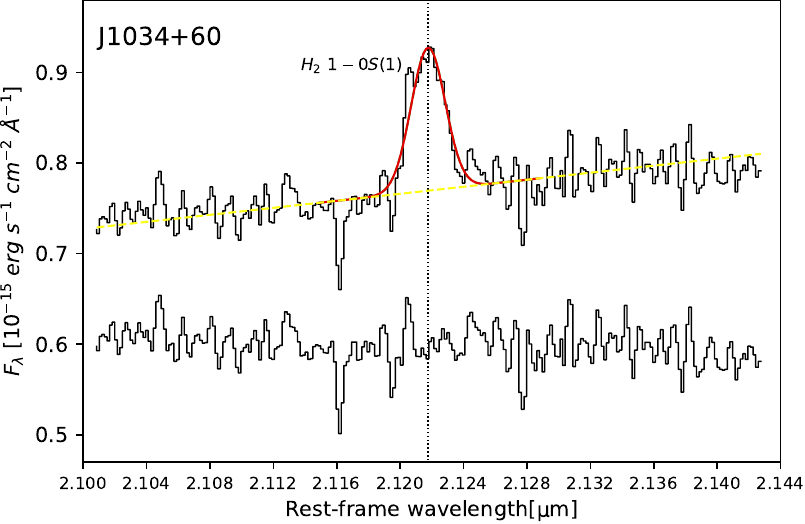}
    \caption{Same as Fig. \ref{fig:3}, but for J1034+60. The spectra were smoothed using a 4-5 pixels boxcar.}
    \label{fig:fit_J1034} 
\end{figure}

\begin{figure}[!htbp]
    \centering
    \includegraphics[width=0.8\linewidth]{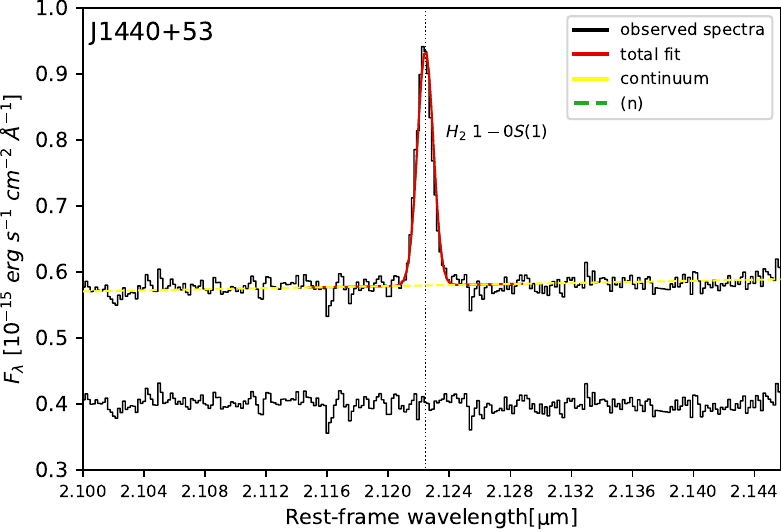}
    \includegraphics[width=0.8\linewidth]{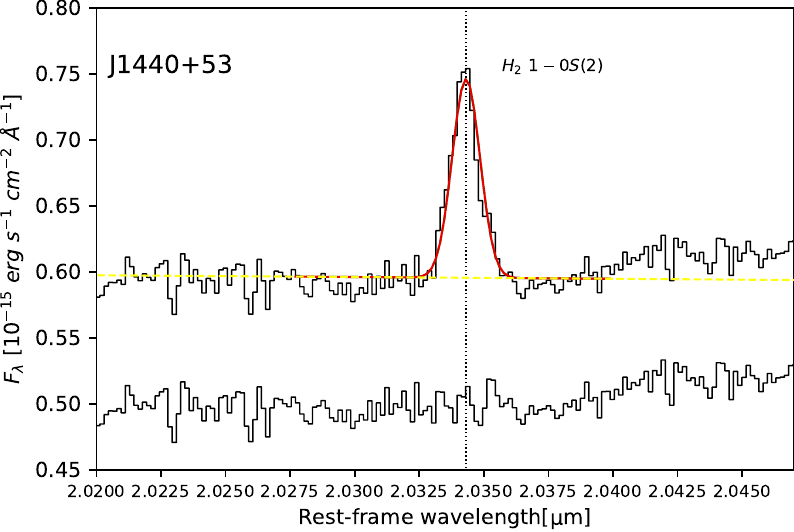}
    \caption{Same as Fig. \ref{fig:3} but for the \hii lines detected in J1440+53. The spectra were smoothed using a 4 pixels boxcar.} 
    \label{fig:fit_J1440}
\end{figure}

\begin{figure}[!htbp]
    \centering
    \includegraphics[width=0.8\linewidth]{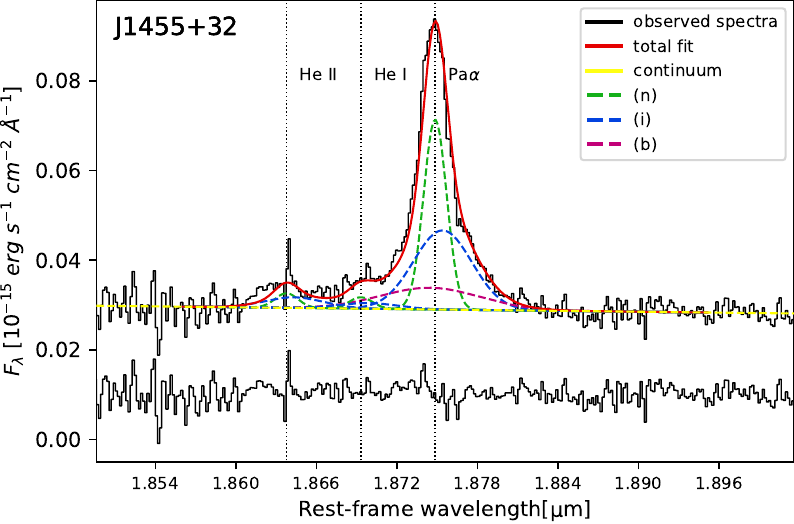}
    \includegraphics[width=0.8\linewidth]{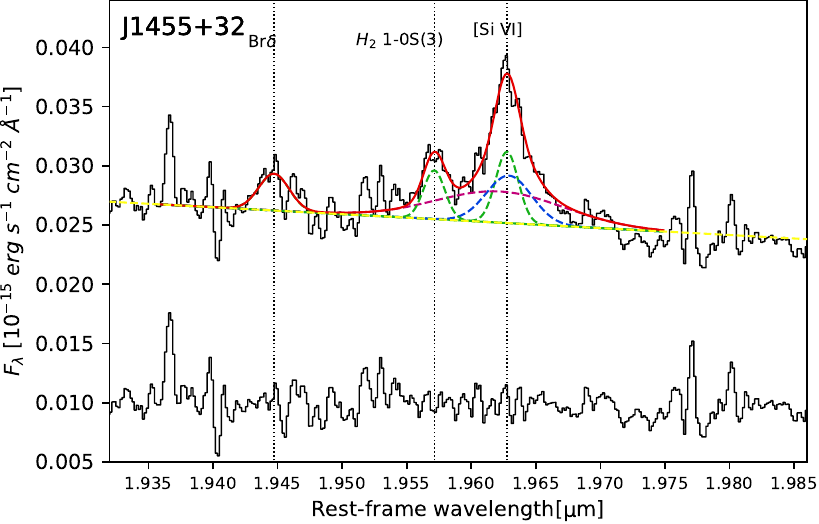}
    \textbf{\includegraphics[width=0.8\linewidth]{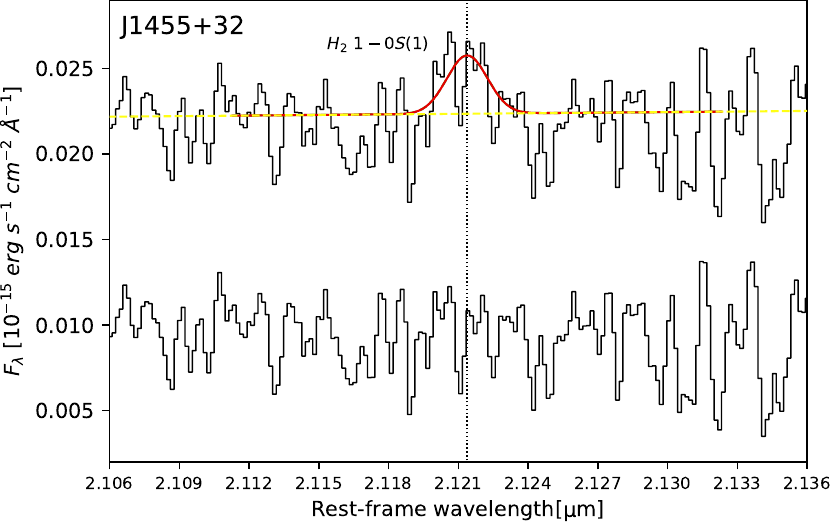}}
    \caption{Same as Fig. \ref{fig:3}, but for J1455+32. The spectra were smoothed using a 4-7 pixels boxcar.}
    \label{fig:fit_J1455}
\end{figure}

\begin{figure}[!htbp]
    \centering
     \includegraphics[width=0.8\linewidth]{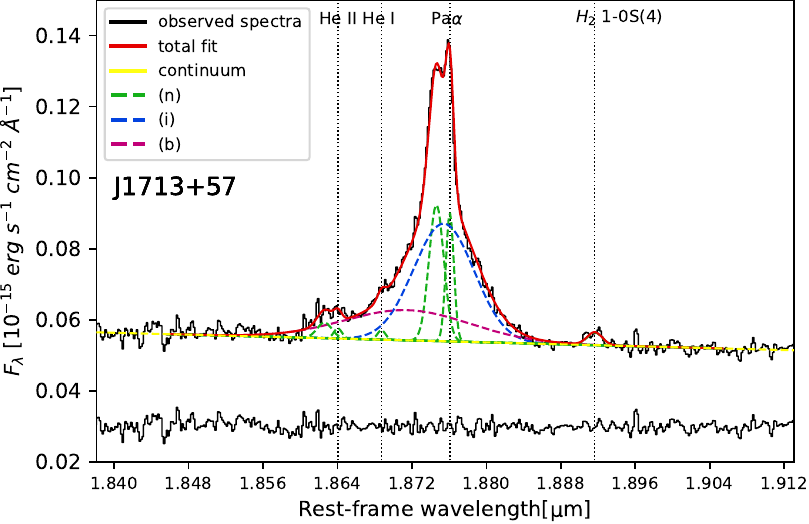}
    \includegraphics[width=0.8\linewidth]{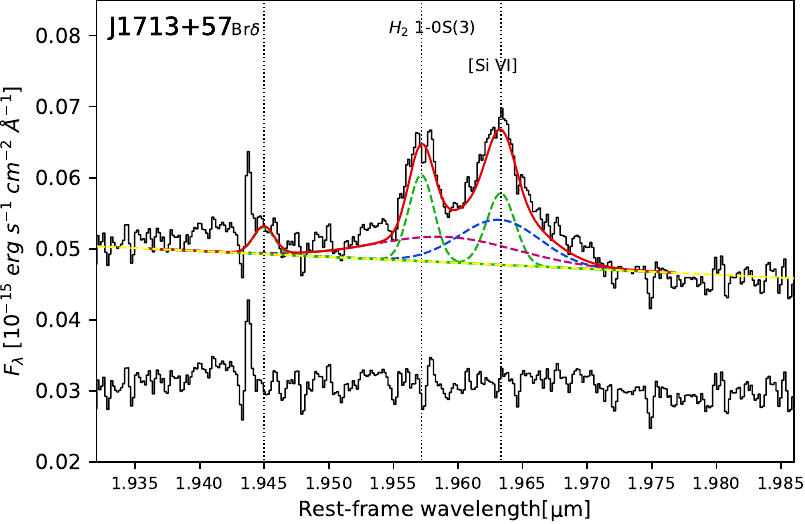}
    \includegraphics[width=0.8\linewidth]{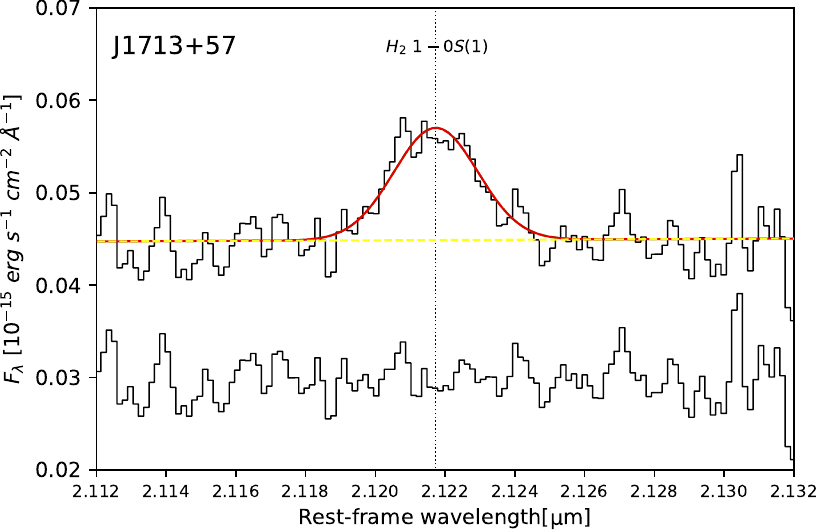}
    \caption{Same as Fig. \ref{fig:3}, but for J1713+57. The spectra were smoothed using a 4-5 pixels boxcar.} 
    \label{fig:fit_J1713}
\end{figure}

The results from the emission line fits of the six QSO2s are shown in Figs. \ref{fig:fit_J0802}-\ref{fig:fit_J1713} and Tables \ref{tab:f_J0802+25}-\ref{tab:f_J1713+57}. For J0802+25, J0945+17, and J1455+32, Pa$\alpha$ and [Si VI] were fitted with three Gaussian components: narrow, intermediate, and broad. For J1713+57, the same components were fitted except for \pa, for which two narrow components were fitted instead of one. Fits of optical emission lines detected in SDSS spectrum were used as a reference for the initial parameters. For J0802+25, J0945+17, and J1713+57, the H$\beta$ fit was the reference for Pa$\alpha$, He I, and He II. Pa$\alpha$ was reference for Br$\delta$, [O III] for [Si VI], and $\rm H_2~1-0S(1)$ for $\rm H_2~1-0S(3)$. For J1455+32, [O III] was used reference for Pa$\alpha$, He I, He II, and [Si VI], and the rest was the same as for the other two QSO2s. 

In the case of the two lower redshift QSO2s, J1034+60 and J1440+53, Br$\gamma$ and [Si VI] were fitted with two narrow components and a broad component. For J1440+53, an intermediate component was needed as well, and the H$\beta$ fit was the reference for Br$\gamma$, Br$\gamma$ for Br$\delta$, [O III] for [Si VI], and $\rm H_2~1-0S(1)$ for $\rm H_2~1-0S(3)$. In the case of J1034+60, the reference fits were the same except for Br$\delta$, whose reference was H$\beta$.



\FloatBarrier
\FloatBarrier

\input{Tables/Tables_flux/J0802+25}



\input{Tables/Tables_flux/J0945+17}



\input{Tables/Tables_flux/J1034+60}



\input{Tables/Tables_flux/J1440+53}



\input{Tables/Tables_flux/J1455+35}


\input{Tables/Tables_flux/j1713+57}

\section{Electron density diagrams}
\label{sec:Ap_elecdens}

\begin{figure}[!htbp]
    \centering
        \includegraphics[width=0.33\textwidth]{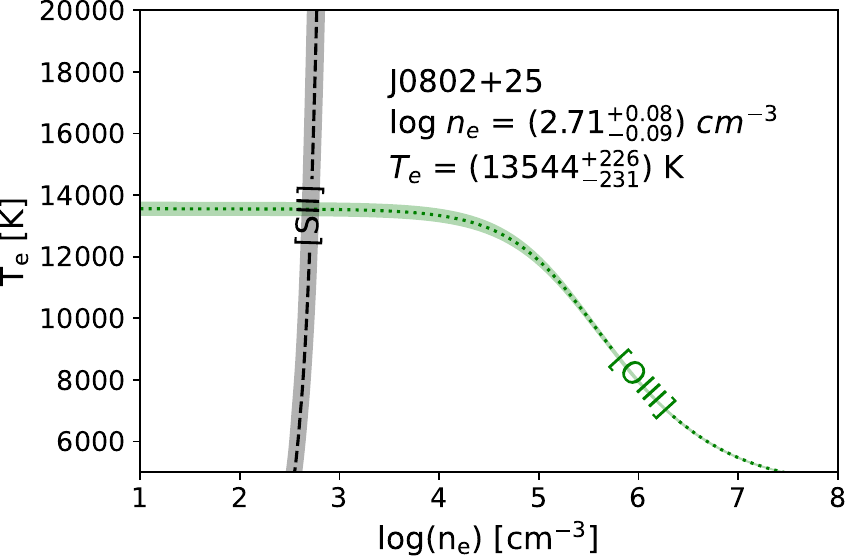}
        \includegraphics[width=0.33\textwidth]{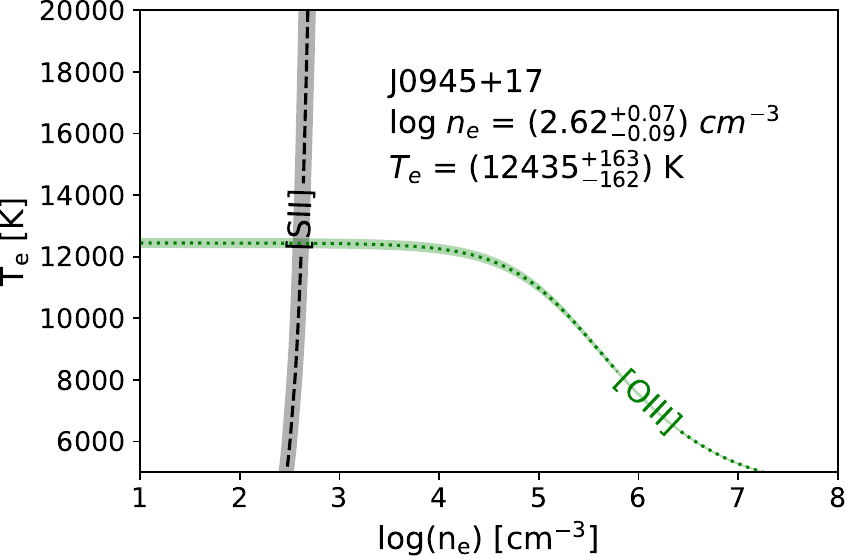}
        \includegraphics[width=0.33\textwidth]{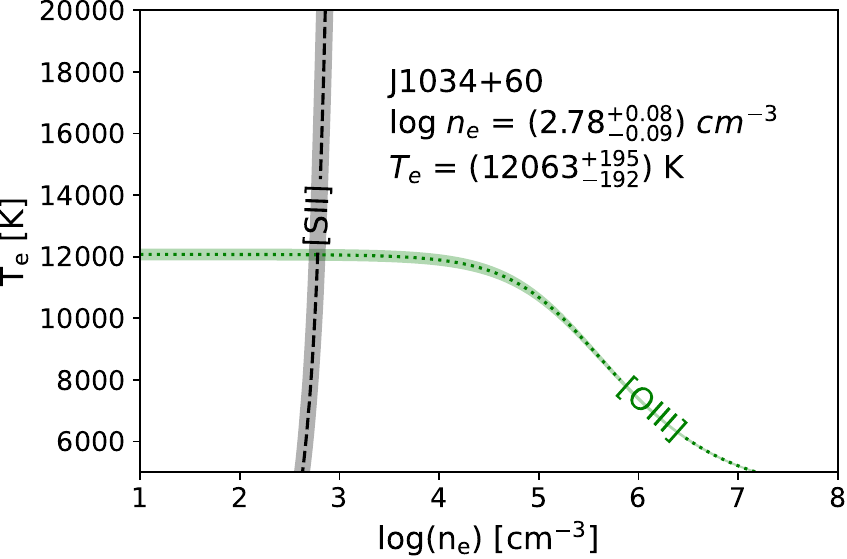} \\
        \includegraphics[width=0.33\textwidth]{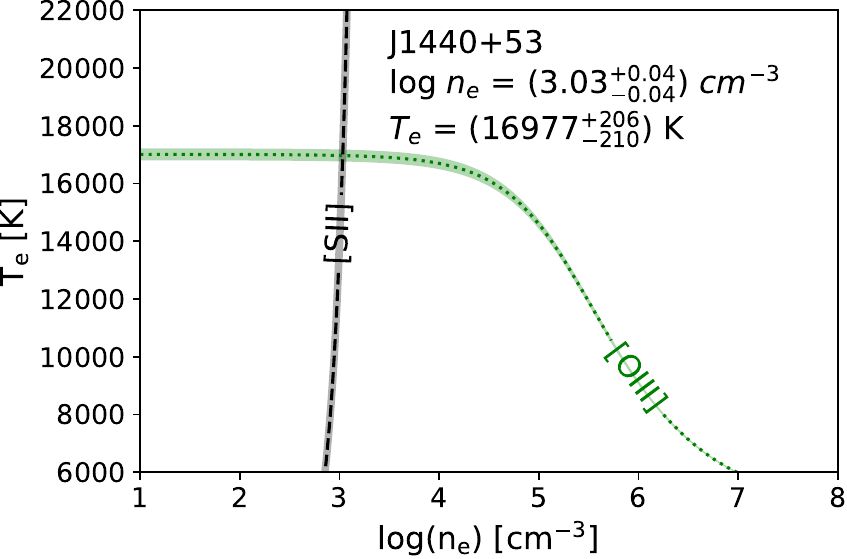}
        \includegraphics[width=0.33\textwidth]{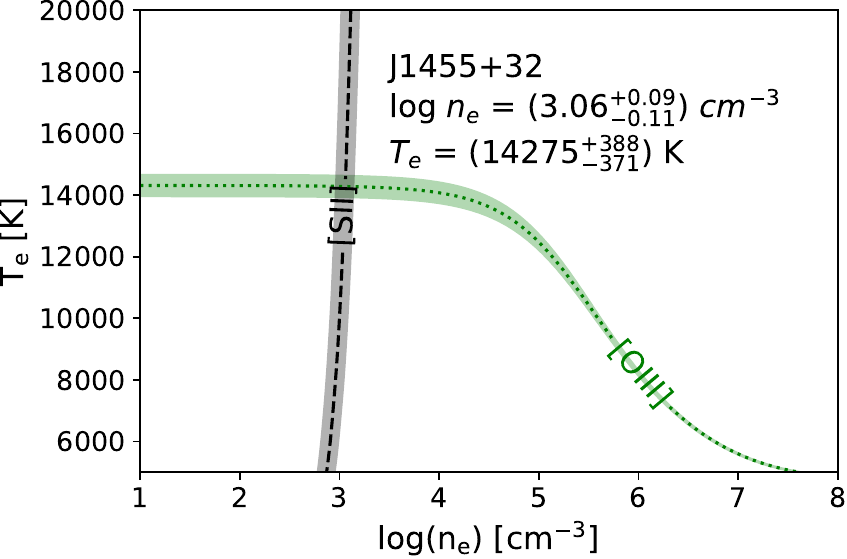}
        \includegraphics[width=0.33\textwidth]{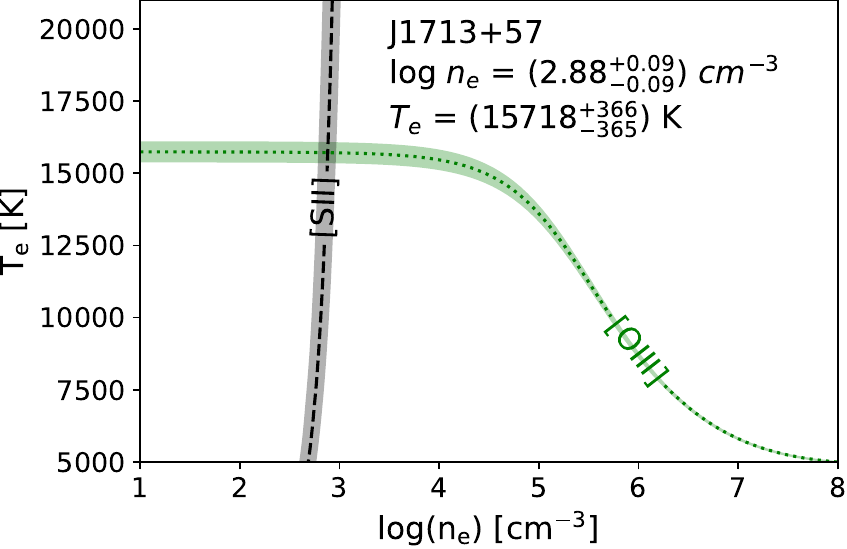}
   \caption{Diagnostic diagrams of the electron density and temperature computed from the [SII] and [OIII] ratios generated with \texttt{Pyneb}.}
    \label{fig:A5}
\end{figure}

Here we present the electron density and temperature diagnostics obtained for the QSO2s based on the [SII] method. As explained in Section \ref{sec:electron_density}, we used the [SII]$\lambda\lambda$6716,6731$\mathring{A}$ doublet and [OIII]$\lambda$4363$\mathring{A}$ and [OIII]$\lambda$5007$\mathring{A}$ ratios to infer both $\rm n_e$ and $\rm T_e$ using \texttt{Pyneb} \citep{Luridiana+15}. The results found for the QSO2s are shown in Fig. \ref{fig:A5} and values are reported in Table \ref{tab:density}.

\FloatBarrier

\section{Outflow extent}
\label{sec:Ap_extent}

\begin{figure}[!htbp]
    \centering
    \begin{minipage}{0.5\linewidth} 
    \centering
    \includegraphics[width=0.99\linewidth]{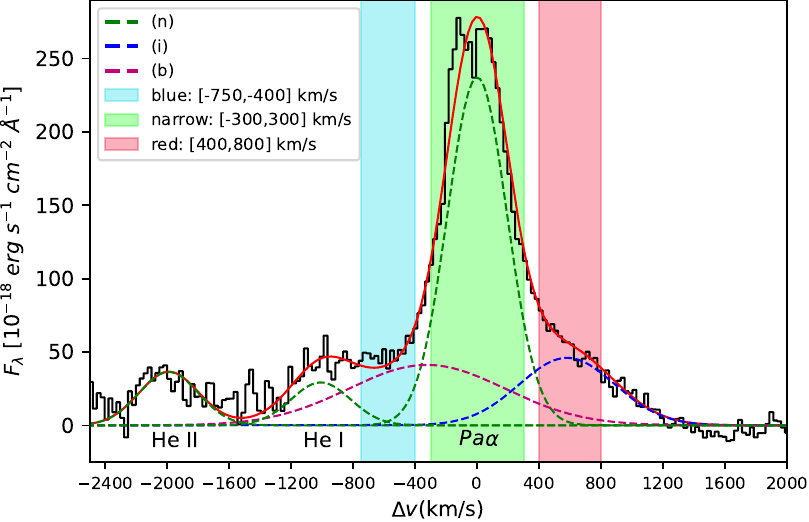}
    \includegraphics[width=0.99\linewidth]{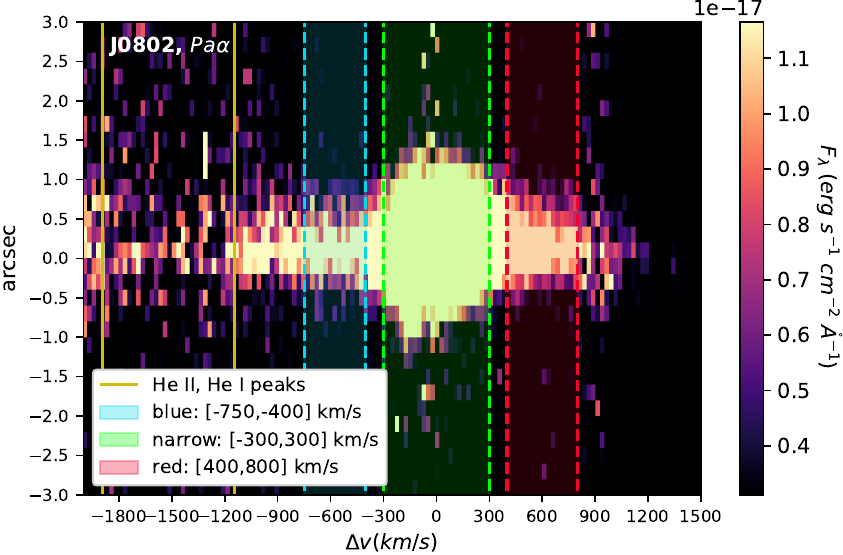}
    \includegraphics[width=0.99\linewidth]{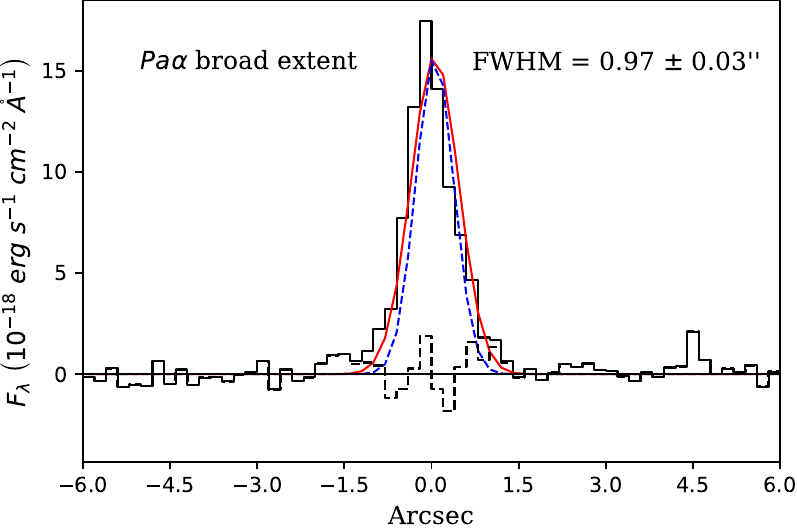}
    \end{minipage}
    \begin{minipage}{0.5\linewidth}
    \centering
    \includegraphics[width=0.99\linewidth]{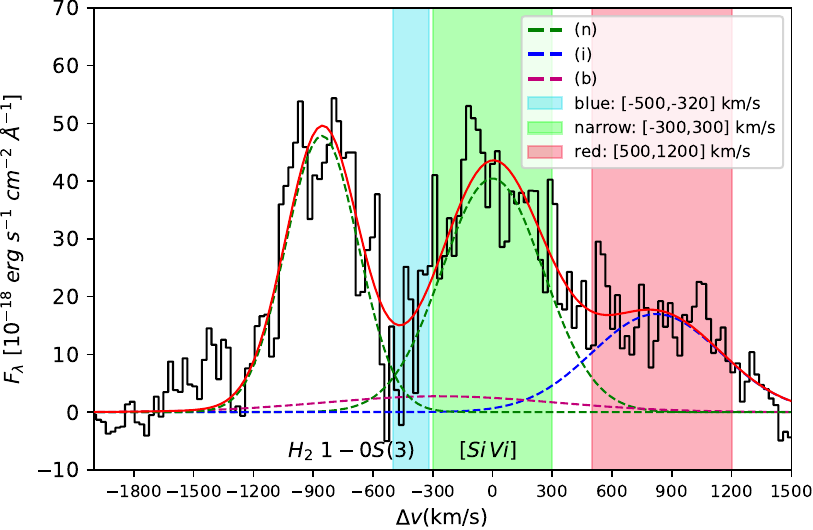}
    \includegraphics[width=0.99\linewidth]{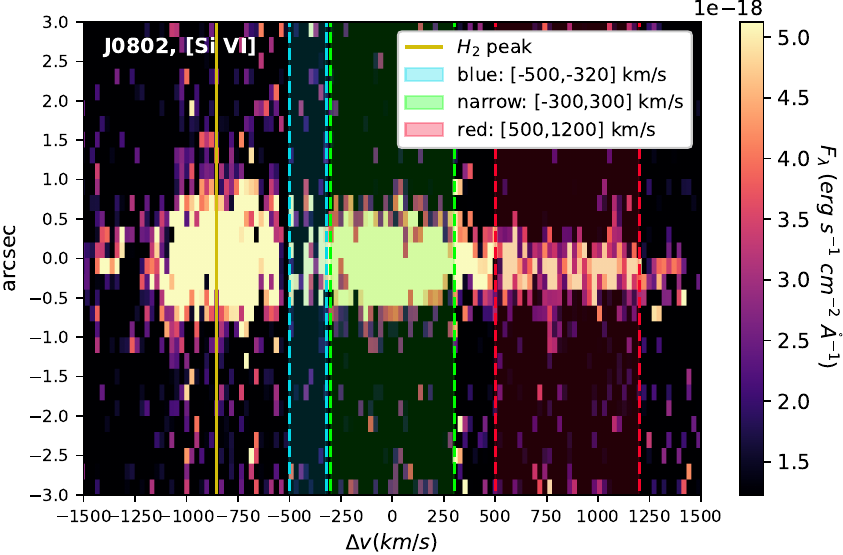}
    \includegraphics[width=0.99\linewidth]{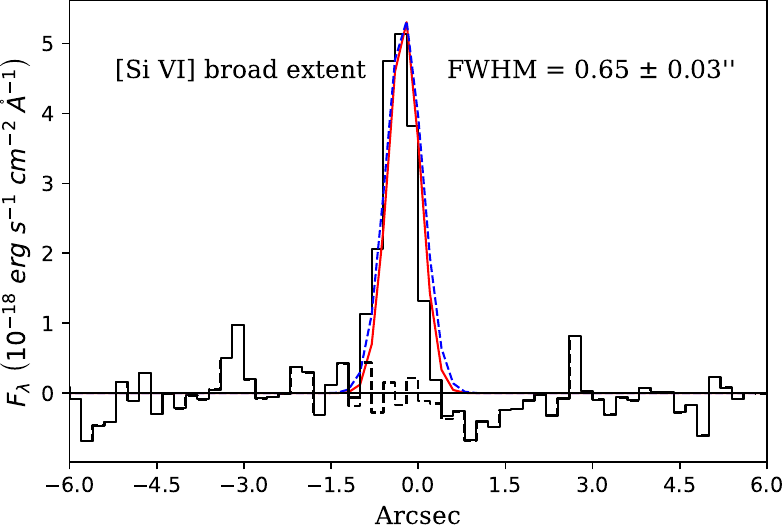}
    \end{minipage}
    \caption{Spectral regions and spatial profiles used to calculate the outflow extent from Pa$\alpha$ (left panels) and [Si VI] (right panels). Top row: Fits of the nuclear spectrum. The red and blue windows correspond to the wings of the lines, which are dominated by the outflows, while the green window captures the bulk of the narrow component. Middle row: Same spectral windows superimposed on the continuum-subtracted maps showing the line emission along the spectral and spatial directions. Bottom row: Average (i.e., weighted mean of the red and blue wings) continuum-subtracted spatial profiles of the Pa$\alpha$ and [Si VI] outflows. The red solid line shows the Gaussian fitted to the line profile, while the blue dashed lines corresponds to the seeing spatial profiles derived from observations of standard stars.}
    \label{fig:out_ext_J0802}
\end{figure}

\begin{figure}[!htbp]
    \centering
    \begin{minipage}{0.5\linewidth} 
    \centering
    \includegraphics[width=0.99\linewidth]{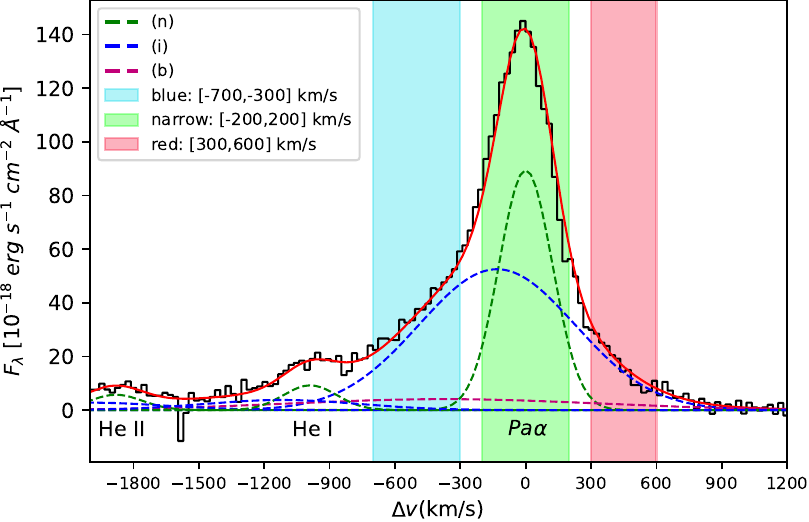}
    \includegraphics[width=0.99\linewidth]{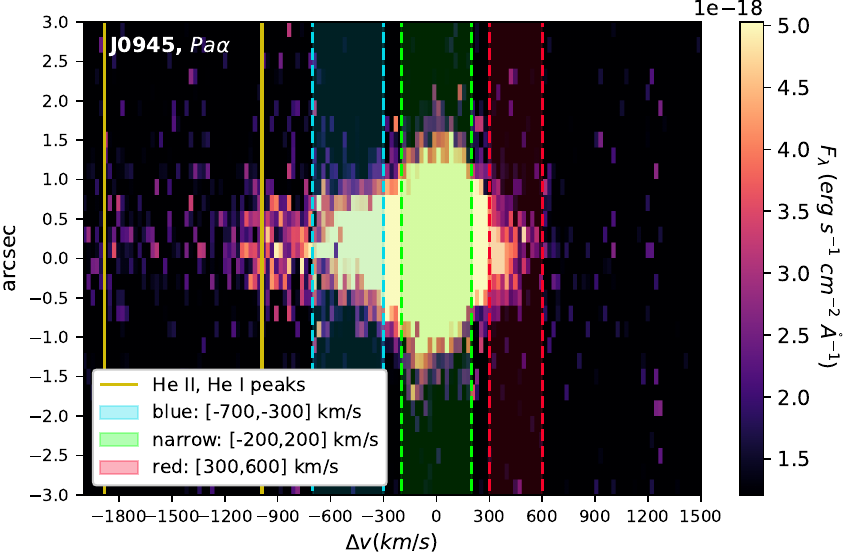}
    \includegraphics[width=0.99\linewidth]{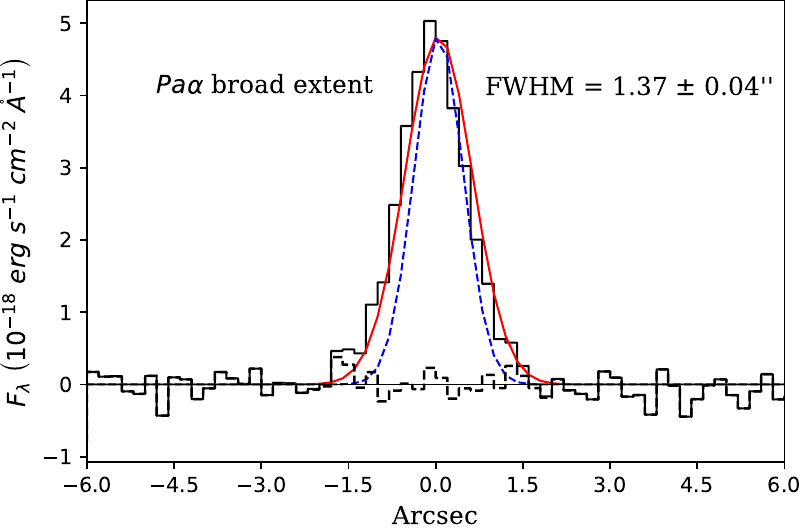}
    \end{minipage}
    \begin{minipage}{0.5\linewidth}
    \centering
    \includegraphics[width=0.88\linewidth]{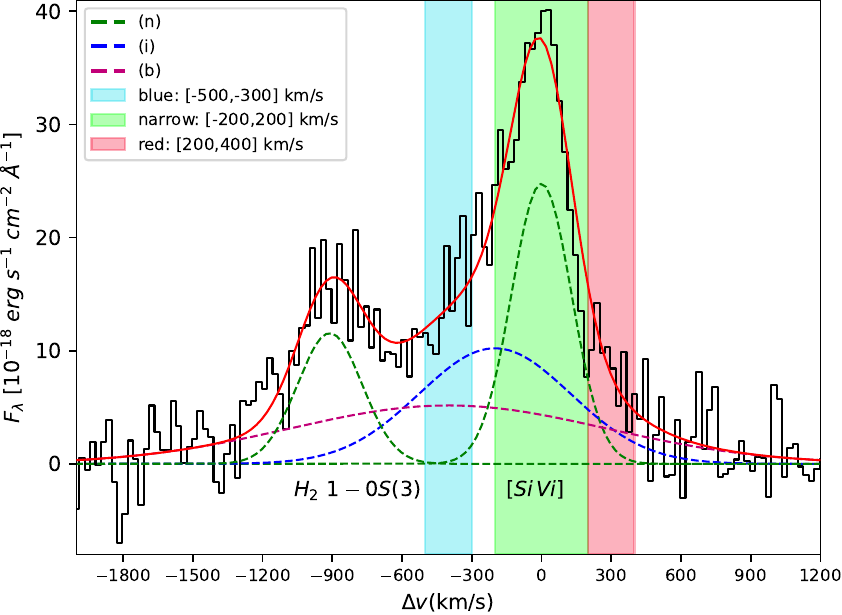}
    \includegraphics[width=0.99\linewidth]{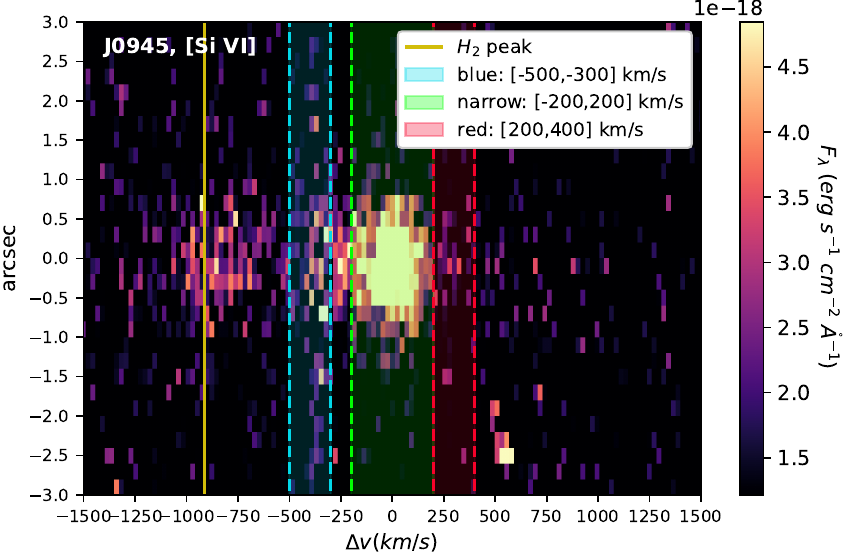}
    \includegraphics[width=0.99\linewidth]{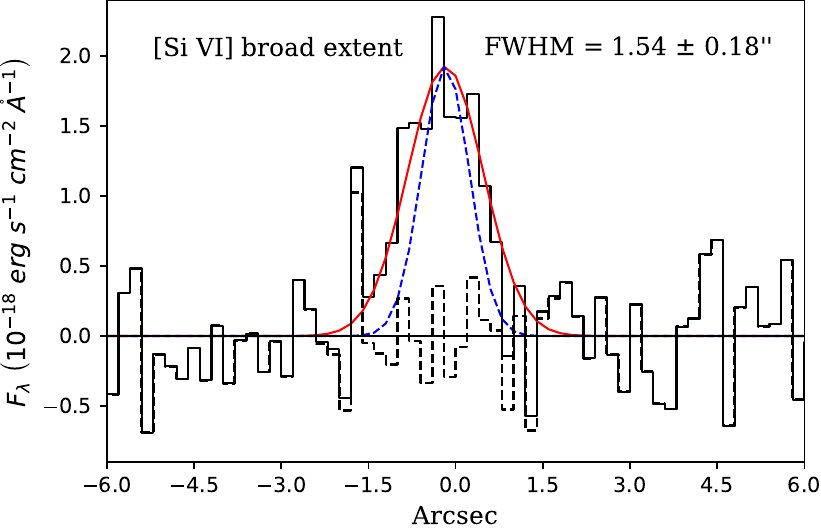}
    \end{minipage}
    \caption{Same as in Fig. \ref{fig:out_ext_J0802}, but for J0945+17.}
    \label{fig:out_ext_J0945}
\end{figure}

\begin{figure}[!htbp]
    \centering
    \begin{minipage}{0.5\linewidth} 
    \centering
    \includegraphics[width=0.99\linewidth]{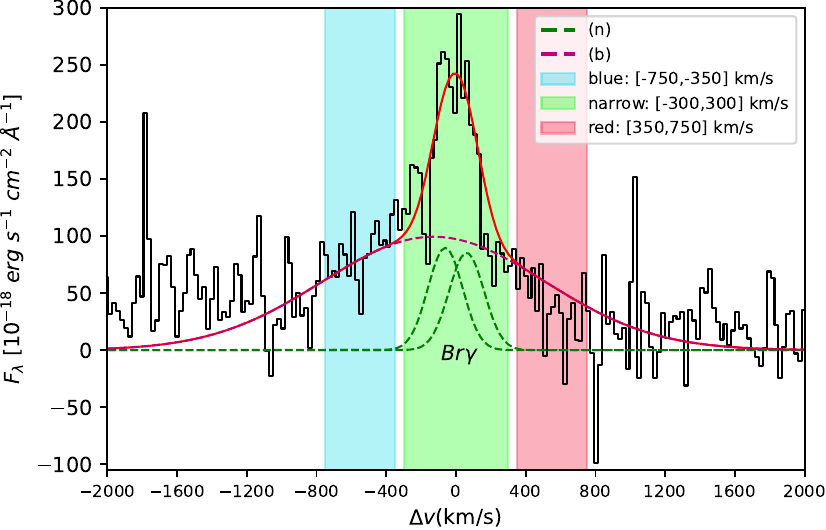}
    \includegraphics[width=0.99\linewidth]{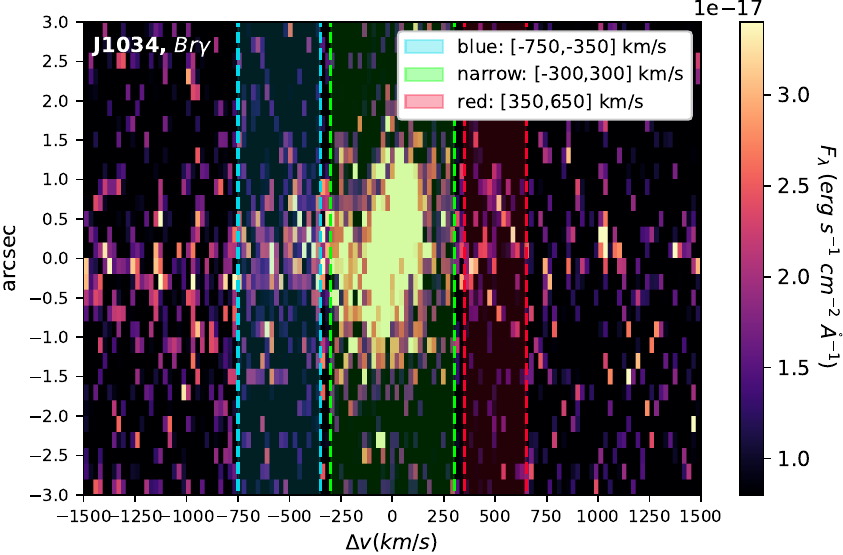}
    \includegraphics[width=0.99\linewidth]{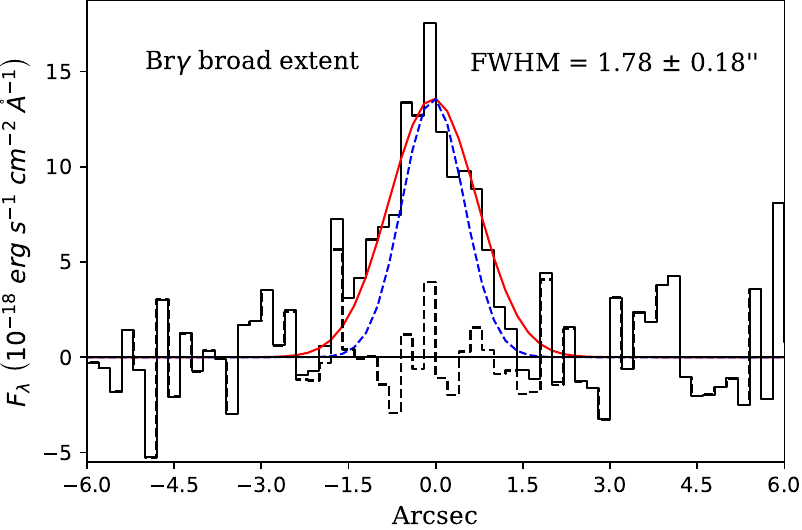}
    \end{minipage}
    \begin{minipage}{0.5\linewidth}
    \centering
    \includegraphics[width=0.99\linewidth]{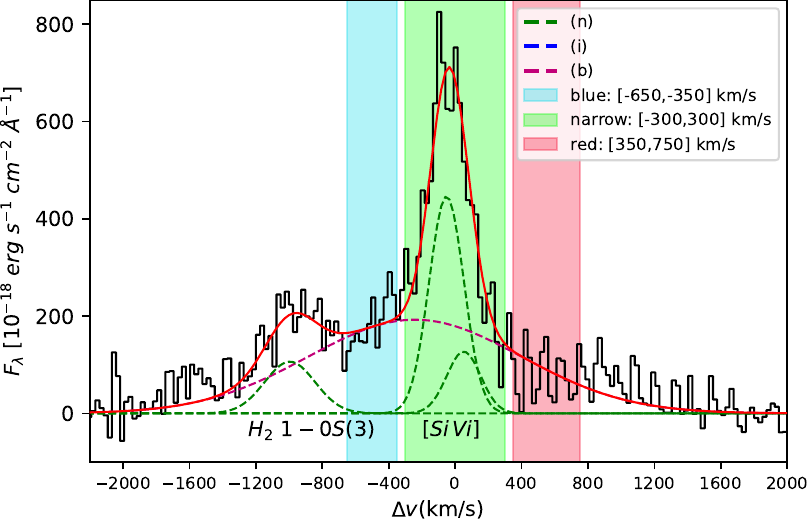}
    \includegraphics[width=0.99\linewidth]{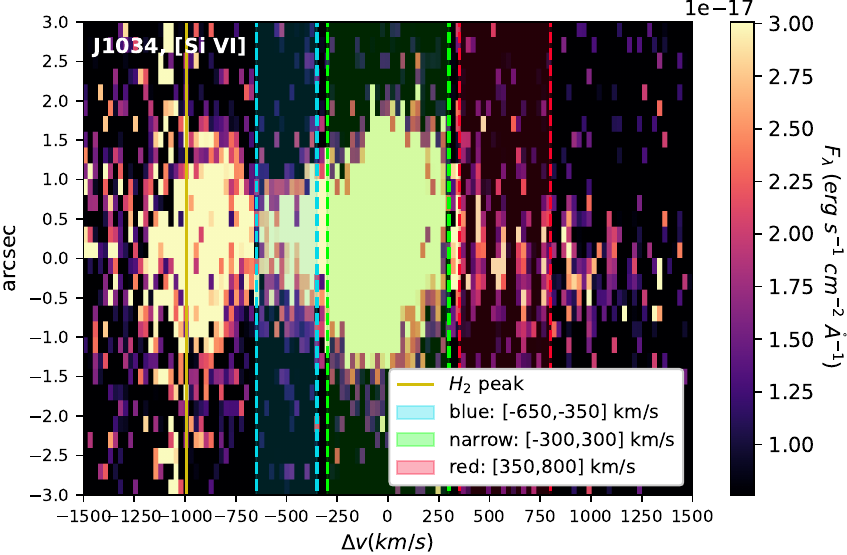}
    \includegraphics[width=0.99\linewidth]{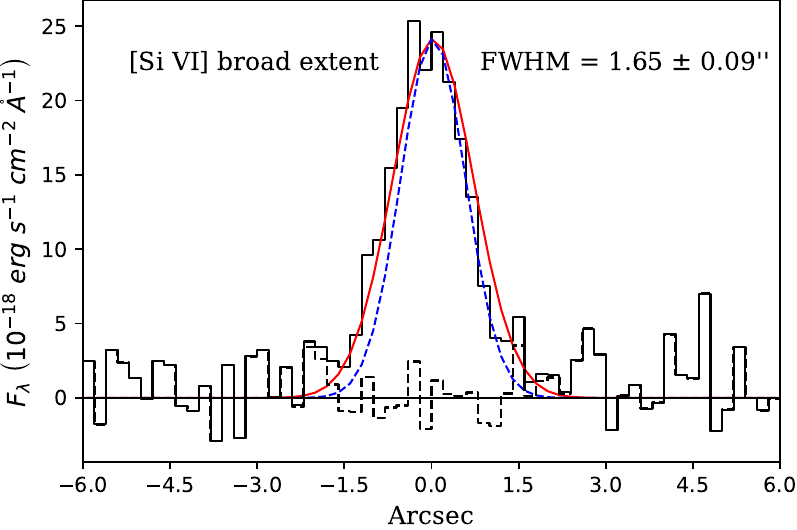}
    \end{minipage}
    \caption{Same as in Fig. \ref{fig:out_ext_J0802}, but for J1034+60 and using \brg~instead of \pa.}
    \label{fig:out_ext_J1034}
\end{figure}

\begin{figure}[!htbp]
    \centering
    \begin{minipage}{0.5\linewidth} 
    \centering
    \includegraphics[width=0.99\linewidth]{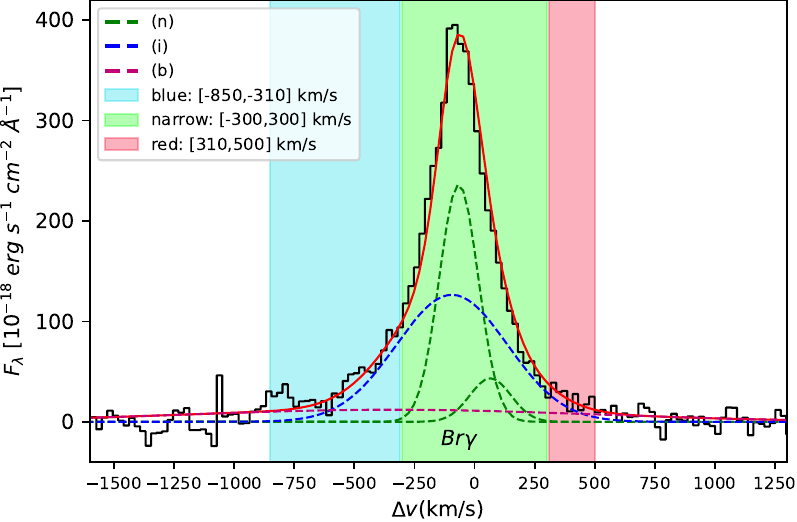}
    \includegraphics[width=0.99\linewidth]{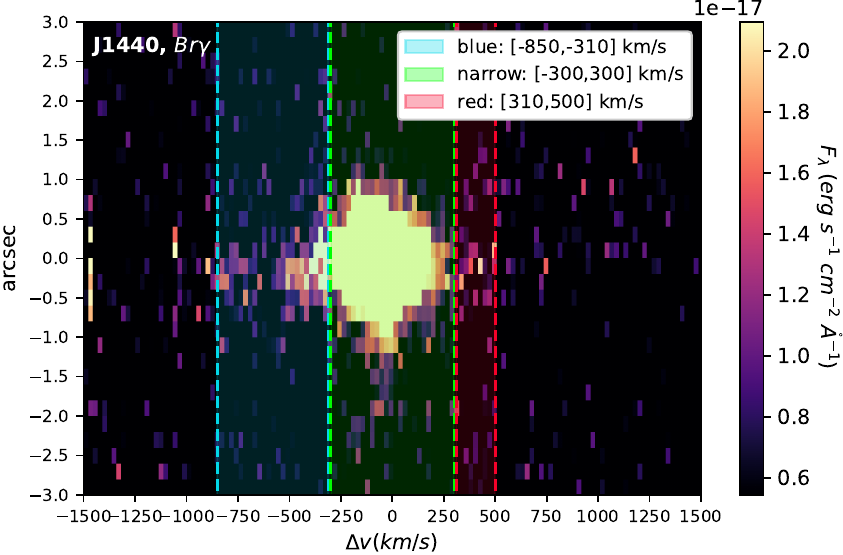}
    \includegraphics[width=0.99\linewidth]{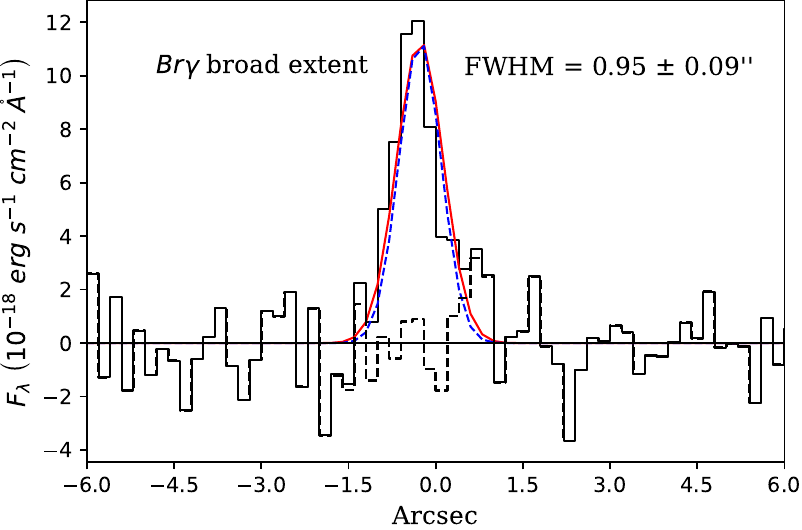}
    \end{minipage}
    \begin{minipage}{0.5\linewidth}
    \centering
    \includegraphics[width=0.99\linewidth]{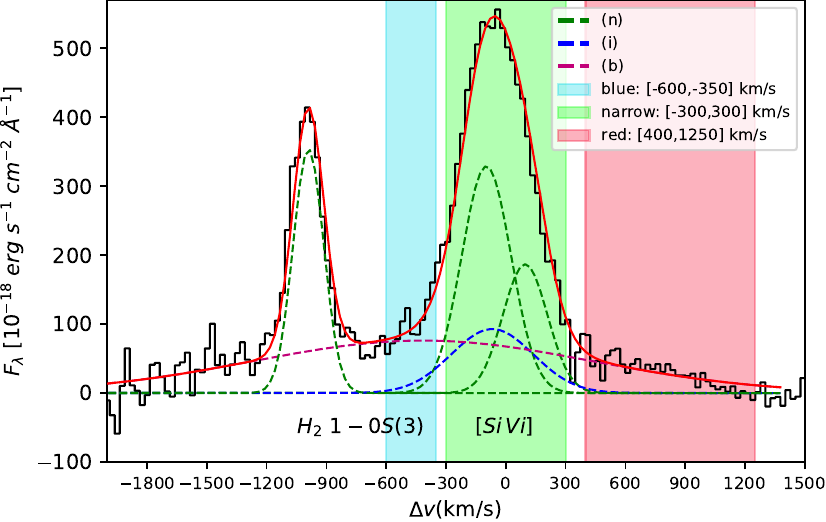}
    \includegraphics[width=0.99\linewidth]{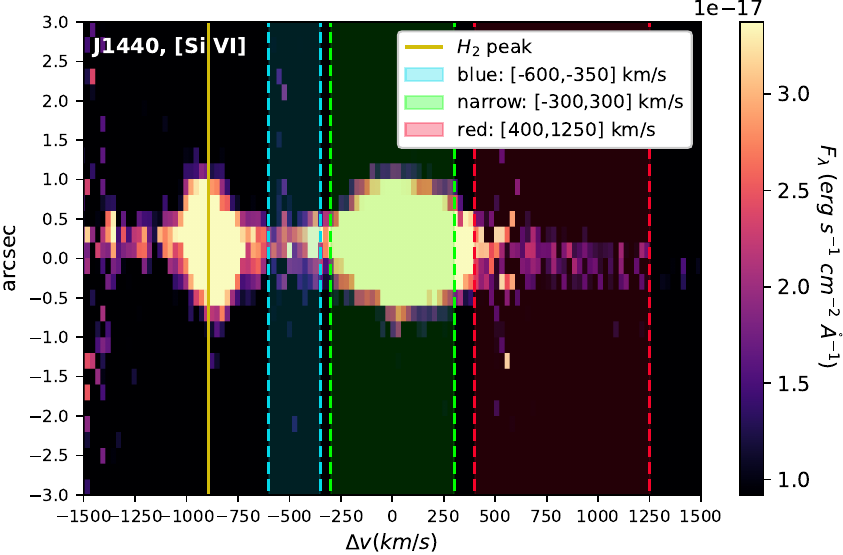}
    \includegraphics[width=0.99\linewidth]{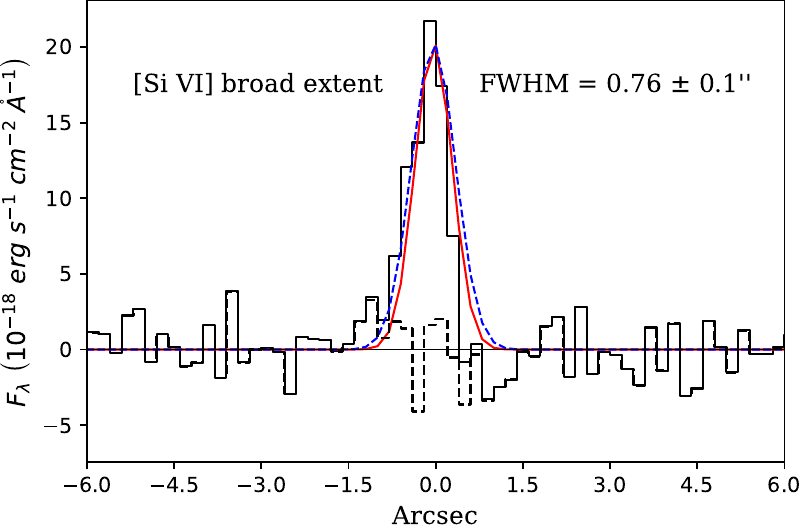}
    \end{minipage}
    \caption{Same as in Fig. \ref{fig:out_ext_J0802}, but for J1440+53 and using \brg~instead of \pa.}
    \label{fig:out_ext_J1440}
\end{figure}

\begin{figure}[!htbp]
    \centering
    \begin{minipage}{0.5\linewidth} 
    \centering
    \includegraphics[width=0.99\linewidth]{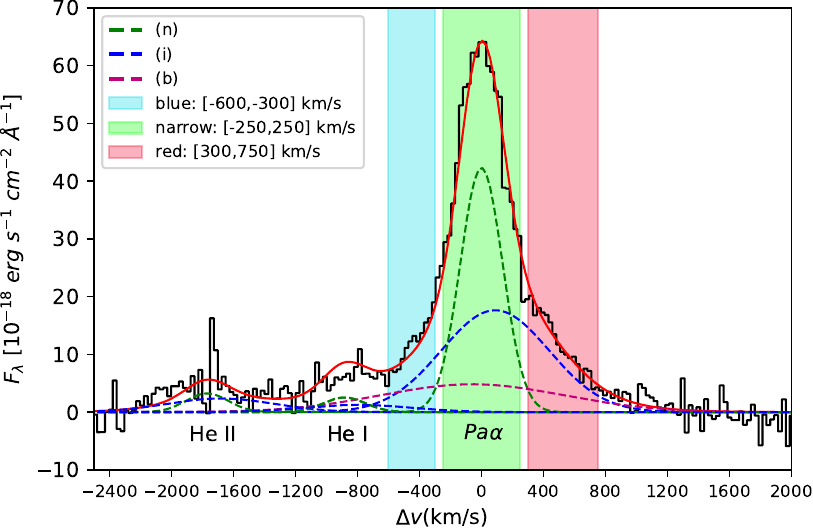}
    \includegraphics[width=0.99\linewidth]{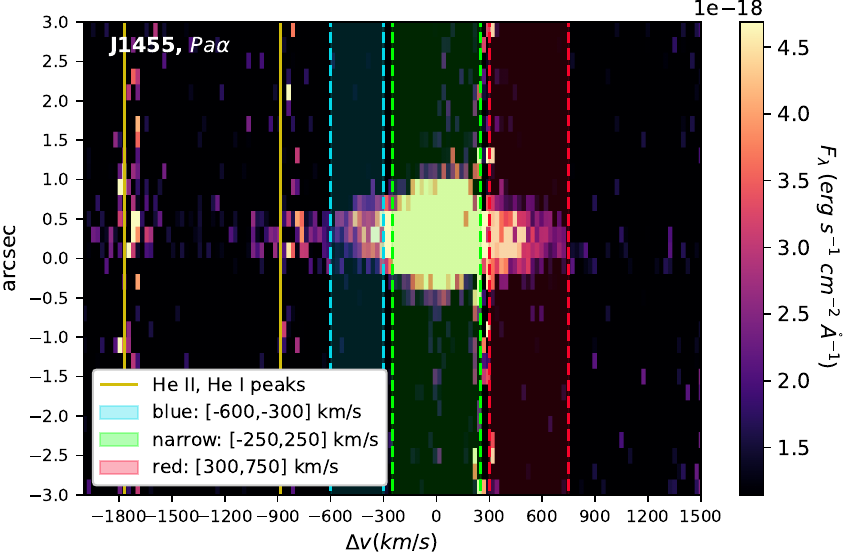}
    \includegraphics[width=0.99\linewidth]{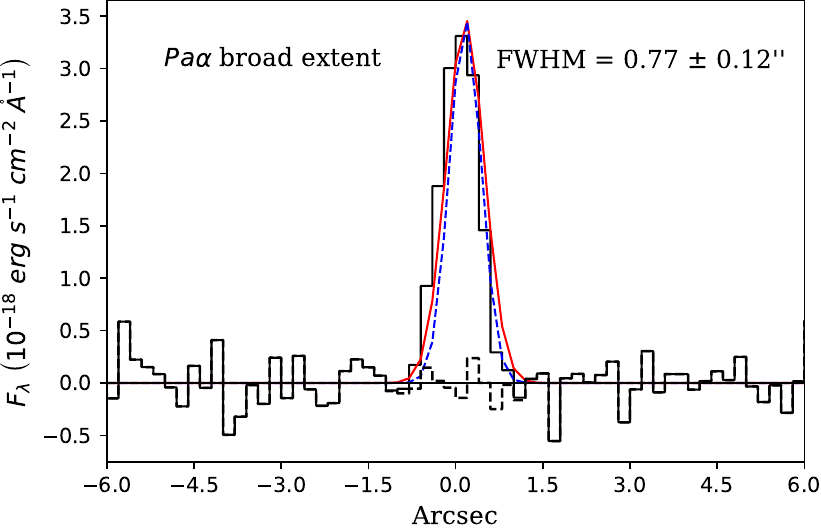}
    \end{minipage}
    \begin{minipage}{0.5\linewidth}
    \centering
    \includegraphics[width=0.99\linewidth]{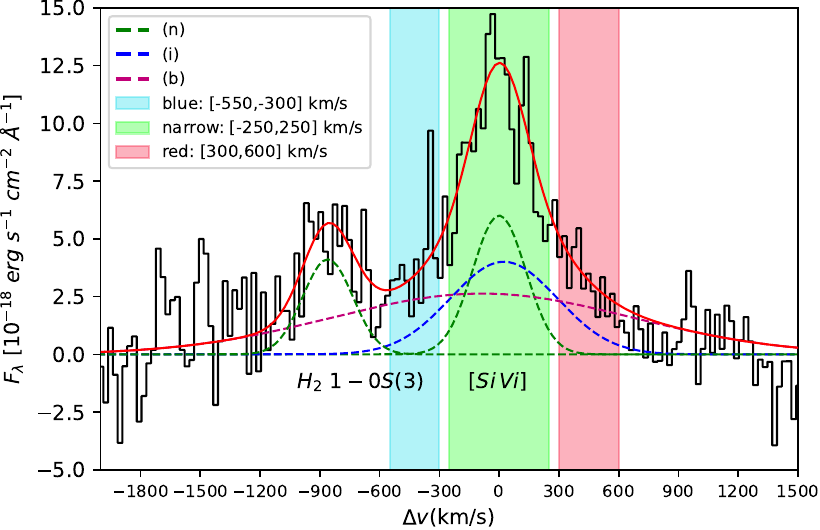}
    \includegraphics[width=0.99\linewidth]{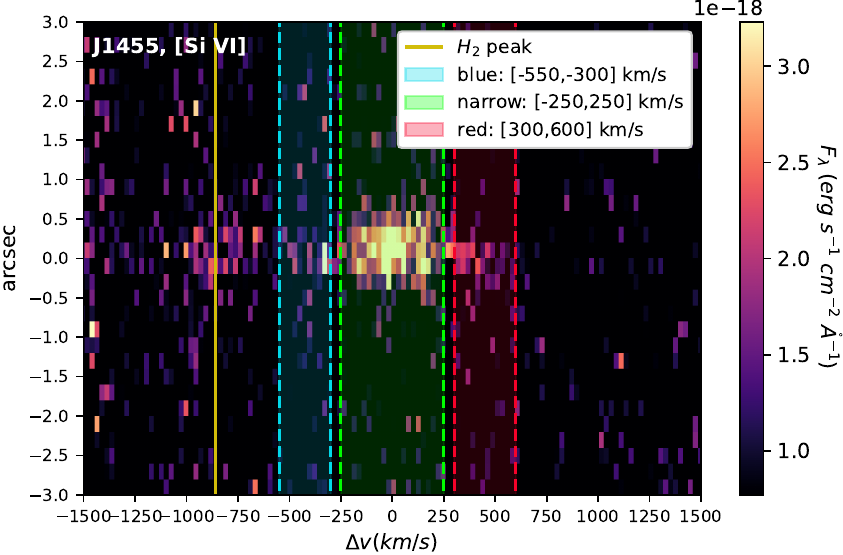}
    \includegraphics[width=0.99\linewidth]{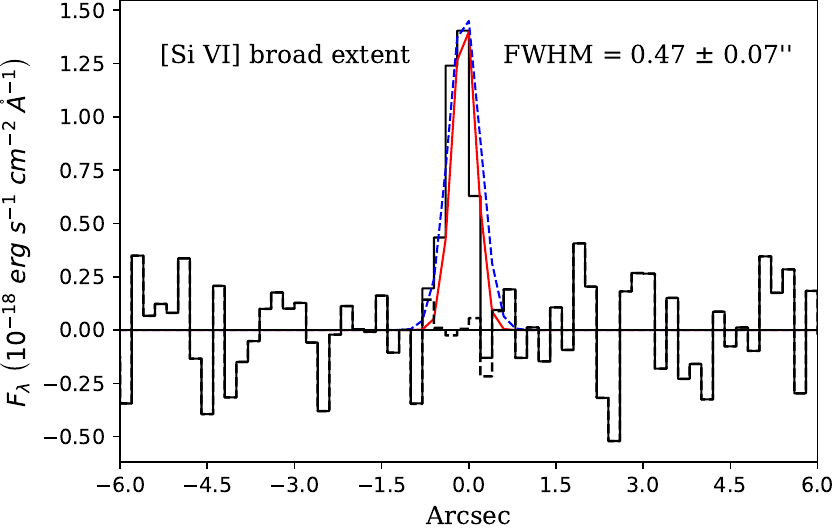}
    \end{minipage}
    \caption{Same as in Fig. \ref{fig:out_ext_J0802}, but for J1455+32.}
    \label{fig:out_ext_J1455}
\end{figure}

\begin{figure}[!htbp]
    \begin{minipage}{0.5\linewidth} 
    \centering
    \includegraphics[width=0.99\linewidth]{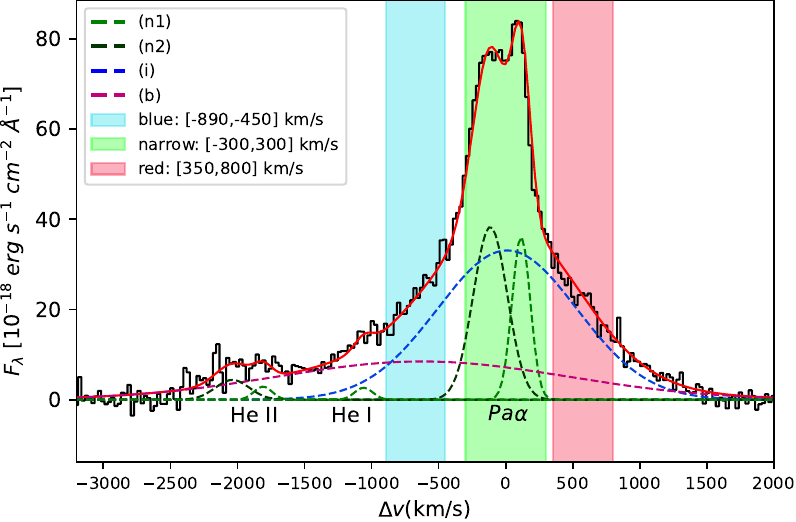}
    \includegraphics[width=0.99\linewidth]{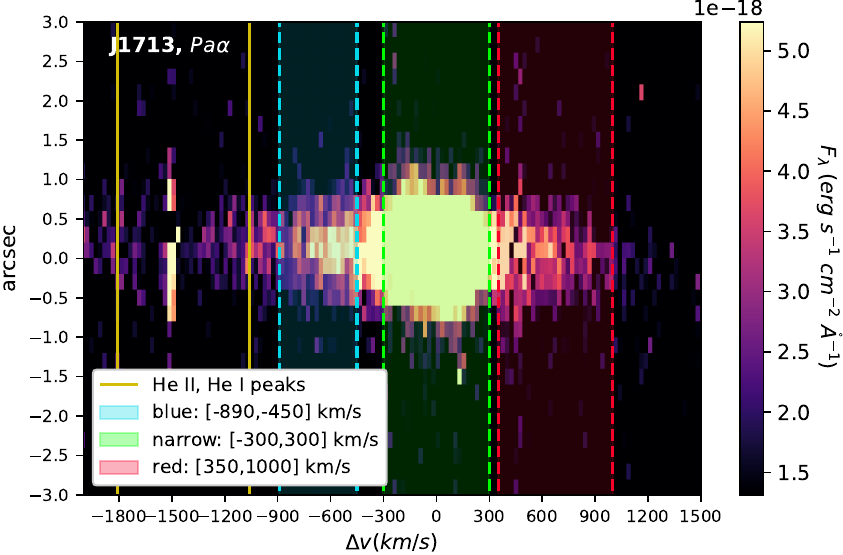}
    \includegraphics[width=0.99\linewidth]{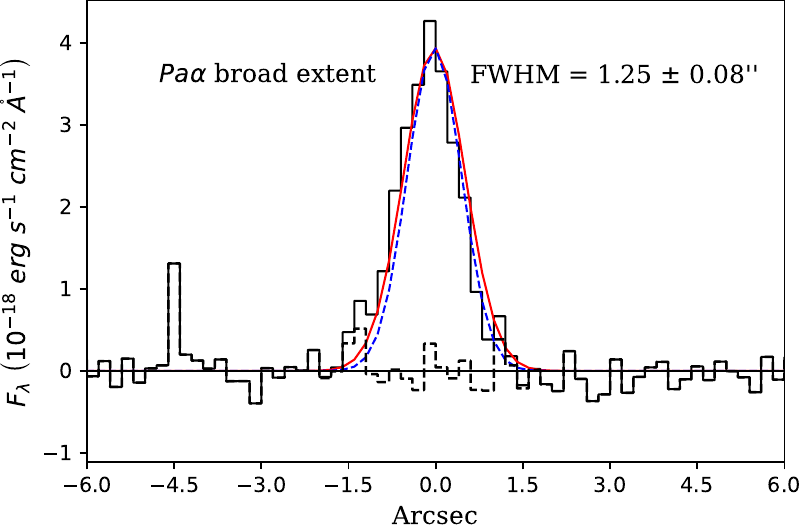}
    \end{minipage}
    \begin{minipage}{0.5\linewidth}
    \centering
    \includegraphics[width=0.99\linewidth]{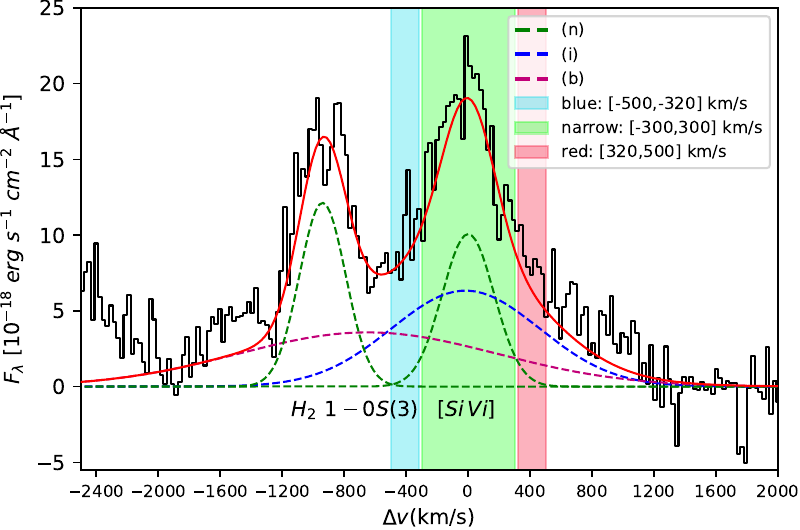}
    \includegraphics[width=0.99\linewidth]{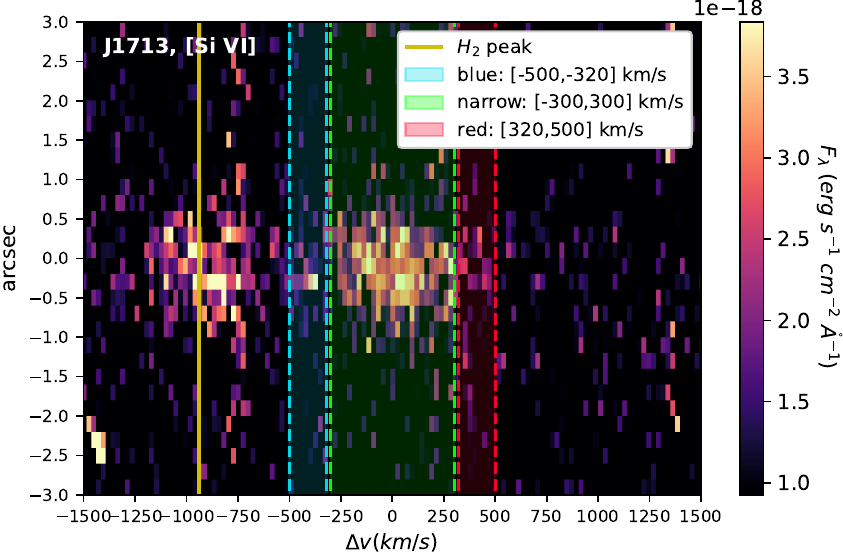}
    \includegraphics[width=0.99\linewidth]{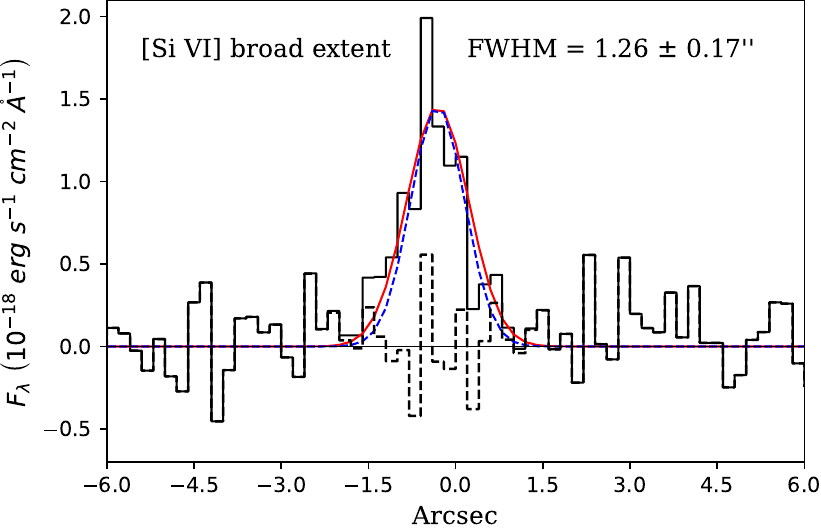}
    \end{minipage}
    \caption{Same as in Fig. \ref{fig:out_ext_J0802}, but for J1713+57.}
    \label{fig:out_ext_J1713}
\end{figure}

Figs. \ref{fig:out_ext_J0802} to \ref{fig:out_ext_J1713} show the spectral regions and spatial profiles used to calculate the outflow extents. While in the case of the \pa~and \brg~all the outflows are spatially resolved, in [Si VI] the outflows are not resolved for J0802+25, J1440+53, and J1455+32.   

\section{Outflow direct and physical properties}

\input{Tables/Table_directproperties}

\input{Tables/Table_summaryenergetics}

\end{appendix}

\end{document}

%% file: Tables/table_sample.tex
\begin{table*}
\caption{Main properties of the QSO2s.}
\centering
\tiny
\begin{tabular}{cccccccccccc}
\hline
SDSS ID & Short & z & $\rm D_L$ & Scale & log $\rm L_{[OIII]}$ & log $\rm L_{bol}$ & $\rm\log L_{\rm 1.4 GHz}$ & log $\rm M_{\rm BH}$ & $\rm \log \frac{L_{\rm bol}}{L_{\rm Edd}}$ & $\rm \log \mbox{M}_*$ & SFR\\ 
    & ID & SDSS & [Mpc] & [kpc/\arcsec]  & [erg s$^{-1}$] & [erg s$^{-1}$] & [W Hz$^{-1}$] & [M$\rm _{\odot}$] & & [M$_{\odot}$] & [M$_{\odot}$yr$^{-1}$]\\
\hline
J080252.92+255255.5	&	J0802+25	&	0.0811	&	369	&	1.529	&	42.84                   &	45.50	&	23.69	&	$	8.2	\pm	0.3	$	&	$	-0.7 \pm	0.3	$	&	11.3	& 44.7\\
J094521.33+173753.2	&	J0945+17	&	0.1280	&	600	&	2.287	&	43.32	                &	45.98	&	24.27	&	$	8.3	\pm	0.8	$	&	$	-0.3	\pm	0.8	$	&	10.9	& 14.1\\
J103408.59+600152.2	&	J1034+60	&	0.0511	&	227	&	0.998	&	42.74	              	&	45.40	&	23.07	&	$	7.8	\pm	0.3	$	&	$	-0.4	\pm	0.3	$	&	11.1	& 12.9\\
J144038.10+533015.9	&	J1440+53	&	0.0370	&	163	&	0.735	&	42.89	              	&	45.55	&	23.27	&	$	7.2	\pm	0.8	$	&	$	0.4	\pm	0.8	$	&	10.6    & 25.1\\
J145519.41+322601.8	&	J1455+32	&	0.0873	&	398	&	1.634	&	42.46	              	&	45.12	&	22.78	&	$	7.7	\pm	0.3	$	&	$	-0.6	\pm	0.4	$	&	10.6    & 12.6\\
J171350.32+572954.9	&	J1713+57	&	0.1128	&	524	&	2.050	&	42.99	              	&	45.65	&	23.37	&	$	7.4	\pm	0.4	$	&	$	0.3	\pm	0.4	$	&	11.1    & 18.2\\
\hline
\end{tabular}
\label{tab:sample} 
\tablefoot{(1-2) SDSS ID and short ID to be used throughout this paper; (3-5) SDSS spectroscopic redshift, luminosity distance, and physical scale; (6-7) extinction corrected [OIII] luminosity from \citet{Kong&Ho+18} and bolometric luminosity obtained from multiplying the [OIII] luminosities by 454 \citep{Lamastra+09}; (8) radio luminosity from \cite{Bessiere+24}; 
(9-10) black holes mass and Eddington ratio from \cite{Kong&Ho+18}; (11) stellar mass from \cite{Pierce+23}; and (12) star formation rate (SFR) over the past 100 Myr from \cite{Bessiere+24}.}
\end{table*}


%% file: Tables/table_observations.tex
\begin{table*}
\tiny
  \caption{Summary of the GTC/EMIR long slit observations.} 
  \begin{center}
\begin{tabular}{c c c  c c c c c c}
\hline
ID    & Exp.   & Slit pos. & Spectral range & PA &  Obs. date & Seeing FWHM & Airmass & Star \\
    & [s]   &  & [$\mathring{A}$] & [$^{\circ}$]&    &  [arcsec] & &  \\
\hline
 J0802+25 & 8 x 240 s & MR & 19780-23300 & -54 & ~~2018 Nov 25 & $\rm 0.8\pm 0.1$ & 1.01-1.03 & HD 76619 (A0*)  \\
          & & && & & & & HD 233594 (G0) \\

J0945+17 & 8 x 240 s & ML & 20810-24340 & 25 & ~~2018 Dec 18& $\rm 1.4\pm 0.2$ &  1.02 & BD+16 2159 (A2)  \\ 
 
J1034+60 &  8 x 240 s & MR & 19780-23300 & -60 & ~~2019 Mar 18 & $1.0\pm0.2$ &  1.19-1.22 & BD+60 1334 (G0)  \\ 

J1440+53	 & 8 x 240 s & MR & 19780-23300 & 44 &  ~~2019 Aug 15 & $0.8\pm0.1$  & 1.36-1.46 & HD 238495 (A5)  \\ 

J1455+32 &  8 x 240 s &  MR & 19780-23300 & 13  & ~~2019 Aug 14 & $0.8\pm0.1$  & 1.27-1.39 & BD+33 2599 (G3)  \\ 
J1713+57 &  
         12 x 240 s & C  & 20285-23820 & 45 & ~~ 2019 Jul 19 & $1.0\pm0.2$& 1.70-2.01 & HD 238926 (A5) \\
         & 12 x 240 s & C  & 20285-23820 & 45 & ~~2019 Jul 21& $0.9\pm0.1$& 1.32-1.46 & \\
\hline
\end{tabular}
\tablefoot{(1) Object ID; (2) exposure time; (3) slit position on the detector: middle left (ML), middle right (MR), and central (C); (4) wavelength coverage; (5) slit position angle (PA); (6) date of observation; (7) seeing measured from the FWHM of stars in the J-band acquisition images; (8) airmass;  (9) standard star name (spectral type). In the case of J0802+25, HD 76619 was used for flux calibration and both HD 233594 and HD 76619 for outflow extent determination. *Double or multiple star.}
\end{center}
\label{tab:obs} 
\end{table*}

%% file: Tables/Table_kinematics.tex
\begin{table}
\centering
\small
\caption{Ranges of velocity shifts and FWHMs measured for the different components fitted to the atomic lines.}
\begin{tabular}{ccccc}
\hline
\multirow{3}{*}{Component} &
  \multicolumn{2}{c}{\pa~/~\brg} &
  \multicolumn{2}{c}{\sivi} \\
             & $\rm v_s$ & FWHM & $\rm v_s$ & FWHM \\
             & [\kms] & [\kms] & [\kms] & [\kms]  \\
             
\hline
narrow       & [-112,118] &  128-443    &  [-48,151]  & 223-593     \\
intermediate & [-132,580] &  537-1225   &  [-248,775] & 475-1119    \\
broad        & [-619,-60] &  1183-2546  &  [-618,-60] & 1361-2012   \\
\hline
\end{tabular}
\tablefoot{All velocity shifts are relative to the central wavelength of the narrow component(s) fitted to either \pa~or \brg. When two narrow components are fitted, v$_s$ is relative to the amplitude weighted mean wavelength of the two components. FWHMs have been corrected for instrumental width.}
\label{tab:kinematic}
\end{table}

%% file: Tables/Table_ne.tex
\begin{table}
\centering
\caption{Electron temperatures and densities obtained from the [SII] and trans-auroral (TR) methods.}
\begin{tabular}{ccccc}
\hline
& \multicolumn{2}{c}{[SII] method}&\multicolumn{2}{c}{TR method}\\   
Quasar & $\rm T_e$ & log $\rm n_e$ & E(B-V) & log $\rm n_e$ \\ 
& [K]&[$\rm cm^{-3}$] & [mag]& [$\rm cm^{-3}$]\\ \hline 
J0802+25  &13544$^{+226}_{-231}$ & 2.71$^{+0.08}_{-0.09}$  & 0.24$^{+0.09}_{-0.07}$ & 3.56$^{+0.12}_{-0.14}$\\
J0945+17 & 12435$^{+163}_{-162}$ & 2.62$^{+0.07}_{-0.09}$  & 0.44$^{+0.04}_{-0.03}$ & 3.39$^{+0.08}_{-0.1}$\\
J1034+60  &12063$^{+195}_{-192}$ & 2.78$^{+0.08}_{-0.09}$  & 0.27$^{+0.07}_{-0.1}$  & 2.99$^{+0.19}_{-0.2}$\\
J1440+53  &16977$^{+206}_{-210}$ & 3.03$^{+0.04}_{-0.04}$  & 0.20$^{+0.08}_{-0.07}$ & 3.92$^{+0.09}_{-0.08}$\\
J1455+32  &14275$^{+388}_{-371}$ & 3.06$^{+0.09}_{-0.11}$  & 0.16$^{+0.06}_{-0.01}$ & 3.88$^{+0.05}_{-0.06}$\\
J1713+57  &15718$^{+366}_{-365}$ & 2.88$^{+0.09}_{-0.09}$  & 0.21$^{+0.07}_{-0.02}$ & 4.06$^{+0.05}_{-0.06}$\\
\hline
\end{tabular}%
\label{tab:density}
\end{table}

%% file: Tables/Table_extent_23Jun25.tex
\begin{table*}
\centering
\tiny
\caption{Spatial extents of the outflow and narrow components measured from Pa$\alpha$/Br$\gamma$ and [Si\,VI].}
\begin{tabular}{lcccccccccc}
\hline
QSO2  & FWHM$_{\rm seeing}$& Line  & FWHM$_{\rm out}$ & \multicolumn{2}{c}{$r_{\rm out}$}  & FWHM$_{\rm nar}$ & \multicolumn{2}{c}{$r_{\rm nar}$}  \\
 &   [arcsec] & &  [arcsec] &  [arcsec] & [kpc] &  [arcsec] &  [arcsec] & [kpc] \\
\hline
J0802+25 & 0.74 & Pa$\alpha$              & 0.97 $\pm$ 0.03     & 0.63 $\pm$ 0.06     & 0.96 $\pm$ 0.09     & 1.16 $\pm$ 0.03     & 0.90 $\pm$ 0.04     & 1.37 $\pm$ 0.06 \\
 & & [Si\,VI]                     & 0.65 $\pm$ 0.03$^*$ & $<$0.738            & $<$1.13             & 0.87 $\pm$ 0.03     & 0.46 $\pm$ 0.07     & 0.70 $\pm$ 0.10 \\
J0945+17  & 1.01 & Pa$\alpha$             & 1.37 $\pm$ 0.04     & 0.93 $\pm$ 0.06     & 2.13 $\pm$ 0.15     & 1.54 $\pm$ 0.02     & 1.17 $\pm$ 0.04     & 2.67 $\pm$ 0.08 \\
 & & [Si\,VI]                     & 1.54 $\pm$ 0.18     & 1.17 $\pm$ 0.25     & 2.67 $\pm$ 0.57     & 1.50 $\pm$ 0.05     & 1.11 $\pm$ 0.08     & 2.55 $\pm$ 0.18 \\
J1034+60  & 1.25 & Br$\gamma$             & 1.78 $\pm$ 0.18     & 1.27 $\pm$ 0.28     & 1.26 $\pm$ 0.28     & 2.25 $\pm$ 0.10     & 1.87 $\pm$ 0.12     & 1.87 $\pm$ 0.12 \\
 & & [Si\,VI]                     & 1.65 $\pm$ 0.09     & 1.08 $\pm$ 0.23     & 1.07 $\pm$ 0.23     & 1.98 $\pm$ 0.03     & 1.54 $\pm$ 0.04     & 1.53 $\pm$ 0.04 \\
J1440+53 & 0.85 & Br$\gamma$              & 0.95 $\pm$ 0.09$^{**}$ & 0.43 $\pm$ 0.23  & 0.32 $\pm$ 0.17     & 1.05 $\pm$ 0.02     & 0.62 $\pm$ 0.04     & 0.46 $\pm$ 0.03 \\
 & & [Si\,VI]                     & 0.76 $\pm$ 0.10     & $<$0.846           & $<$0.62             & 0.94 $\pm$ 0.02     & 0.41 $\pm$ 0.05     & 0.30 $\pm$ 0.04 \\
J1455+35 & 0.64 & Pa$\alpha$             & 0.77 $\pm$ 0.12     & 0.43 $\pm$ 0.22     & 0.70 $\pm$ 0.35     & 0.90 $\pm$ 0.03     & 0.63 $\pm$ 0.04     & 1.03 $\pm$ 0.07 \\
 & & [Si\,VI]                     & 0.47 $\pm$ 0.07     & $<$0.64             & $<$1.04             & 0.86 $\pm$ 0.05     & 0.58 $\pm$ 0.08     & 0.94 $\pm$ 0.13 \\
J1713+57  & 1.10  & Pa$\alpha$           & 1.25 $\pm$ 0.08     & 0.59 $\pm$ 0.18     & 1.21 $\pm$ 0.36     & 1.25 $\pm$ 0.03     & 0.59 $\pm$ 0.07     & 1.21 $\pm$ 0.15 \\
 & & [Si\,VI]                     & 1.26 $\pm$ 0.17     & 0.61 $\pm$ 0.45     & 1.25 $\pm$ 0.93     & 1.40 $\pm$ 0.09     & 0.86 $\pm$ 0.15     & 1.77 $\pm$ 0.31 \\
\hline
\end{tabular}
\tablefoot{Column 2 is the seeing FWHM measured from the K-band spectra of the standard stars shown in Table \ref{tab:obs}. In the case of J0802+25 FWHM$_{\rm seeing}$ is the average value measured from the stars HD 76619 and HD 233594. Columns 3-5 are the FWHM of the continuum-subtracted spatial profile of the outflow component and corresponding seeing deconvolved spatial extent (in arcsec and kpc). Columns 6-8 are the same, but for the narrow component. * Using only the red wing of the line because of blend with $\rm H_2$ 1-0S(3) on the blue side of the line.
** Using only the blue wing of the line because the redshifted outflow component is not spatially resolved.}
\label{tab:extent}
\end{table*}

%% file: Tables/Table_comparison.tex
\begin{table*}
\caption{Comparison between the outflow properties of different samples of QSO2s.}
\centering
\tiny
\begin{tabular}{lcccccccccc}
\hline
\multirow{2}{*}{Properties} & \multicolumn{2}{c}{6 QSO2s EMIR} & \multicolumn{2}{c}{6 QSO2s SDSS} & \multicolumn{2}{c}{48 QSO2s SDSS} & \multicolumn{2}{c}{18 z$\sim$0.3-0.4 QSO2s} & \multicolumn{2}{c}{4 z$\sim$2-3.5 QSO2s}\\ 
                            & range           & median          & range             & median            & range           & median          & range       & median &  range       & median     \\ \hline
log $\rm M_{Hion}$ [$\rm M_{\odot}$]                            &   4.90-7.30              &    5.81             &     4.51-5.59              &  4.91                 &        3.75-5.88         &        5.04         & 4.44-6.43 &5.35 & 5.85-7.88 & 7.30 \\ 
$\rm \dot{M}_{Hion}$  [$\rm M_{\odot}~yr^{-1}$]                     &     0.03-6.2            &       0.6           &       0.1-1.5            &  0.35                 &      0.04-1.90           &      0.30           & 0.02-0.72 & 0.14 & 0.2-27 & 6.2\\
log $\rm \dot{E}_{Hion}$  [$\rm erg~s^{-1}$]                     &  37.8-40.8               &       40.2          &   40.2-41.8                &   41.1                &      39.3-41.8           &  40.6     & 37.2-41.0 & 39.2 & 38.7-41.7 & 40.3\\

$\rm \xi_{Hion}$ [$\%$]                       &    (0.05-20)$\rm\times10^{-4}$ &  4$\rm \times10^{-4}$              &  (1-8)$\rm \times10^{-3}$    & 3$\rm\times10^{-3}$                  &       (0.1-10)$\rm \times10^{-3}$ &    2$\rm \times10^{-3}$            & (0.01-100)$\rm \times10^{-4}$ & 0.8$\rm \times10^{-4}$ & (0.02-3)$\rm \times10^{-3}$ & 10$\rm^{-3}$\\
$\rm \eta_{Hion}$                        &      0.002-0.482           &     0.022            &      0.008-0.106             &       0.013            &        0.001-1.408         &      0.019           &       \dots     &  \dots &  \dots &  \dots \\
\hline
\end{tabular} 
\tablefoot{Columns 2-3 are the range and median outflow properties measured here from the near-infrared hydrogen recombination lines (Hion). Columns 4-7 are the values reported by \cite{Bessiere+24} for the same six QSO2s and for the whole sample of 48 QSO2s. The authors used [OIII] measurements from SDSS data, adopted a non-parametric approach and TR-based electron densities, and assumed an outflow extent of $\rm r_{out}=$ 0.62 kpc. Columns 8-9 are the values reported for 18 QSO2s at z = 0.3-0.41 by \cite{Hervella+23}, using [OIII] from Gemini/GMOS-S data, adopting a parametric approach, and assuming $\rm r_{out}$ = 1 kpc. Since they assumed $\rm n_e= 200~cm^{-3}$, here we recalculated their outflow properties using the median TR-based electron density of the QSOFEED sample ($\rm n_e\sim 2570~cm^{-3}$; \citealt{Bessiere+24}).  Columns 10-11 are the values reported for four QSO2s at z = 2.0-3.5 from \cite{Bertola+25} and \cite{Perna+25}, observed with JWST NIRSpec. In \cite{Bertola+25} the \ha outflow properties were derived from parametric fits, using an outflow extent of $\rm r_{out}\sim$ 1 kpc, $\rm v_{out} = max(v_{10},v_{90})$, and assuming $\rm n_e$ = 1000 $\rm cm^{-3}$. In \cite{Perna+25} they used a parametric fit of the \hb line, assumed $\rm r_{out}\sim$ 3 kpc, $\rm n_e$ = 1000 $\rm cm^{-3}$, and $\rm v_{out} =$ 1000 \kms. Here we recomputed their outflow properties assuming $\rm n_e\sim 2570~cm^{-3}$ \citep{Bessiere+24} and $\rm v_{out} = v_s$. \cite{Bertola+25} defined the outflow velocity as $\rm v_{out} = max(v_{10},v_{90})$, with $\rm v_{max}\sim 1.2\times max(v_{10},v_{90})$. To infer $\rm v_s$, we assumed that $\rm v_s\sim 0.1\times v_{max}$ \citep{Hervella+23}.}
\label{tab:NIROPCOMP}
\end{table*}


%% file: Tables/Table_H2.tex
\begin{table*}[!htbp]
\centering
\tiny
\caption{QSO2s with NIR \hii measurements reported in the literature.}
\begin{tabular}{ccccccccc}
\hline
QSO2 & log L$\rm _{bol}$ & log L$\rm _{1.4~GHz}$ &On source&$\rm H_2$ outflow  &M$\rm _{H_2}^{out}$& M$\rm _{H_2}^{nuc}$ & M$\rm _{H_2}^{tot}$ & Ref.\\
& [erg s$\rm ^{-1}$]& [W Hz$\rm ^{-1}$] &exp. time [s] &detected & [$\rm 10^3$ M$_{\odot}$]& [$\rm 10^3$ M$_{\odot}$] & [$\rm 10^3$ M$_{\odot}$]&\\
\hline 
J0802+25        & 45.50 & 23.69  &1920&   No    &-& 10.7 $\pm$ 1.6   &    20.1 $\pm$ 3.1&  a  \\
J0945+17        & 45.98 & 24.27  &1920&   No    &-& 8.9 $\pm$ 4.9    &   11.9 $\pm$ 16.5 &  a  \\
J1034+60        & 45.40 & 23.07  &1920&   No    &-& 12.9 $\pm$ 2.4   &     31.9 $\pm$ 5.8 & a \\ 
J1440+53        & 45.55 & 23.27  &1920&   No    &-& 7.0 $\pm$ 1.3    &     12.7  $\pm$ 2.7 & a \\
J1455+32        & 45.12 & 22.78  &1920&   No    &-& 0.7 $\pm$ 0.4    &    1.1 $\pm$ 0.5 & a  \\
J1713+57        & 45.65 & 23.37  &5760&   No    &-& 6.1 $\pm$ 2.6    &     8.8 $\pm$ 4.0 & a \\
J1347+12        & 45.52 & 26.25  &26880&  No    &-&  -               & 21  &b   \\
J1430+13        & 45.82 & 23.67  &3711& Yes     &2.6& 5.9            &  45 & c\\
J1509+04        & 46.03 & 23.81  &1920& Yes     &10&     19          &  26.0  $\pm$ 5.5 & d,a \\
J1356+10        & 45.56 & 24.36  &3528& Yes     &1.5&     5.5        &  13  & c \\
F08572+3915 NW  & 45.62 & 22.59  &7200& Yes     &52&  $>$52          &  $>$52 & e  \\
\hline
\end{tabular}%
\tablefoot{Columns 2-4 list bolometric luminosity, radio luminosity at 1.4 GHz, and total on-source exposure time of the corresponding NIR observations. Columns 6-9 list the H$_2$ outflow masses when detected, nuclear and total H$_2$ masses, and their corresponding references. (a) This work, with nuclear masses computed from the H$_2$ 1-0S(1) line detected in the nuclear spectra shown in Figs. \ref{fig:fit_J0802}-\ref{fig:fit_J1713}, extracted with apertures of 0.6-3.2 kpc and total masses from spectra extracted with apertures of 2.1–6.1 kpc. 
(b) \citet{VillarMartin+23}, with total \hii mass was measured from the central 2.6 $\rm \times$ 8.8 kpc$\rm ^{2}$. (c) \citet{Zanchettin+25}, with nuclear and total H$_2$ masses measured in apertures of 1.2  and 4.8 kpc for J1430+13 and 1.8 kpc and 5.8 kpc for J1356+10. 
(d) \citet{RamosAlmeida+19}, with nuclear H$_2$ mass in the central 1.6 kpc and total in 6 kpc, the latter calculated here using the NIR spectrum. 
(e) \citet{Rupke+13b}, with the \hii mass in the outflow measured in the central 400 pc. L$\rm _{bol}$ and L$\rm _{1.4~GHz}$ from \cite{Bessiere+24}, except for F08572+3915, for which L$\rm _{bol}$ is from \citet{Herrera-Camus+20} and L$\rm _{1.4~GHz}$ was calculated 
using the flux of 4.89 mJy 
from FIRST \citep{Becker+95} and $\rm \alpha=0.7$.} 
\label{tab:H2}
\end{table*}

%% file: Tables/Tables_flux/J0802+25.tex
\begin{table}[!htbp]
\caption{Parameters derived from the fits of the lines of J0802+25.}
\centering
\tiny
\begin{tabular}{ccccc}
\hline
Line & FWHM & $\rm v_s$ & Flux x$\rm \,10^{-15}$ & Flux fraction \\
 & [$\rm km\,s^{-1}$] & [$\rm km\,s^{-1}$] & [erg $\rm cm^{-2}\,s^{-1}$)]& [$\rm \%$] \\
\hline
Pa$\alpha$ (n) & 443$\pm$12 & 0$\pm$5 & 7.74$\pm$1.11 & 56 \\
Pa$\alpha$ (i) & 719$\pm$48 & 580$\pm$32 & 2.41$\pm$0.39 & 17 \\
Pa$\alpha$ (b) & 1183 & -325 & 3.52$\pm$0.58 & 25 \\
He II (n) & 443 & -165$\pm$19 & 1.18$\pm$0.18 & 100 \\
He I (n) & 443 & 42$\pm$30 & 0.95$\pm$0.16 & 100 \\
Br$\delta$ (n) & 445 & -84$\pm$81 & 0.3$\pm$0.07 & 100 \\
\sivi~(n) & 593$\pm$106 & -47$\pm$41 & 1.83$\pm$0.45 & 59 \\
\sivi~(i) & 753$\pm$186 & 775$\pm$151 & 0.97$\pm$0.3 & 31 \\
\sivi~(b) & 1361 & -313 & 0.28$\pm$0.3 & 9 \\
\hii 1-0S(1) (n) & 421$\pm$19 & 17$\pm$7 & 1.43$\pm$0.21 & 100 \\
\hii 1-0S(2) (n) & 453$\pm$81 & -94$\pm$26 & 0.67$\pm$0.17 & 100 \\
\hii 1-0S(3) (n) & 420$\pm$30 & -16$\pm$11 & 1.55$\pm$0.26 & 100 \\
\hline
\end{tabular}
\tablefoot{Emission line and component fitted: narrow (n), intermediate (i), and broad (b). (2) FWHM corrected for instrumental broadening. (3) Velocity relative to the narrow component(s) of either Pa$\alpha$ or Br$\gamma$. (4) Integrated line flux. Since the measurements were done in the rest-frame spectrum, the fluxes include a multiplicative factor of (1+z). Errors were computed as the quadratic sum of the errors from the fit and the flux calibration error estimated from the standard star spectrum. Parameters without errors have been fixed to their initial guess.}
\label{tab:f_J0802+25} 
\end{table}

%% file: Tables/Tables_flux/J0945+17.tex
\begin{table}[!htbp]
\caption{Parameters derived from the fits of the lines of J0945+17.}
\centering
\tiny
\begin{tabular}{ccccc}
\hline
Line & FWHM & $\rm v_s$ & Flux x$\rm \,10^{-15}$ & Flux fraction \\
 & [$\rm km\,s^{-1}$] & [$\rm km\,s^{-1}$] & [erg $\rm cm^{-2}\,s^{-1}$)]& [$\%$] \\
\hline
Pa$\alpha$ (n) & 273$\pm$11 & 0$\pm$4 & 1.94$\pm$0.72 & 33 \\
Pa$\alpha$ (i) & 837$\pm$41 & -132$\pm$14 & 3.32$\pm$1.26 & 57 \\
Pa$\alpha$ (b) & 1663 & -374$\pm$129 & 0.51$\pm$0.33 & 8 \\
He II (n) & 272 & -53$\pm$32 & 0.12$\pm$0.05 & 41 \\
He II (i) & 837 & -184$\pm$34* & 0.18$\pm$0.07 & 58 \\
He I (n) & 272 & 62$\pm$29 & 0.2$\pm$0.08 & 45 \\
He I (i) & 837 & -69$\pm$39* & 0.24$\pm$0.14 & 54 \\
Br$\delta$ (n) & 540$\pm$198 & -79$\pm$44 & 0.26$\pm$0.14 & 100 \\
\sivi~(n) & 290 & -48$\pm$7 & 0.59$\pm$0.22 & 31 \\
\sivi~(i) & 762 & -248$\pm$41 & 0.62$\pm$0.29 & 33 \\
\sivi~(b) & 1601$\pm$340 & -449$\pm$288 & 0.65$\pm$0.36 & 34 \\
\hii 1-0S(1) (n) & 361$\pm$104 & -9$\pm$45 & 0.42$\pm$0.23 & 100 \\
\hii 1-0S(2) (n) & 202$\pm$198 & -129$\pm$78 & 0.11$\pm$0.12 & 100 \\
\hii 1-0S(3) (n) & 317$\pm$61 & -71$\pm$20 & 0.3$\pm$0.13 & 100 \\
\hii 1-0S(4) (n) & 280$\pm$188 & -23$\pm$38 & 0.1$\pm$0.08 & 100 \\
\hline
\end{tabular}
\tablefoot{Same as in Table \ref{tab:f_J0802+25}, but for J0945+17. 
* The shift between the n and i components was tied following the fit of Pa$\alpha$.}
\label{tab:f_J0945+17} 
\end{table}

%% file: Tables/Tables_flux/J1034+60.tex
\begin{table}[!htbp]
\caption{Parameters derived from the fits of the lines of J1034+60.}
\centering
\tiny
\begin{tabular}{ccccc}
\hline
Line & FWHM & $\rm v_s$ & Flux x$\rm \,10^{-15}$ & Flux fraction \\
 & [$\rm km\,s^{-1}$] & [$\rm km\,s^{-1}$] & [erg $\rm cm^{-2}\,s^{-1}$)]& [$\rm \%$] \\
\hline
Br$\gamma$ (n1) & 221 & 63 & 1.63$\pm$0.46 & 10 \\
Br$\gamma$ (n2) & 219$\pm$86 & -60 & 1.7$\pm$0.79 & 11 \\
Br$\gamma$ (b) & 1496 & -127$\pm$10 & 12.03$\pm$2.16 & 78 \\
Br$\delta$ (n) & 270$\pm$99 & -6$\pm$35 & 2.7$\pm$1.14 & 100 \\
\sivi~ (n1) & 223$\pm$59 & 100 & 2.24$\pm$2.26 & 7 \\
\sivi~ (n2) & 230$\pm$46 & -3$\pm$23 & 8.08$\pm$2.15 & 25 \\
\sivi~ (b) & 1508 & -194$\pm$48 & 21.3$\pm$3.78 & 67 \\
\hii 1-0S(1) (n) & 348$\pm$20 & 23$\pm$13 & 4.48$\pm$0.82 & 100 \\
\hii 1-0S(2) (n) & 452$\pm$40 & -16$\pm$19 & 2.84$\pm$0.57 & 100 \\
\hii 1-0S(3) (n) & 346 & -60$\pm$27 & 2.78$\pm$0.68 & 100 \\
\hline
\end{tabular}
\tablefoot{Same as in Table \ref{tab:f_J0802+25}, but for J1034+60.
}
\label{tab:f_J1034+60} 
\end{table}

%% file: Tables/Tables_flux/J1440+53.tex
\begin{table}[!htbp]
\caption{Parameters derived from the fits of the lines of J1440+53.}
\centering
\tiny
\begin{tabular}{ccccc}
\hline
Line & FWHM & $\rm v_s$ & Flux x$\rm \,10^{-15}$ & Flux fraction \\
 & [$\rm km\,s^{-1}$] & [$\rm km\,s^{-1}$] & [erg $\rm cm^{-2}\,s^{-1}$)]& [$\rm \%$] \\
\hline
Br$\gamma$ (n1) & 173$\pm$23 & -20$\pm$8 & 3.61$\pm$0.85 & 30 \\
Br$\gamma$ (n2) & 184$\pm$48 & 111 & 0.7$\pm$0.32 & 5 \\
Br$\gamma$ (i) & 537$\pm$58 & -49$\pm$22 & 5.48$\pm$1.76 & 46 \\
Br$\gamma$ (b) & 1991 & -300$\pm$131 & 1.92$\pm$0.4 & 16 \\
Br$\delta$ (n) & 173 & -56$\pm$10 & 4.53$\pm$0.84 & 100\\
$\rm[Si~VI]$ (n1) & 277$\pm$120 & -44$\pm$66 & 6.93$\pm$4.35 & 27 \\
$\rm[Si~VI]$ (n2) & 247$\pm$147 & 151$\pm$133 & 3.56$\pm$3.34 & 14 \\
$\rm[Si~VI]$ (i) & 475 & -14$\pm$320 & 3.25$\pm$3.7 & 13 \\
$\rm[Si~VI]$ (b) & 2012$\pm$832 & -364$\pm$407 & 11.07$\pm$5.77 & 44 \\
\hii 1-0S(0) (n) & 123$\pm$21 & -36$\pm$9 & 1.62$\pm$0.39 & 100 \\
\hii 1-0S(1) (n) & 152$\pm$7 & -27$\pm$7 & 4.83$\pm$0.88 & 100 \\
\hii 1-0S(2) (n) & 165$\pm$12 & -41$\pm$8 & 2.11$\pm$0.41 & 100 \\
\hii 1-0S(3) (n) & 157$\pm$24 & -50$\pm$10 & 4.64$\pm$1.11 & 100 \\
\hline
\end{tabular}
\tablefoot{Same as in Table \ref{tab:f_J0802+25}, but for J1440+53.
}
\label{tab:f_J1440+53} 
\end{table}

%% file: Tables/Tables_flux/J1455+35.tex
\begin{table}[!htbp]
\caption{Parameters derived from the fits of the lines of J1455+32.}
\centering
\tiny
\begin{tabular}{ccccc}
\hline
Line & FWHM & $\rm v_s$ & Flux x$\rm \,10^{-15}$ & Flux fraction \\
 & [$\rm km\,s^{-1}$] & [$\rm km\,s^{-1}$] & [erg $\rm cm^{-2}\,s^{-1}$)]& [$\rm \%$] \\
\hline
Pa$\alpha$ (n) & 308$\pm$54 & 0$\pm$35 & 0.99$\pm$0.41 & 39 \\
Pa$\alpha$ (i) & 812$\pm$231 & 89$\pm$181 & 1.04$\pm$0.63 & 41 \\
Pa$\alpha$ (b) & 1386$\pm$492 & -60 & 0.49$\pm$1.01 & 19 \\
He II (n) & 308 & 57$\pm$99 & 0.08$\pm$0.05 & 35 \\
He II (i) & 811 & 146$\pm$162* & 0.14$\pm$0.11 & 64 \\
He I (n) & 308 & 167$\pm$143 & 0.06$\pm$0.05 & 44 \\
He I (i) & 811 & 256$\pm$227* & 0.07$\pm$0.12 & 55 \\
Br$\delta$ (n) & 416$\pm$101 & 68$\pm$50 & 0.1$\pm$0.04 & 100 \\
$\rm[Si~VI]$ (n) & 301 & 25$\pm$39 & 0.14$\pm$0.07 & 20 \\
$\rm[Si~VI]$ (i) & 625$\pm$213 & 48$\pm$242 & 0.19$\pm$0.12 & 27 \\
$\rm[Si~VI]$ (b) & 1772$\pm$2023 & -60 & 0.35$\pm$0.44 & 51 \\
\hii 1-0S(1) (n) & 279$\pm$98 & 64$\pm$63 & 0.08$\pm$0.05 & 100 \\
\hii 1-0S(3) (n) & 282$\pm$239 & 52$\pm$61 & 0.09$\pm$0.08 & 100 \\
\hline
\end{tabular}
\tablefoot{Same as in Table \ref{tab:f_J0802+25}, but for J1455+32. * The shift between the n and i components was tied following the fit of Pa$\alpha$.
}
\label{tab:f_J1455+35} 
\end{table}

%% file: Tables/Tables_flux/j1713+57.tex
\begin{table}[!htbp]
\caption{Parameters derived from the fits of the lines of J1713+57.}
\centering
\tiny
\begin{tabular}{ccccc}
\hline
Line & FWHM & $\rm v_s$ & Flux x$\rm \,10^{-15}$ & Flux fraction \\
 & [$\rm km\,s^{-1}$] & [$\rm km\,s^{-1}$] & [erg $\rm cm^{-2}\,s^{-1}$)]& [$\rm \%$] \\
\hline
Pa$\alpha$ (n1) & 128$\pm$28 & 118$\pm$15 & 0.43$\pm$0.19 & 7 \\
Pa$\alpha$ (n2) & 284$\pm$47 & -112$\pm$24 & 0.85$\pm$0.37 & 14 \\
Pa$\alpha$ (i) & 1225$\pm$85 & 14$\pm$50 & 3.01$\pm$1.26 & 51 \\
Pa$\alpha$ (b) & 2546$\pm$803 & -619$\pm$370 & 1.6$\pm$0.97 & 27 \\
He II (n1) & 128 & 21$\pm$50 & 0.03$\pm$0.02 & 26 \\
He II (n2) & 284 & -209$\pm$50* & 0.1$\pm$0.05 & 73 \\
He I (n) & 128 & -5$\pm$48 & 0.03$\pm$0.02 & 100 \\
Br$\delta$ (n) & 278$\pm$115 & 28$\pm$56 & 0.09$\pm$0.05 & 100 \\
$\rm[Si~VI]$ (n) & 384$\pm$87 & 31$\pm$21 & 0.31$\pm$0.15 & 21 \\
$\rm[Si~VI]$ (i) & 1119$\pm$122 & 14 & 0.55$\pm$0.29 & 39 \\
$\rm[Si~VI]$ (b) & 1966$\pm$104 & -618 & 0.55$\pm$0.25 & 38 \\
\hii 1-0S(1) (n) & 392$\pm$35 & 36$\pm$23 & 0.41$\pm$0.17 & 100 \\
\hii 1-0S(2) (n) & 462$\pm$46 & 5$\pm$30 & 0.19$\pm$0.08 & 100 \\
\hii 1-0S(3) (n) & 341$\pm$23 & -21$\pm$17 & 0.33$\pm$0.14 & 100 \\
\hii 1-0S(4) (n) & 284$\pm$71 & -11$\pm$34 & 0.08$\pm$0.04 & 100 \\
\hii 1-0S(5) (n) & 424$\pm$61 & -9$\pm$23 & 0.29$\pm$0.13 & 100 \\
\hline
\end{tabular}
\tablefoot{Same as in Table \ref{tab:f_J0802+25}, but for J1713+57.* The shift between the n and i components was tied following the fit of Pa$\alpha$. 
}
\label{tab:f_J1713+57} 
\end{table}

%% file: Tables/Table_directproperties.tex
\begin{table*}
\centering
\tiny
\caption{Direct outflow properties of the QSO2s.}
\begin{tabular}{cccccccccc}
\hline
QSO2 / Line & $\rm A_V$ & $\rm log(n_e)_{[SII]}$ & $\rm log(n_e)_{TR}$ & Flux x$\rm \,10^{-15}$ & $\rm log(L_{out})$ & $\rm v_{out}$ & $\rm v_{max}$ & FWHM & $\rm r_{out}$ \\
           & [mag] & [$\rm cm^{-3}$] & [$\rm cm^{-3}$] & [erg $\rm cm^{-2}\,s^{-1}$)] & [$\rm erg~s^{-1}$] & [$\rm km s^{-1}$]& [$\rm km~s^{-1}$] & [$\rm km~s^{-1}$] & [kpc] \\
\hline
J0802+25 &0.74$^{+0.28}_{-0.22}$ & 2.71$^{+0.08}_{-0.09}$ & 3.56$^{+0.12}_{-0.14}$ &\multicolumn{6}{c}{}\\
\pa~ (i) &&  &  & 2.41$\pm$0.39 &  41.09  & 580$\pm$32 & 1191$\pm$52 & 719$\pm$48 & 0.96$\pm$0.09 \\
 (b) &&&& 3.52$\pm$0.58 &  41.25 & -325 & 1330 &  1183  & 0.96$\pm$0.09 \\
$\rm[Si~VI]$ (i) &&&& 0.97$\pm$0.3 & 40.24 &  775$\pm$151 & 1415$\pm$219 & 753$\pm$186 & $<$1.13 \\
(b) &&&& 0.28$\pm$0.3 & 39.7 & -313 & 1469 &  1361 & $<$1.13 \\
\hline

J0945+17 &1.36$^{+0.12}_{-0.09}$& 2.62$^{+0.07}_{-0.09}$ & 3.39$^{+0.08}_{-0.1}$ &\multicolumn{6}{c}{}\\ 
\pa~ (i) & &  & & 3.32$\pm$1.26  & 41.69  & -132$\pm$14 & 843$\pm$38 & 837$\pm$41 & 2.13$\pm$0.15 \\
(b) &&&& 0.51$\pm$0.33 &   40.88 & -374$\pm$129 & 1786$\pm$129 & 1663 & 2.13$\pm$0.15 \\
$\rm[Si~VI]$ (i) &&&& 0.62$\pm$0.29 &  40.5  & -248$\pm$41 & 895$\pm$41 & 762 & 2.67$\pm$0.57 \\
(b) &&&& 0.65$\pm$0.36 & 40.52 & -449$\pm$288 & 1809$\pm$408 & 1601$\pm$340 & 2.67$\pm$0.57 \\
\hline

J1034+60 & 0.84$^{+0.22}_{-0.31}$ & 2.78$^{+0.08}_{-0.09}$ & 2.99$^{+0.19}_{-0.2}$ & \multicolumn{6}{c}{}\\
\brg~ (b) & && & 12.03$\pm$2.16  & 42.46  & -127$\pm$10 & 1398$\pm$11 & 1496 & 1.26$\pm$0.28 \\
$\rm[Si~VI]$ (b) &&&& 21.3$\pm$3.78 & 41.17 & -194$\pm$48 & 1475$\pm$48 & 1508 & 1.07$\pm$0.23 \\
 \hline

 J1440+53 &0.62$^{+0.25}_{-0.22}$ &3.03$^{+0.04}_{-0.04}$ & 3.92$^{+0.09}_{-0.08}$& \multicolumn{6}{c}{}\\
\brg~ (i) & && & 5.48$\pm$1.76 & 41.82 & -49$\pm$22 & 505$\pm$54  & 537$\pm$58 & 0.32$\pm$0.17 \\
(b) &&&& 1.92$\pm$0.4  & 41.37 & -300$\pm$131 & 1991$\pm$131 & 1991 & 0.32$\pm$0.17 \\
$\rm[Si~VI]$ (i) &&&& 3.25$\pm$3.7  & 40.05 & -14$\pm$320 & 417$\pm$320 &  475 & $<$0.62  \\
(b) &&&& 11.07$\pm$5.77 &  40.58 &  -364$\pm$407 & 2073$\pm$815 & 2012$\pm$832 & $<$0.62  \\
\hline

J1455+32 &0.50$^{+0.19}_{-0.03}$ & 3.06$^{+0.09}_{-0.11}$ & 3.88$^{+0.05}_{-0.06}$ & \multicolumn{6}{c}{}\\ 
\pa~ (i) & &&  & 1.04$\pm$0.63  & 40.78 & 89$\pm$181 & 779$\pm$267 & 812$\pm$231 & 0.7$\pm$0.35 \\
 (b) &&&& 0.49$\pm$1.01 &   40.45 & -60 & 1237$\pm$419 & 1386$\pm$492 & 0.7$\pm$0.35  \\
$\rm[Si~VI]$ (i) &&&& 0.19$\pm$0.12  & 39.59 & 48$\pm$242 & 579$\pm$302 & 625$\pm$213 & $<$1.04 \\
(b) &&&& 0.35$\pm$0.44 & 39.85 & -60 & 1565$\pm$1718 & 1772$\pm$2023 & $<$1.04  \\
\hline

J1713+57 & 0.65$^{+0.22}_{-0.06}$ & 2.88$^{+0.09}_{-0.09}$ & 4.06$^{+0.05}_{-0.06}$&\multicolumn{6}{c}{}\\
\pa~ (i) &&& & 3.01$\pm$1.26 & 41.49  & 14$\pm$50 & 1054$\pm$88 & 1225$\pm$85 & 1.21$\pm$0.36  \\
 (b) &&&& 1.6$\pm$0.97 & 41.21 & -619$\pm$370 & 2781$\pm$776 & 2546$\pm$803 & 1.21$\pm$0.36  \\
$\rm[Si~VI]$ (i) &&&& 0.55$\pm$0.29 &   40.29 & 14 & 964$\pm$104 & 1119$\pm$122 & 1.25$\pm$0.93  \\
(b) &&&& 0.55$\pm$0.25   & 40.29 & -618 & 2288$\pm$89 & 1966$\pm$104 & 1.25$\pm$0.93  \\
\hline

\end{tabular}
\tablefoot{(1) QSO2 ID and emission line component. (2) $\rm A_V = 3.1\times E(B-V)$ obtained from the trans-auroral density estimates from \cite{Bessiere+24} and shown in Table \ref{tab:density}. (3-4) Electron densities from Table \ref{tab:density}. (5) Emission line fluxes from Tables \ref{tab:f_J0802+25}-\ref{tab:f_J1713+57} of Appendix \ref{sec:ApA}. (6) Outflow luminosities ($\rm L_{H\beta}$, $\rm L_{[Si VI]}$). (7-9) Velocity shift ($\rm v_s$), $\rm v_{max} = |v_{s}|+2\sigma$ (with $\rm \sigma \approx FWHM/2.355$), and FWHM of the intermediate and broad components shown in Tables \ref{tab:f_J0802+25}-\ref{tab:f_J1713+57} of Appendix \ref{sec:ApA}. (10) Ouflow extent from Table \ref{tab:extent}.}
\label{tab:outflowprop}
\end{table*}

%% file: Tables/Table_summaryenergetics.tex
\begin{table*}
\centering
\tiny
\caption{Physical outflow properties obtained from the direct properties reported in Table \ref{tab:outflowprop}.}
\begin{tabular}{ccccccc|cccc}
\hline
QSO2 / Line & $\rm M_{out} \times 10^{6}$ & $\rm M_{out}/M_{tot}$ & $\rm \dot{M}_{out}$ & $\rm log(\dot{E}_{kin})$ & $\rm \xi$ & $\rm \eta$ & $\rm \dot{M}_{max}$ & $\rm log(\dot{E}_{kin,max})$ & $\rm \xi_{max}$ & $\rm \eta_{max}$\\
           & [$\rm M_{\odot}$] &  & [$\rm M_{\odot}~yr^{-1}$] & [$\rm erg~s^{-1}$] & [$\rm \%$] &  & [$\rm M_{\odot}~yr^{-1}$] & [$\rm erg~s^{-1}$] & [$\rm \%$] & \\
\hline
\multicolumn{11}{c}{$\rm n_e$ - [SII] method}\\
\hline
J0802+25  &\multicolumn{10}{c}{}\\ 
Hion & 4.01$\pm$0.74 & 0.43 & 5.5$\pm$1.1 & 41.6 & 0.01 & 0.123 & 16.3$\pm$3.3 & 42.92 & 0.3 & 0.364 \\
$\rm[Si~VI]$ & 0.64$\pm$0.24 & 0.41 & $>$1.2 & $>$41.3 & $>$0.006 & $>$0.026 & $>$2.5 & $>$42.2 & $>$0.05 & $>$0.055 \\
\hline
J0945+17  &\multicolumn{10}{c}{}\\
Hion & 9.2$\pm$3.44 & 0.66 & 2.2$\pm$0.8 & 40.57 & 0.0004 & 0.154 & 12.8$\pm$4.6 & 42.73 & 0.06 & 0.909 \\
$\rm[Si~VI]$ & 2.28$\pm$0.88 & 0.68 & 0.9$\pm$0.6 & 40.65 & 0.0005 & 0.065 & 3.6$\pm$1.7 & 42.45 & 0.03 & 0.254 \\
\hline
J1034+60   &\multicolumn{10}{c}{}\\ 
Hion & 32.66$\pm$8.72 & 0.78 & 10.1$\pm$3.6 & 40.71 & 0.002 & 0.781 & 111.1$\pm$38.6 & 43.83 & 2.7 & 8.612 \\
$\rm[Si~VI]$ & 3.56$\pm$0.95 & 0.67 & 2.0$\pm$0.8 & 40.37 & 0.0009 & 0.153 & 15.1$\pm$5.2 & 43.01 & 0.4 & 1.167 \\
 \hline
J1440+53   &\multicolumn{10}{c}{}\\ 
Hion & 5.67$\pm$1.45 & 0.63 & 6.2$\pm$3.4 & 41.08 & 0.003 & 0.247 & 48.4$\pm$20.8 & 43.56 & 1.0 & 1.928 \\
$\rm[Si~VI]$ & 0.67$\pm$0.33 & 0.58 & $>$1.0 & $>$40.59 & $>$0.001 & $>$0.038 & $>$5.6 & $>$42.86 & $>$0.2 & $>$0.225 \\
\hline
J1455+32  &\multicolumn{10}{c}{}\\ 
Hion & 0.52$\pm$0.42 & 0.61 & 0.2$\pm$0.3 & 38.59 & 0.00003 & 0.014 & 2.1$\pm$2.2 & 41.82 & 0.05 & 0.168 \\
$\rm[Si~VI]$ & 0.14$\pm$0.12 & 0.79 & $>$0.02 & $>$37.37 & $>$0.000002 & $>$0.002 & $>$0.5 & $>$41.52 & $>$0.03 & $>$0.04 \\
\hline
J1713+57  &\multicolumn{10}{c}{}\\ 
Hion & 4.2$\pm$1.59 & 0.78 & 2.4$\pm$2.1 & 41.44 & 0.006 & 0.131 & 17.6$\pm$8.8 & 43.44 & 0.6 & 0.967 \\
$\rm[Si~VI]$ & 0.76$\pm$0.29 & 0.78 & 0.6$\pm$0.5 & 40.84 & 0.002 & 0.032 & 3.0$\pm$2.1 & 42.57 & 0.08 & 0.165 \\
\hline
\multicolumn{11}{c}{$\rm n_e$ - TR method}\\
\hline
J0802+25  &\multicolumn{10}{c}{}\\ 
Hion & 0.57$\pm$0.14 & 0.43 & 0.8$\pm$0.2 & 40.75 & 0.002 & 0.017 & 2.3$\pm$0.6 & 42.07 & 0.04 & 0.052 \\
$\rm[Si~VI]$ & 0.09$\pm$0.04 & 0.41 & $>$0.2 & $>$40.45 & $>$0.0009 & $>$0.004 & $>$0.4 & $>$41.35 & $>$0.007 & $>$0.008\\
\hline
J0945+17  &\multicolumn{10}{c}{}\\ 
Hion & 1.56$\pm$0.6 & 0.66 & 0.4$\pm$0.1 & 39.8 & 0.00007 & 0.026 & 2.2$\pm$0.8 & 41.96 & 0.01 & 0.155 \\
$\rm[Si~VI]$ & 0.39$\pm$0.15 & 0.68 & 0.2$\pm$0.1 & 39.88 & 0.00008 & 0.011 & 0.6$\pm$0.3 & 41.68 & 0.005 & 0.043 \\
\hline
J1034+60   &\multicolumn{10}{c}{}\\ 
Hion & 20.14$\pm$9.95 & 0.78 & 6.2$\pm$3.4 & 40.5 & 0.001 & 0.482 & 68.5$\pm$37.1 & 43.62 & 1.7 & 5.309 \\
$\rm[Si~VI]$ & 2.2$\pm$1.08 & 0.67 & 1.2$\pm$0.7 & 40.16 & 0.0006 & 0.095 & 9.3$\pm$5.0 & 42.8 & 0.3 & 0.72 \\
\hline
J1440+53   &\multicolumn{10}{c}{}\\ 
Hion & 0.73$\pm$0.21 & 0.63 & 0.8$\pm$0.5 & 40.19 & 0.0004 & 0.032 & 6.2$\pm$2.8 & 42.67 & 0.1 & 0.248 \\
$\rm[Si~VI]$ & 0.09$\pm$0.04 & 0.58 & $>$0.1 & $>$39.7 & $>$0.0001 & $>$0.005 & $>$0.7 & $>$41.97 & $>$0.03 & $>$0.029 \\
\hline
J1455+32  &\multicolumn{10}{c}{}\\ 
Hion & 0.08$\pm$0.06 & 0.61 & 0.03$\pm$0.05 & 37.77 & 0.000005 & 0.002 & 0.3$\pm$0.3 & 41.0 & 0.008 & 0.025 \\
$\rm[Si~VI]$ & 0.02$\pm$0.02 & 0.79 & $>$0.004 & $>$36.55 & $>$0.0000003 & $>$0.0003 & $>$0.08 & $>$40.7 & $>$0.004 & $>$0.006\\
\hline
J1713+57  &\multicolumn{10}{c}{}\\ 
 Hion & 0.28$\pm$0.1 & 0.78 & 0.2$\pm$0.1 & 40.26 & 0.0004 & 0.009 & 1.2$\pm$0.6 & 42.26 & 0.04 & 0.064 \\
$\rm[Si~VI]$ & 0.05$\pm$0.02 & 0.78 & 0.04$\pm$0.03 & 39.66 & 0.0001 & 0.002 & 0.2$\pm$0.1 & 41.39 & 0.01 & 0.011 \\
\hline
\end{tabular}
\tablefoot{(1) QSO2 ID and emission line; (2) mass in the outflow, calculated using the [SII]-based density for the upper part of the table and the TR-based density for the lower part; (3) fraction of the mass in the outflow over total mass (narrow + outflow components); (4) mass outflow rate; (5) kinetic power, computed as  $\rm \dot{E}_{kin} = 1/2~\dot{M}_{out} \times v_{out}^2$; (6) coupling efficiency defined as $\rm \xi = \dot{E}_{kin}/L_{Bol}$; (7) mass-loading factor defined as $\rm \eta = \dot{M}/SFR$; (8-11) same quantities as in (4-7), but computed using the maximum outflow velocities  $\rm v_{max} = |v_{s}|+2\sigma$.}
\label{tab:outfenergetics}
\end{table*}

%% file: References.bib
@inproceedings{Garzon+06,
author = {F. {Garz{\'o}n} and D. Abreu and S. Barrera and S. Becerril and L. M. Cair{\'o}s and J. J. D{\'i}az and A. B. Fragoso and F. Gago and R. Grange and C. Gonz{\'a}lez and P. L{\'o}pez and J. Patr{\'o}n and J. P{\'e}rez and J. L. Rasilla and P. Redondo and R. Restrepo and P. Saavedra and V. S{\'a}nchez and F. Tenegi and M. Vallb{\'e}},
title = {{EMIR: the GTC NIR multi-object imager-spectrograph}},
volume = {6269},
booktitle = {Ground-based and Airborne Instrumentation for Astronomy},
editor = {Ian S. McLean and Masanori Iye},
organization = {International Society for Optics and Photonics},
publisher = {SPIE},
pages = {626918},
year = {2006},
doi = {10.1117/12.671302},
URL = {https://doi.org/10.1117/12.671302}
}

@ARTICLE{RamosAlmeida+26,
       author = {{Ramos Almeida}, C. and {Asensio Ramos}, A. and {Westerdorp Plaza}, C. and {Garc{\'\i}a-Bernete}, I. and {Lopez-Rodriguez}, E. and {H{\"o}nig}, S. and {Audibert}, A. and {Garc{\'\i}a-Burillo}, S. and {Pereira-Santaella}, M. and {Donnan}, F. and {Alonso-Herrero}, A. and {Gonz{\'a}lez-Mart{\'\i}n}, O. and {Levenson}, N. and {Rigopoulou}, D. and {Tadhunter}, C. and {Speranza}, G.},
        title = "{Silicate emission in a type-2 quasar: JWST/MIRI constraints on torus geometry and radiative feedback}",
      journal = {arXiv e-prints},
         year = 2026,
        month = dec,
          eid = {arXiv:2512.02629},
        pages = {arXiv:2512.02629},
          doi = {10.48550/arXiv.2512.02629},
archivePrefix = {arXiv},
       eprint = {2512.02629},
 primaryClass = {astro-ph.GA},
       adsurl = {https://ui.adsabs.harvard.edu/abs/2025arXiv251202629R}
}

@inproceedings{Garzon+14,
author = {F. {Garz{\'o}n} and N. Castro-Rodr{\'i}guez and M. Insausti and L. L{\'o}pez-Mart{\'i}n and Peter Hammersley and M. Barreto and P. Fern{\'a}ndez and E. Joven and P. L{\'o}pez and A. Mato and H. Moreno and M. N{\'u}{\~n}ez and J. Patr{\'o}n and J. L. Rasilla and P. Redondo and J. Rosich and S. Pascual and R. Grange},
title = {{Results of the verification of the NIR MOS EMIR}},
volume = {9147},
booktitle = {Ground-based and Airborne Instrumentation for Astronomy V},
editor = {Suzanne K. Ramsay and Ian S. McLean and Hideki Takami},
organization = {International Society for Optics and Photonics},
publisher = {SPIE},
pages = {91470U},
year = {2014},
doi = {10.1117/12.2054804},
URL = {https://doi.org/10.1117/12.2054804}
}

@ARTICLE{Garzon+22,
       author = {{Garz{\'o}n}, F. and {Balcells}, M. and {Gallego}, J. and {Gry}, C. and {Guzm{\'a}n}, R. and {Hammersley}, P. and {Herrero}, A. and {Mu{\~n}oz-Tu{\~n}{\'o}n}, C. and {Pell{\'o}}, R. and {Prieto}, M. and {Bourrec}, {\'E}. and {Cabello}, C. and {Cardiel}, N. and {Gonz{\'a}lez-Fern{\'a}ndez}, C. and {Laporte}, N. and {Milliard}, B. and {Pascual}, S. and {Patrick}, L.~R. and {Patr{\'o}n}, J. and {Ram{\'\i}rez-Alegr{\'\i}a}, S. and {Streblyanska}, A.},
        title = "{EMIR, the near-infrared camera and multi-object spectrograph for the GTC. EMIR at GTC}",
      journal = {\aap},
         year = 2022,
        month = nov,
       volume = {667},
          eid = {A107},
        pages = {A107},
          doi = {10.1051/0004-6361/202244729},
archivePrefix = {arXiv},
       eprint = {2209.15395},
 primaryClass = {astro-ph.IM},
       adsurl = {https://ui.adsabs.harvard.edu/abs/2022A&A...667A.107G}
}

@article{Reyes+08,
doi = {10.1088/0004-6256/136/6/2373},
url = {https://dx.doi.org/10.1088/0004-6256/136/6/2373},
year = {2008},
month = {nov},
publisher = {The American Astronomical Society},
volume = {136},
number = {6},
pages = {2373},
author = {Reyes, Reinabelle and Zakamska, Nadia L. and Strauss, Michael A. and Green, Joshua and Krolik, Julian H. and Shen, Yue and Richards, Gordon T. and Anderson, Scott F. and Schneider, Donald P.},
title = {SPACE DENSITY OF OPTICALLY SELECTED TYPE 2 QUASARS},
journal = {The Astronomical Journal},
}

@ARTICLE{Zakamska14,
       author = {{Zakamska}, Nadia L. and {Greene}, Jenny E.},
        title = "{Quasar feedback and the origin of radio emission in radio-quiet quasars}",
      journal = {\mnras},
         year = "2014",
        month = "Jul",
       volume = {442},
       number = {1},
        pages = {784-804},
          doi = {10.1093/mnras/stu842},
archivePrefix = {arXiv},
       eprint = {1402.6736},
 primaryClass = {astro-ph.GA},
       adsurl = {https://ui.adsabs.harvard.edu/abs/2014MNRAS.442..784Z}
}

@ARTICLE{Mullaney13,
       author = {{Mullaney}, J.~R. and {Alexander}, D.~M. and {Fine}, S. and
         {Goulding}, A.~D. and {Harrison}, C.~M. and {Hickox}, R.~C.},
        title = "{Narrow-line region gas kinematics of 24 264 optically selected AGN: the radio connection}",
      journal = {\mnras},
         year = "2013",
        month = "Jul",
       volume = {433},
       number = {1},
        pages = {622-638},
          doi = {10.1093/mnras/stt751},
archivePrefix = {arXiv},
       eprint = {1305.0263},
 primaryClass = {astro-ph.CO},
       adsurl = {https://ui.adsabs.harvard.edu/abs/2013MNRAS.433..622M}
}

@ARTICLE{RamosAlmeida+19,
       author = {{Ramos Almeida}, C. and {Acosta-Pulido}, J.~A. and {Tadhunter}, C.~N. and {Gonz{\'a}lez-Fern{\'a}ndez}, C. and {Cicone}, C. and {Fern{\'a}ndez-Torreiro}, M.},
        title = "{A near-infrared study of the multiphase outflow in the type-2 quasar J1509+0434}",
      journal = {\mnras},
         year = 2019,
        month = jul,
       volume = {487},
       number = {1},
        pages = {L18-L23},
          doi = {10.1093/mnrasl/slz072},
archivePrefix = {arXiv},
       eprint = {1905.06288},
 primaryClass = {astro-ph.GA},
       adsurl = {https://ui.adsabs.harvard.edu/abs/2019MNRAS.487L..18R}
}

@ARTICLE{RamosAlmeida+22,
       author = {{Ramos Almeida}, C. and {Bischetti}, M. and {Garc{\'\i}a-Burillo}, S. and {Alonso-Herrero}, A. and {Audibert}, A. and {Cicone}, C. and {Feruglio}, C. and {Tadhunter}, C.~N. and {Pierce}, J.~C.~S. and {Pereira-Santaella}, M. and {Bessiere}, P.~S.},
        title = "{The diverse cold molecular gas contents, morphologies, and kinematics of type-2 quasars as seen by ALMA}",
      journal = {\aap},
         year = 2022,
        month = feb,
       volume = {658},
          eid = {A155},
        pages = {A155},
          doi = {10.1051/0004-6361/202141906},
archivePrefix = {arXiv},
       eprint = {2111.13578},
 primaryClass = {astro-ph.GA},
       adsurl = {https://ui.adsabs.harvard.edu/abs/2022A&A...658A.155R}
}

@ARTICLE{Bessiere+24,
       author = {{Bessiere}, P.~S. and {Ramos Almeida}, C. and {Holden}, L.~R. and {Tadhunter}, C.~N. and {Canalizo}, G.},
        title = "{QSOFEED: Relationship between star formation and active galactic nuclei feedback}",
      journal = {\aap},
         year = 2024,
        month = sep,
       volume = {689},
          eid = {A271},
        pages = {A271},
          doi = {10.1051/0004-6361/202348795},
archivePrefix = {arXiv},
       eprint = {2405.06421},
 primaryClass = {astro-ph.GA},
       adsurl = {https://ui.adsabs.harvard.edu/abs/2024A&A...689A.271B}
}

@ARTICLE{Holt+11,
       author = {{Holt}, J. and {Tadhunter}, C.~N. and {Morganti}, R. and {Emonts}, B.~H.~C.},
        title = "{The impact of the warm outflow in the young (GPS) radio source and ULIRG PKS 1345+12 (4C 12.50)}",
      journal = {\mnras},
         year = 2011,
        month = jan,
       volume = {410},
       number = {3},
        pages = {1527-1536},
          doi = {10.1111/j.1365-2966.2010.17535.x},
archivePrefix = {arXiv},
       eprint = {1008.2846},
 primaryClass = {astro-ph.CO},
       adsurl = {https://ui.adsabs.harvard.edu/abs/2011MNRAS.410.1527H}
}

@ARTICLE{Astropy+22,
       author = {{Astropy Collaboration} and {Price-Whelan}, Adrian M. and {Lim}, Pey Lian and {Earl}, Nicholas and {Starkman}, Nathaniel and {Bradley}, Larry and {Shupe}, David L. and {Patil}, Aarya A. and {Corrales}, Lia and {Brasseur}, C.~E. and {N{\"o}the}, Maximilian and {Donath}, Axel and {Tollerud}, Erik and {Morris}, Brett M. and {Ginsburg}, Adam and {Vaher}, Eero and {Weaver}, Benjamin A. and {Tocknell}, James and {Jamieson}, William and {van Kerkwijk}, Marten H. and {Robitaille}, Thomas P. and {Merry}, Bruce and {Bachetti}, Matteo and {G{\"u}nther}, H. Moritz and {Aldcroft}, Thomas L. and {Alvarado-Montes}, Jaime A. and {Archibald}, Anne M. and {B{\'o}di}, Attila and {Bapat}, Shreyas and {Barentsen}, Geert and {Baz{\'a}n}, Juanjo and {Biswas}, Manish and {Boquien}, M{\'e}d{\'e}ric and {Burke}, D.~J. and {Cara}, Daria and {Cara}, Mihai and {Conroy}, Kyle E. and {Conseil}, Simon and {Craig}, Matthew W. and {Cross}, Robert M. and {Cruz}, Kelle L. and {D'Eugenio}, Francesco and {Dencheva}, Nadia and {Devillepoix}, Hadrien A.~R. and {Dietrich}, J{\"o}rg P. and {Eigenbrot}, Arthur Davis and {Erben}, Thomas and {Ferreira}, Leonardo and {Foreman-Mackey}, Daniel and {Fox}, Ryan and {Freij}, Nabil and {Garg}, Suyog and {Geda}, Robel and {Glattly}, Lauren and {Gondhalekar}, Yash and {Gordon}, Karl D. and {Grant}, David and {Greenfield}, Perry and {Groener}, Austen M. and {Guest}, Steve and {Gurovich}, Sebastian and {Handberg}, Rasmus and {Hart}, Akeem and {Hatfield-Dodds}, Zac and {Homeier}, Derek and {Hosseinzadeh}, Griffin and {Jenness}, Tim and {Jones}, Craig K. and {Joseph}, Prajwel and {Kalmbach}, J. Bryce and {Karamehmetoglu}, Emir and {Ka{\l}uszy{\'n}ski}, Miko{\l}aj and {Kelley}, Michael S.~P. and {Kern}, Nicholas and {Kerzendorf}, Wolfgang E. and {Koch}, Eric W. and {Kulumani}, Shankar and {Lee}, Antony and {Ly}, Chun and {Ma}, Zhiyuan and {MacBride}, Conor and {Maljaars}, Jakob M. and {Muna}, Demitri and {Murphy}, N.~A. and {Norman}, Henrik and {O'Steen}, Richard and {Oman}, Kyle A. and {Pacifici}, Camilla and {Pascual}, Sergio and {Pascual-Granado}, J. and {Patil}, Rohit R. and {Perren}, Gabriel I. and {Pickering}, Timothy E. and {Rastogi}, Tanuj and {Roulston}, Benjamin R. and {Ryan}, Daniel F. and {Rykoff}, Eli S. and {Sabater}, Jose and {Sakurikar}, Parikshit and {Salgado}, Jes{\'u}s and {Sanghi}, Aniket and {Saunders}, Nicholas and {Savchenko}, Volodymyr and {Schwardt}, Ludwig and {Seifert-Eckert}, Michael and {Shih}, Albert Y. and {Jain}, Anany Shrey and {Shukla}, Gyanendra and {Sick}, Jonathan and {Simpson}, Chris and {Singanamalla}, Sudheesh and {Singer}, Leo P. and {Singhal}, Jaladh and {Sinha}, Manodeep and {Sip{\H{o}}cz}, Brigitta M. and {Spitler}, Lee R. and {Stansby}, David and {Streicher}, Ole and {{\v{S}}umak}, Jani and {Swinbank}, John D. and {Taranu}, Dan S. and {Tewary}, Nikita and {Tremblay}, Grant R. and {de Val-Borro}, Miguel and {Van Kooten}, Samuel J. and {Vasovi{\'c}}, Zlatan and {Verma}, Shresth and {de Miranda Cardoso}, Jos{\'e} Vin{\'\i}cius and {Williams}, Peter K.~G. and {Wilson}, Tom J. and {Winkel}, Benjamin and {Wood-Vasey}, W.~M. and {Xue}, Rui and {Yoachim}, Peter and {Zhang}, Chen and {Zonca}, Andrea and {Astropy Project Contributors}},
        title = "{The Astropy Project: Sustaining and Growing a Community-oriented Open-source Project and the Latest Major Release (v5.0) of the Core Package}",
      journal = {\apj},
         year = 2022,
        month = aug,
       volume = {935},
       number = {2},
          eid = {167},
        pages = {167},
          doi = {10.3847/1538-4357/ac7c74},
archivePrefix = {arXiv},
       eprint = {2206.14220},
 primaryClass = {astro-ph.IM},
       adsurl = {https://ui.adsabs.harvard.edu/abs/2022ApJ...935..167A}
}

@book{Osterbrock+06,
       author = {{Osterbrock}, Donald E. and {Ferland}, Gary J.},
        title = "{Astrophysics of gaseous nebulae and active galactic nuclei}",
         year = 2006,
       adsurl = {https://ui.adsabs.harvard.edu/abs/2006agna.book.....O},
      publisher = {University Science Books}
}

@ARTICLE{Luridiana+15,
       author = {{Luridiana}, V. and {Morisset}, C. and {Shaw}, R.~A.},
        title = "{PyNeb: a new tool for analyzing emission lines. I. Code description and validation of results}",
      journal = {\aap},
         year = 2015,
        month = jan,
       volume = {573},
          eid = {A42},
        pages = {A42},
          doi = {10.1051/0004-6361/201323152},
archivePrefix = {arXiv},
       eprint = {1410.6662},
 primaryClass = {astro-ph.IM},
       adsurl = {https://ui.adsabs.harvard.edu/abs/2015A&A...573A..42L}
}

@ARTICLE{Rose+18,
       author = {{Rose}, Marvin and {Tadhunter}, Clive and {Ramos Almeida}, Cristina and {Rodr{\'\i}guez Zaur{\'\i}n}, Javier and {Santoro}, Francesco and {Spence}, Robert},
        title = "{Quantifying the AGN-driven outflows in ULIRGs (QUADROS) - I: VLT/Xshooter observations of nine nearby objects}",
      journal = {\mnras},
         year = 2018,
        month = feb,
       volume = {474},
       number = {1},
        pages = {128-156},
          doi = {10.1093/mnras/stx2590},
archivePrefix = {arXiv},
       eprint = {1710.06600},
 primaryClass = {astro-ph.GA},
       adsurl = {https://ui.adsabs.harvard.edu/abs/2018MNRAS.474..128R}
}

@ARTICLE{Ferland+13,
       author = {{Ferland}, G.~J. and {Porter}, R.~L. and {van Hoof}, P.~A.~M. and {Williams}, R.~J.~R. and {Abel}, N.~P. and {Lykins}, M.~L. and {Shaw}, G. and {Henney}, W.~J. and {Stancil}, P.~C.},
        title = "{The 2013 Release of Cloudy}",
      journal = {\rmxaa},
         year = 2013,
        month = apr,
       volume = {49},
        pages = {137-163},
          doi = {10.48550/arXiv.1302.4485},
archivePrefix = {arXiv},
       eprint = {1302.4485},
 primaryClass = {astro-ph.GA},
       adsurl = {https://ui.adsabs.harvard.edu/abs/2013RMxAA..49..137F}
}

@ARTICLE{Carniani+15,
       author = {{Carniani}, S. and {Marconi}, A. and {Maiolino}, R. and {Balmaverde}, B. and {Brusa}, M. and {Cano-D{\'\i}az}, M. and {Cicone}, C. and {Comastri}, A. and {Cresci}, G. and {Fiore}, F. and {Feruglio}, C. and {La Franca}, F. and {Mainieri}, V. and {Mannucci}, F. and {Nagao}, T. and {Netzer}, H. and {Piconcelli}, E. and {Risaliti}, G. and {Schneider}, R. and {Shemmer}, O.},
        title = "{Ionised outflows in z \raisebox{-0.5ex}\textasciitilde 2.4 quasar host galaxies}",
      journal = {\aap},
         year = 2015,
        month = aug,
       volume = {580},
          eid = {A102},
        pages = {A102},
          doi = {10.1051/0004-6361/201526557},
archivePrefix = {arXiv},
       eprint = {1506.03096},
 primaryClass = {astro-ph.GA},
       adsurl = {https://ui.adsabs.harvard.edu/abs/2015A&A...580A.102C}
}

@ARTICLE{Belli+24,
       author = {{Belli}, Sirio and {Park}, Minjung and {Davies}, Rebecca L. and {Mendel}, J. Trevor and {Johnson}, Benjamin D. and {Conroy}, Charlie and {Benton}, Chlo{\"e} and {Bugiani}, Letizia and {Emami}, Razieh and {Leja}, Joel and {Li}, Yijia and {Maheson}, Gabriel and {Mathews}, Elijah P. and {Naidu}, Rohan P. and {Nelson}, Erica J. and {Tacchella}, Sandro and {Terrazas}, Bryan A. and {Weinberger}, Rainer},
        title = "{Star formation shut down by multiphase gas outflow in a galaxy at a redshift of 2.45}",
      journal = {\nat},
         year = 2024,
        month = jun,
       volume = {630},
       number = {8015},
        pages = {54-58},
          doi = {10.1038/s41586-024-07412-1},
archivePrefix = {arXiv},
       eprint = {2308.05795},
 primaryClass = {astro-ph.GA},
       adsurl = {https://ui.adsabs.harvard.edu/abs/2024Natur.630...54B}
}

@ARTICLE{Fiore+17,
       author = {{Fiore}, F. and {Feruglio}, C. and {Shankar}, F. and {Bischetti}, M. and {Bongiorno}, A. and {Brusa}, M. and {Carniani}, S. and {Cicone}, C. and {Duras}, F. and {Lamastra}, A. and {Mainieri}, V. and {Marconi}, A. and {Menci}, N. and {Maiolino}, R. and {Piconcelli}, E. and {Vietri}, G. and {Zappacosta}, L.},
        title = "{AGN wind scaling relations and the co-evolution of black holes and galaxies}",
      journal = {\aap},
         year = 2017,
        month = may,
       volume = {601},
          eid = {A143},
        pages = {A143},
          doi = {10.1051/0004-6361/201629478},
archivePrefix = {arXiv},
       eprint = {1702.04507},
 primaryClass = {astro-ph.GA},
       adsurl = {https://ui.adsabs.harvard.edu/abs/2017A&A...601A.143F}
}

@ARTICLE{Asplund+21,
       author = {{Asplund}, M. and {Amarsi}, A.~M. and {Grevesse}, N.},
        title = "{The chemical make-up of the Sun: A 2020 vision}",
      journal = {\aap},
         year = 2021,
        month = sep,
       volume = {653},
          eid = {A141},
        pages = {A141},
          doi = {10.1051/0004-6361/202140445},
archivePrefix = {arXiv},
       eprint = {2105.01661},
 primaryClass = {astro-ph.SR},
       adsurl = {https://ui.adsabs.harvard.edu/abs/2021A&A...653A.141A}
}

@ARTICLE{Dere+97,
       author = {{Dere}, K.~P. and {Landi}, E. and {Mason}, H.~E. and {Monsignori Fossi}, B.~C. and {Young}, P.~R.},
        title = "{CHIANTI - an atomic database for emission lines}",
      journal = {\aaps},
         year = 1997,
        month = oct,
       volume = {125},
        pages = {149-173},
          doi = {10.1051/aas:1997368},
       adsurl = {https://ui.adsabs.harvard.edu/abs/1997A&AS..125..149D}
}

@ARTICLE{DelZanna+20,
       author = {{Del Zanna}, G. and {Dere}, K.~P. and {Young}, P.~R. and {Landi}, E.},
        title = "{CHIANTI{\textemdash}An Atomic Database for Emission Lines. XVI. Version 10, Further Extensions}",
      journal = {\apj},
         year = 2021,
        month = mar,
       volume = {909},
       number = {1},
          eid = {38},
        pages = {38},
          doi = {10.3847/1538-4357/abd8ce},
archivePrefix = {arXiv},
       eprint = {2011.05211},
 primaryClass = {physics.atom-ph},
       adsurl = {https://ui.adsabs.harvard.edu/abs/2021ApJ...909...38D}
}

@ARTICLE{Hervella+23,
       author = {{Hervella Seoane}, K. and {Ramos Almeida}, C. and {Acosta-Pulido}, J.~A. and {Speranza}, G. and {Tadhunter}, C.~N. and {Bessiere}, P.~S.},
        title = "{Investigating the impact of quasar-driven outflows on galaxies at z {\ensuremath{\sim}} 0.3-0.4}",
      journal = {\aap},
         year = 2023,
        month = dec,
       volume = {680},
          eid = {A71},
        pages = {A71},
          doi = {10.1051/0004-6361/202347756},
archivePrefix = {arXiv},
       eprint = {2309.10572},
 primaryClass = {astro-ph.GA},
       adsurl = {https://ui.adsabs.harvard.edu/abs/2023A&A...680A..71H}
}

@ARTICLE{Oliva+90,
       author = {{Oliva}, E. and {Moorwood}, A.~F.~M.},
        title = "{Detection of SI VI 1.962 Microns and New Observations of Infrared H, Fe ii, and H 2 Line Emission in the Seyfert Galaxy NGC 1068}",
      journal = {\apjl},
         year = 1990,
        month = jan,
       volume = {348},
        pages = {L5},
          doi = {10.1086/185617},
       adsurl = {https://ui.adsabs.harvard.edu/abs/1990ApJ...348L...5O}
}

@ARTICLE{Ardila+25,
       author = {{Rodr{\'\i}guez-Ardila}, Alberto and {Cerqueira-Campos}, Fernando},
        title = "{The coronal line region of active galactic nuclei}",
      journal = {Frontiers in Astronomy and Space Sciences},
         year = 2025,
        month = feb,
       volume = {12},
          eid = {1548632},
        pages = {1548632},
          doi = {10.3389/fspas.2025.1548632},
archivePrefix = {arXiv},
       eprint = {2502.06442},
 primaryClass = {astro-ph.GA},
       adsurl = {https://ui.adsabs.harvard.edu/abs/2025FrASS..1248632R}
}

@ARTICLE{RamosAlmeida+06,
       author = {{Ramos Almeida}, C. and {P{\'e}rez Garc{\'\i}a}, A.~M. and {Acosta-Pulido}, J.~A. and {Rodr{\'\i}guez Espinosa}, J.~M. and {Barrena}, R. and {Manchado}, A.},
        title = "{The Narrow-Line Region of the Seyfert 2 Galaxy Mrk 78: An Infrared View}",
      journal = {\apj},
         year = 2006,
        month = jul,
       volume = {645},
       number = {1},
        pages = {148-159},
          doi = {10.1086/504284},
archivePrefix = {arXiv},
       eprint = {astro-ph/0603420},
 primaryClass = {astro-ph},
       adsurl = {https://ui.adsabs.harvard.edu/abs/2006ApJ...645..148R}
}

@ARTICLE{Mazzalay+13,
       author = {{Mazzalay}, X. and {Saglia}, R.~P. and {Erwin}, Peter and {Fabricius}, M.~H. and {Rusli}, S.~P. and {Thomas}, J. and {Bender}, R. and {Opitsch}, M. and {Nowak}, N. and {Williams}, Michael J.},
        title = "{Molecular gas in the centre of nearby galaxies from VLT/SINFONI integral field spectroscopy - I. Morphology and mass inventory}",
      journal = {\mnras},
         year = 2013,
        month = jan,
       volume = {428},
       number = {3},
        pages = {2389-2406},
          doi = {10.1093/mnras/sts204},
archivePrefix = {arXiv},
       eprint = {1210.4171},
 primaryClass = {astro-ph.CO},
       adsurl = {https://ui.adsabs.harvard.edu/abs/2013MNRAS.428.2389M}
}

@ARTICLE{Speranza+24,
       author = {{Speranza}, G. and {Ramos Almeida}, C. and {Acosta-Pulido}, J.~A. and {Audibert}, A. and {Holden}, L.~R. and {Tadhunter}, C.~N. and {Lapi}, A. and {Gonz{\'a}lez-Mart{\'\i}n}, O. and {Brusa}, M. and {L{\'o}pez}, I.~E. and {Musiimenta}, B. and {Shankar}, F.},
        title = "{Multiphase characterization of AGN winds in five local type-2 quasars}",
      journal = {\aap},
         year = 2024,
        month = jan,
       volume = {681},
          eid = {A63},
        pages = {A63},
          doi = {10.1051/0004-6361/202347715},
archivePrefix = {arXiv},
       eprint = {2311.10132},
 primaryClass = {astro-ph.GA},
       adsurl = {https://ui.adsabs.harvard.edu/abs/2024A&A...681A..63S}
}

@ARTICLE{Speranza+22,
       author = {{Speranza}, G. and {Ramos Almeida}, C. and {Acosta-Pulido}, J.~A. and {Riffel}, R.~A. and {Tadhunter}, C. and {Pierce}, J.~C.~S. and {Rodr{\'\i}guez-Ardila}, A. and {Coloma Puga}, M. and {Brusa}, M. and {Musiimenta}, B. and {Alexander}, D.~M. and {Lapi}, A. and {Shankar}, F. and {Villforth}, C.},
        title = "{Warm molecular and ionized gas kinematics in the type-2 quasar J0945+1737}",
      journal = {\aap},
         year = 2022,
        month = sep,
       volume = {665},
          eid = {A55},
        pages = {A55},
          doi = {10.1051/0004-6361/202243585},
archivePrefix = {arXiv},
       eprint = {2206.15347},
 primaryClass = {astro-ph.GA},
       adsurl = {https://ui.adsabs.harvard.edu/abs/2022A&A...665A..55S}
}

@ARTICLE{Pierce+23,
       author = {{Pierce}, J.~C.~S. and {Tadhunter}, C. and {Ramos Almeida}, C. and {Bessiere}, P. and {Heaton}, J.~V. and {Ellison}, S.~L. and {Speranza}, G. and {Gordon}, Y. and {O'Dea}, C. and {Grimmett}, L. and {Makrygianni}, L.},
        title = "{Galaxy interactions are the dominant trigger for local type 2 quasars}",
      journal = {\mnras},
         year = 2023,
        month = jun,
       volume = {522},
       number = {2},
        pages = {1736-1751},
          doi = {10.1093/mnras/stad455},
archivePrefix = {arXiv},
       eprint = {2303.15506},
 primaryClass = {astro-ph.GA},
       adsurl = {https://ui.adsabs.harvard.edu/abs/2023MNRAS.522.1736P}
}

@ARTICLE{Kong&Ho+18,
       author = {{Kong}, Minzhi and {Ho}, Luis C.},
        title = "{The Black Hole Masses and Eddington Ratios of Type 2 Quasars}",
      journal = {\apj},
         year = 2018,
        month = jun,
       volume = {859},
       number = {2},
          eid = {116},
        pages = {116},
          doi = {10.3847/1538-4357/aabe2a},
archivePrefix = {arXiv},
       eprint = {1804.09852},
 primaryClass = {astro-ph.GA},
       adsurl = {https://ui.adsabs.harvard.edu/abs/2018ApJ...859..116K}
}

@ARTICLE{Lamastra+09,
       author = {{Lamastra}, A. and {Bianchi}, S. and {Matt}, G. and {Perola}, G.~C. and {Barcons}, X. and {Carrera}, F.~J.},
        title = "{The bolometric luminosity of type 2 AGN from extinction-corrected [OIII]. No evidence of Eddington-limited sources}",
      journal = {\aap},
         year = 2009,
        month = sep,
       volume = {504},
       number = {1},
        pages = {73-79},
          doi = {10.1051/0004-6361/200912023},
archivePrefix = {arXiv},
       eprint = {0905.4439},
 primaryClass = {astro-ph.CO},
       adsurl = {https://ui.adsabs.harvard.edu/abs/2009A&A...504...73L}
}

@ARTICLE{Becker+95,
       author = {{Becker}, Robert H. and {White}, Richard L. and {Helfand}, David J.},
        title = "{The FIRST Survey: Faint Images of the Radio Sky at Twenty Centimeters}",
      journal = {\apj},
         year = 1995,
        month = sep,
       volume = {450},
        pages = {559},
          doi = {10.1086/176166},
       adsurl = {https://ui.adsabs.harvard.edu/abs/1995ApJ...450..559B}
}

@ARTICLE{Noll+12,
       author = {{Noll}, S. and {Kausch}, W. and {Barden}, M. and {Jones}, A.~M. and {Szyszka}, C. and {Kimeswenger}, S. and {Vinther}, J.},
        title = "{An atmospheric radiation model for Cerro Paranal. I. The optical spectral range}",
      journal = {\aap},
         year = 2012,
        month = jul,
       volume = {543},
          eid = {A92},
        pages = {A92},
          doi = {10.1051/0004-6361/201219040},
archivePrefix = {arXiv},
       eprint = {1205.2003},
 primaryClass = {astro-ph.IM},
       adsurl = {https://ui.adsabs.harvard.edu/abs/2012A&A...543A..92N}
}

@ARTICLE{Moehler+14,
       author = {{Moehler}, S. and {Modigliani}, A. and {Freudling}, W. and {Giammichele}, N. and {Gianninas}, A. and {Gonneau}, A. and {Kausch}, W. and {Lan{\c{c}}on}, A. and {Noll}, S. and {Rauch}, T. and {Vinther}, J.},
        title = "{Flux calibration of medium-resolution spectra from 300 nm to 2500 nm: Model reference spectra and telluric correction}",
      journal = {\aap},
         year = 2014,
        month = aug,
       volume = {568},
          eid = {A9},
        pages = {A9},
          doi = {10.1051/0004-6361/201423790},
archivePrefix = {arXiv},
       eprint = {1408.1797},
 primaryClass = {astro-ph.IM},
       adsurl = {https://ui.adsabs.harvard.edu/abs/2014A&A...568A...9M}
}

@ARTICLE{Jones+13,
       author = {{Jones}, A. and {Noll}, S. and {Kausch}, W. and {Szyszka}, C. and {Kimeswenger}, S.},
        title = "{An advanced scattered moonlight model for Cerro Paranal}",
      journal = {\aap},
         year = 2013,
        month = dec,
       volume = {560},
          eid = {A91},
        pages = {A91},
          doi = {10.1051/0004-6361/201322433},
archivePrefix = {arXiv},
       eprint = {1310.7030},
 primaryClass = {astro-ph.IM},
       adsurl = {https://ui.adsabs.harvard.edu/abs/2013A&A...560A..91J}
}

@INPROCEEDINGS{Tody+86,
       author = {{Tody}, Doug},
        title = "{The IRAF Data Reduction and Analysis System}",
    booktitle = {Instrumentation in astronomy VI},
         year = 1986,
       editor = {{Crawford}, David L.},
       series = {Society of Photo-Optical Instrumentation Engineers (SPIE) Conference Series},
       volume = {627},
        month = jan,
        pages = {733},
          doi = {10.1117/12.968154},
       adsurl = {https://ui.adsabs.harvard.edu/abs/1986SPIE..627..733T}
}

@INPROCEEDINGS{Tody+93,
       author = {{Tody}, Doug},
        title = "{IRAF in the Nineties}",
    booktitle = {Astronomical Data Analysis Software and Systems II},
         year = 1993,
       editor = {{Hanisch}, R.~J. and {Brissenden}, R.~J.~V. and {Barnes}, J.},
       series = {Astronomical Society of the Pacific Conference Series},
       volume = {52},
        month = jan,
        pages = {173},
       adsurl = {https://ui.adsabs.harvard.edu/abs/1993ASPC...52..173T}
}

@ARTICLE{Rupke+13b,
       author = {{Rupke}, David S.~N. and {Veilleux}, Sylvain},
        title = "{Breaking the Obscuring Screen: A Resolved Molecular Outflow in a Buried QSO}",
      journal = {\apjl},
         year = 2013,
        month = sep,
       volume = {775},
       number = {1},
          eid = {L15},
        pages = {L15},
          doi = {10.1088/2041-8205/775/1/L15},
archivePrefix = {arXiv},
       eprint = {1308.4988},
 primaryClass = {astro-ph.CO},
       adsurl = {https://ui.adsabs.harvard.edu/abs/2013ApJ...775L..15R}
}

@ARTICLE{Zanchettin+25,
       author = {{Zanchettin}, M.~V. and {Ramos Almeida}, C. and {Audibert}, A. and {Acosta-Pulido}, J.~A. and {Cezar}, P.~H. and {Hicks}, E. and {Lapi}, A. and {Mullaney}, J.},
        title = "{Unveiling the warm molecular outflow component of type-2 quasars with SINFONI}",
      journal = {\aap},
         year = 2025,
        month = mar,
       volume = {695},
          eid = {A185},
        pages = {A185},
          doi = {10.1051/0004-6361/202453224},
archivePrefix = {arXiv},
       eprint = {2502.12800},
 primaryClass = {astro-ph.GA},
       adsurl = {https://ui.adsabs.harvard.edu/abs/2025A&A...695A.185Z}
}

@ARTICLE{Herrera-Camus+20,
       author = {{Herrera-Camus}, R. and {Janssen}, A. and {Sturm}, E. and {Lutz}, D. and {Veilleux}, S. and {Davies}, R. and {Shimizu}, T. and {Gonz{\'a}lez-Alfonso}, E. and {Rupke}, D.~S.~N. and {Tacconi}, L. and {Genzel}, R. and {Cicone}, C. and {Maiolino}, R. and {Contursi}, A. and {Graci{\'a}-Carpio}, J.},
        title = "{AGN feedback in a galaxy merger: multi-phase, galaxy-scale outflows with a fast molecular gas blob {\ensuremath{\sim}}6 kpc away from IRAS F08572+3915}",
      journal = {\aap},
         year = 2020,
        month = mar,
       volume = {635},
          eid = {A47},
        pages = {A47},
          doi = {10.1051/0004-6361/201936434},
archivePrefix = {arXiv},
       eprint = {1911.06326},
 primaryClass = {astro-ph.GA},
       adsurl = {https://ui.adsabs.harvard.edu/abs/2020A&A...635A..47H}
}

@ARTICLE{RamosAlmeida+17,
       author = {{Ramos Almeida}, C. and {Piqueras L{\'o}pez}, J. and {Villar-Mart{\'\i}n}, M. and {Bessiere}, P.~S.},
        title = "{An infrared view of AGN feedback in a type-2 quasar: the case of the Teacup galaxy}",
      journal = {\mnras},
         year = 2017,
        month = sep,
       volume = {470},
       number = {1},
        pages = {964-976},
          doi = {10.1093/mnras/stx1287},
archivePrefix = {arXiv},
       eprint = {1705.07631},
 primaryClass = {astro-ph.GA},
       adsurl = {https://ui.adsabs.harvard.edu/abs/2017MNRAS.470..964R}
}

@ARTICLE{VillarMartin+23,
       author = {{Villar Mart{\'\i}n}, M. and {Castro-Rodr{\'\i}guez}, N. and {Pereira Santaella}, M. and {Lamperti}, I. and {Tadhunter}, C. and {Emonts}, B. and {Colina}, L. and {Alonso Herrero}, A. and {Cabrera-Lavers}, A. and {Bellocchi}, E.},
        title = "{Limited impact of jet-induced feedback in the multi-phase nuclear interstellar medium of 4C12.50}",
      journal = {\aap},
         year = 2023,
        month = may,
       volume = {673},
          eid = {A25},
        pages = {A25},
          doi = {10.1051/0004-6361/202245418},
archivePrefix = {arXiv},
       eprint = {2303.00291},
 primaryClass = {astro-ph.GA},
       adsurl = {https://ui.adsabs.harvard.edu/abs/2023A&A...673A..25V}
}

@ARTICLE{RodriguezArdila+17,
       author = {{Rodr{\'\i}guez-Ardila}, A. and {Prieto}, M.~A. and {Mazzalay}, X. and {Fern{\'a}ndez-Ontiveros}, J.~A. and {Luque}, R. and {M{\"u}ller-S{\'a}nchez}, F.},
        title = "{Powerful outflows in the central parsecs of the low-luminosity active galactic nucleus NGC 1386}",
      journal = {\mnras},
         year = 2017,
        month = sep,
       volume = {470},
       number = {3},
        pages = {2845-2860},
          doi = {10.1093/mnras/stx1401},
archivePrefix = {arXiv},
       eprint = {1706.01370},
 primaryClass = {astro-ph.GA},
       adsurl = {https://ui.adsabs.harvard.edu/abs/2017MNRAS.470.2845R}
}

@ARTICLE{May+18,
       author = {{May}, D. and {Rodr{\'\i}guez-Ardila}, A. and {Prieto}, M.~A. and {Fern{\'a}ndez-Ontiveros}, J.~A. and {Diaz}, Y. and {Mazzalay}, X.},
        title = "{Powerful mechanical-driven outflows in the central parsecs of the low-luminosity active galactic nucleus ESO 428-G14}",
      journal = {\mnras},
         year = 2018,
        month = nov,
       volume = {481},
       number = {1},
        pages = {L105-L109},
          doi = {10.1093/mnrasl/sly155},
archivePrefix = {arXiv},
       eprint = {1811.03985},
 primaryClass = {astro-ph.GA},
       adsurl = {https://ui.adsabs.harvard.edu/abs/2018MNRAS.481L.105M}
}

@ARTICLE{Muller+11,
       author = {{M{\"u}ller-S{\'a}nchez}, F. and {Prieto}, M.~A. and {Hicks}, E.~K.~S. and {Vives-Arias}, H. and {Davies}, R.~I. and {Malkan}, M. and {Tacconi}, L.~J. and {Genzel}, R.},
        title = "{Outflows from Active Galactic Nuclei: Kinematics of the Narrow-line and Coronal-line Regions in Seyfert Galaxies}",
      journal = {\apj},
         year = 2011,
        month = oct,
       volume = {739},
       number = {2},
          eid = {69},
        pages = {69},
          doi = {10.1088/0004-637X/739/2/69},
archivePrefix = {arXiv},
       eprint = {1107.3140},
 primaryClass = {astro-ph.CO},
       adsurl = {https://ui.adsabs.harvard.edu/abs/2011ApJ...739...69M}
}

@ARTICLE{Harrison+18,
       author = {{Harrison}, C.~M. and {Costa}, T. and {Tadhunter}, C.~N. and {Fl{\"u}tsch}, A. and {Kakkad}, D. and {Perna}, M. and {Vietri}, G.},
        title = "{AGN outflows and feedback twenty years on}",
      journal = {Nature Astronomy},
         year = 2018,
        month = feb,
       volume = {2},
        pages = {198-205},
          doi = {10.1038/s41550-018-0403-6},
archivePrefix = {arXiv},
       eprint = {1802.10306},
 primaryClass = {astro-ph.GA},
       adsurl = {https://ui.adsabs.harvard.edu/abs/2018NatAs...2..198H}
}

@ARTICLE{Harrison+24,
       author = {{Harrison}, Chris M. and {Ramos Almeida}, Cristina},
        title = "{Observational Tests of Active Galactic Nuclei Feedback: An Overview of Approaches and Interpretation}",
      journal = {Galaxies},
         year = 2024,
        month = apr,
       volume = {12},
       number = {2},
          eid = {17},
        pages = {17},
          doi = {10.3390/galaxies12020017},
archivePrefix = {arXiv},
       eprint = {2404.08050},
 primaryClass = {astro-ph.GA},
       adsurl = {https://ui.adsabs.harvard.edu/abs/2024Galax..12...17H}
}

@ARTICLE{Bessiere+22,
       author = {{Bessiere}, P.~S. and {Ramos Almeida}, C.},
        title = "{Spatially resolved evidence of the impact of quasar-driven outflows on recent star formation: the case of Mrk 34}",
      journal = {\mnras},
         year = 2022,
        month = may,
       volume = {512},
       number = {1},
        pages = {L54-L59},
          doi = {10.1093/mnrasl/slac016},
archivePrefix = {arXiv},
       eprint = {2202.06788},
 primaryClass = {astro-ph.GA},
       adsurl = {https://ui.adsabs.harvard.edu/abs/2022MNRAS.512L..54B}
}

@ARTICLE{CostaSouza+24,
       author = {{Costa-Souza}, J.~H. and {Riffel}, Rogemar A. and {Souza-Oliveira}, Gabriel L. and {Zakamska}, Nadia L. and {Bianchin}, Marina and {Storchi-Bergmann}, Thaisa and {Riffel}, Rog{\'e}rio},
        title = "{Blowing Star Formation Away in Active Galactic Nuclei Hosts. I. Observation of Warm Molecular Outflows with JWST MIRI}",
      journal = {\apj},
         year = 2024,
        month = oct,
       volume = {974},
       number = {1},
          eid = {127},
        pages = {127},
          doi = {10.3847/1538-4357/ad702a},
archivePrefix = {arXiv},
       eprint = {2408.06100},
 primaryClass = {astro-ph.GA},
       adsurl = {https://ui.adsabs.harvard.edu/abs/2024ApJ...974..127C}
}

@ARTICLE{Dale+05,
       author = {{Dale}, Daniel A. and {Sheth}, Kartik and {Helou}, George and {Regan}, Michael W. and {H{\"u}ttemeister}, Susanne},
        title = "{Warm and Cold Molecular Gas in Galaxies}",
      journal = {\aj},
         year = 2005,
        month = may,
       volume = {129},
       number = {5},
        pages = {2197-2202},
          doi = {10.1086/429134},
archivePrefix = {arXiv},
       eprint = {astro-ph/0501449},
 primaryClass = {astro-ph},
       adsurl = {https://ui.adsabs.harvard.edu/abs/2005AJ....129.2197D}
}

@ARTICLE{Emonts+17,
       author = {{Emonts}, B.~H.~C. and {Colina}, L. and {Piqueras-L{\'o}pez}, J. and {Garcia-Burillo}, S. and {Pereira-Santaella}, M. and {Arribas}, S. and {Labiano}, A. and {Alonso-Herrero}, A.},
        title = "{Outflows of hot molecular gas in ultra-luminous infrared galaxies mapped with VLT-SINFONI}",
      journal = {\aap},
         year = 2017,
        month = nov,
       volume = {607},
          eid = {A116},
        pages = {A116},
          doi = {10.1051/0004-6361/201731508},
archivePrefix = {arXiv},
       eprint = {1708.09503},
 primaryClass = {astro-ph.GA},
       adsurl = {https://ui.adsabs.harvard.edu/abs/2017A&A...607A.116E}
}

@ARTICLE{York+00,
       author = {{York}, Donald G. and {Adelman}, J. and {Anderson}, Jr., John E. and {Anderson}, Scott F. and {Annis}, James and {Bahcall}, Neta A. and {Bakken}, J.~A. and {Barkhouser}, Robert and {Bastian}, Steven and {Berman}, Eileen and {Boroski}, William N. and {Bracker}, Steve and {Briegel}, Charlie and {Briggs}, John W. and {Brinkmann}, J. and {Brunner}, Robert and {Burles}, Scott and {Carey}, Larry and {Carr}, Michael A. and {Castander}, Francisco J. and {Chen}, Bing and {Colestock}, Patrick L. and {Connolly}, A.~J. and {Crocker}, J.~H. and {Csabai}, Istv{\'a}n and {Czarapata}, Paul C. and {Davis}, John Eric and {Doi}, Mamoru and {Dombeck}, Tom and {Eisenstein}, Daniel and {Ellman}, Nancy and {Elms}, Brian R. and {Evans}, Michael L. and {Fan}, Xiaohui and {Federwitz}, Glenn R. and {Fiscelli}, Larry and {Friedman}, Scott and {Frieman}, Joshua A. and {Fukugita}, Masataka and {Gillespie}, Bruce and {Gunn}, James E. and {Gurbani}, Vijay K. and {de Haas}, Ernst and {Haldeman}, Merle and {Harris}, Frederick H. and {Hayes}, J. and {Heckman}, Timothy M. and {Hennessy}, G.~S. and {Hindsley}, Robert B. and {Holm}, Scott and {Holmgren}, Donald J. and {Huang}, Chi-hao and {Hull}, Charles and {Husby}, Don and {Ichikawa}, Shin-Ichi and {Ichikawa}, Takashi and {Ivezi{\'c}}, {\v{Z}}eljko and {Kent}, Stephen and {Kim}, Rita S.~J. and {Kinney}, E. and {Klaene}, Mark and {Kleinman}, A.~N. and {Kleinman}, S. and {Knapp}, G.~R. and {Korienek}, John and {Kron}, Richard G. and {Kunszt}, Peter Z. and {Lamb}, D.~Q. and {Lee}, B. and {Leger}, R. French and {Limmongkol}, Siriluk and {Lindenmeyer}, Carl and {Long}, Daniel C. and {Loomis}, Craig and {Loveday}, Jon and {Lucinio}, Rich and {Lupton}, Robert H. and {MacKinnon}, Bryan and {Mannery}, Edward J. and {Mantsch}, P.~M. and {Margon}, Bruce and {McGehee}, Peregrine and {McKay}, Timothy A. and {Meiksin}, Avery and {Merelli}, Aronne and {Monet}, David G. and {Munn}, Jeffrey A. and {Narayanan}, Vijay K. and {Nash}, Thomas and {Neilsen}, Eric and {Neswold}, Rich and {Newberg}, Heidi Jo and {Nichol}, R.~C. and {Nicinski}, Tom and {Nonino}, Mario and {Okada}, Norio and {Okamura}, Sadanori and {Ostriker}, Jeremiah P. and {Owen}, Russell and {Pauls}, A. George and {Peoples}, John and {Peterson}, R.~L. and {Petravick}, Donald and {Pier}, Jeffrey R. and {Pope}, Adrian and {Pordes}, Ruth and {Prosapio}, Angela and {Rechenmacher}, Ron and {Quinn}, Thomas R. and {Richards}, Gordon T. and {Richmond}, Michael W. and {Rivetta}, Claudio H. and {Rockosi}, Constance M. and {Ruthmansdorfer}, Kurt and {Sandford}, Dale and {Schlegel}, David J. and {Schneider}, Donald P. and {Sekiguchi}, Maki and {Sergey}, Gary and {Shimasaku}, Kazuhiro and {Siegmund}, Walter A. and {Smee}, Stephen and {Smith}, J. Allyn and {Snedden}, S. and {Stone}, R. and {Stoughton}, Chris and {Strauss}, Michael A. and {Stubbs}, Christopher and {SubbaRao}, Mark and {Szalay}, Alexander S. and {Szapudi}, Istvan and {Szokoly}, Gyula P. and {Thakar}, Anirudda R. and {Tremonti}, Christy and {Tucker}, Douglas L. and {Uomoto}, Alan and {Vanden Berk}, Dan and {Vogeley}, Michael S. and {Waddell}, Patrick and {Wang}, Shu-i. and {Watanabe}, Masaru and {Weinberg}, David H. and {Yanny}, Brian and {Yasuda}, Naoki and {SDSS Collaboration}},
        title = "{The Sloan Digital Sky Survey: Technical Summary}",
      journal = {\aj},
         year = 2000,
        month = sep,
       volume = {120},
       number = {3},
        pages = {1579-1587},
          doi = {10.1086/301513},
archivePrefix = {arXiv},
       eprint = {astro-ph/0006396},
 primaryClass = {astro-ph},
       adsurl = {https://ui.adsabs.harvard.edu/abs/2000AJ....120.1579Y}
}

@ARTICLE{Abazajian+09,
       author = {{Abazajian}, Kevork N. and {Adelman-McCarthy}, Jennifer K. and {Ag{\"u}eros}, Marcel A. and {Allam}, Sahar S. and {Allende Prieto}, Carlos and {An}, Deokkeun and {Anderson}, Kurt S.~J. and {Anderson}, Scott F. and {Annis}, James and {Bahcall}, Neta A. and {Bailer-Jones}, C.~A.~L. and {Barentine}, J.~C. and {Bassett}, Bruce A. and {Becker}, Andrew C. and {Beers}, Timothy C. and {Bell}, Eric F. and {Belokurov}, Vasily and {Berlind}, Andreas A. and {Berman}, Eileen F. and {Bernardi}, Mariangela and {Bickerton}, Steven J. and {Bizyaev}, Dmitry and {Blakeslee}, John P. and {Blanton}, Michael R. and {Bochanski}, John J. and {Boroski}, William N. and {Brewington}, Howard J. and {Brinchmann}, Jarle and {Brinkmann}, J. and {Brunner}, Robert J. and {Budav{\'a}ri}, Tam{\'a}s and {Carey}, Larry N. and {Carliles}, Samuel and {Carr}, Michael A. and {Castander}, Francisco J. and {Cinabro}, David and {Connolly}, A.~J. and {Csabai}, Istv{\'a}n and {Cunha}, Carlos E. and {Czarapata}, Paul C. and {Davenport}, James R.~A. and {de Haas}, Ernst and {Dilday}, Ben and {Doi}, Mamoru and {Eisenstein}, Daniel J. and {Evans}, Michael L. and {Evans}, N.~W. and {Fan}, Xiaohui and {Friedman}, Scott D. and {Frieman}, Joshua A. and {Fukugita}, Masataka and {G{\"a}nsicke}, Boris T. and {Gates}, Evalyn and {Gillespie}, Bruce and {Gilmore}, G. and {Gonzalez}, Belinda and {Gonzalez}, Carlos F. and {Grebel}, Eva K. and {Gunn}, James E. and {Gy{\"o}ry}, Zsuzsanna and {Hall}, Patrick B. and {Harding}, Paul and {Harris}, Frederick H. and {Harvanek}, Michael and {Hawley}, Suzanne L. and {Hayes}, Jeffrey J.~E. and {Heckman}, Timothy M. and {Hendry}, John S. and {Hennessy}, Gregory S. and {Hindsley}, Robert B. and {Hoblitt}, J. and {Hogan}, Craig J. and {Hogg}, David W. and {Holtzman}, Jon A. and {Hyde}, Joseph B. and {Ichikawa}, Shin-ichi and {Ichikawa}, Takashi and {Im}, Myungshin and {Ivezi{\'c}}, {\v{Z}}eljko and {Jester}, Sebastian and {Jiang}, Linhua and {Johnson}, Jennifer A. and {Jorgensen}, Anders M. and {Juri{\'c}}, Mario and {Kent}, Stephen M. and {Kessler}, R. and {Kleinman}, S.~J. and {Knapp}, G.~R. and {Konishi}, Kohki and {Kron}, Richard G. and {Krzesinski}, Jurek and {Kuropatkin}, Nikolay and {Lampeitl}, Hubert and {Lebedeva}, Svetlana and {Lee}, Myung Gyoon and {Lee}, Young Sun and {French Leger}, R. and {L{\'e}pine}, S{\'e}bastien and {Li}, Nolan and {Lima}, Marcos and {Lin}, Huan and {Long}, Daniel C. and {Loomis}, Craig P. and {Loveday}, Jon and {Lupton}, Robert H. and {Magnier}, Eugene and {Malanushenko}, Olena and {Malanushenko}, Viktor and {Mandelbaum}, Rachel and {Margon}, Bruce and {Marriner}, John P. and {Mart{\'\i}nez-Delgado}, David and {Matsubara}, Takahiko and {McGehee}, Peregrine M. and {McKay}, Timothy A. and {Meiksin}, Avery and {Morrison}, Heather L. and {Mullally}, Fergal and {Munn}, Jeffrey A. and {Murphy}, Tara and {Nash}, Thomas and {Nebot}, Ada and {Neilsen}, Jr., Eric H. and {Newberg}, Heidi Jo and {Newman}, Peter R. and {Nichol}, Robert C. and {Nicinski}, Tom and {Nieto-Santisteban}, Maria and {Nitta}, Atsuko and {Okamura}, Sadanori and {Oravetz}, Daniel J. and {Ostriker}, Jeremiah P. and {Owen}, Russell and {Padmanabhan}, Nikhil and {Pan}, Kaike and {Park}, Changbom and {Pauls}, George and {Peoples}, Jr., John and {Percival}, Will J. and {Pier}, Jeffrey R. and {Pope}, Adrian C. and {Pourbaix}, Dimitri and {Price}, Paul A. and {Purger}, Norbert and {Quinn}, Thomas and {Raddick}, M. Jordan and {Re Fiorentin}, Paola and {Richards}, Gordon T. and {Richmond}, Michael W. and {Riess}, Adam G. and {Rix}, Hans-Walter and {Rockosi}, Constance M. and {Sako}, Masao and {Schlegel}, David J. and {Schneider}, Donald P. and {Scholz}, Ralf-Dieter and {Schreiber}, Matthias R. and {Schwope}, Axel D. and {Seljak}, Uro{\v{s}} and {Sesar}, Branimir and {Sheldon}, Erin and {Shimasaku}, Kazu and {Sibley}, Valena C. and {Simmons}, A.~E. and {Sivarani}, Thirupathi and {Allyn Smith}, J. and {Smith}, Martin C. and {Smol{\v{c}}i{\'c}}, Vernesa and {Snedden}, Stephanie A. and {Stebbins}, Albert and {Steinmetz}, Matthias and {Stoughton}, Chris and {Strauss}, Michael A. and {SubbaRao}, Mark and {Suto}, Yasushi and {Szalay}, Alexander S. and {Szapudi}, Istv{\'a}n and {Szkody}, Paula and {Tanaka}, Masayuki and {Tegmark}, Max and {Teodoro}, Luis F.~A. and {Thakar}, Aniruddha R. and {Tremonti}, Christy A. and {Tucker}, Douglas L. and {Uomoto}, Alan and {Vanden Berk}, Daniel E. and {Vandenberg}, Jan and {Vidrih}, S. and {Vogeley}, Michael S. and {Voges}, Wolfgang and {Vogt}, Nicole P. and {Wadadekar}, Yogesh and {Watters}, Shannon and {Weinberg}, David H. and {West}, Andrew A. and {White}, Simon D.~M. and {Wilhite}, Brian C. and {Wonders}, Alainna C. and {Yanny}, Brian and {Yocum}, D.~R.},
        title = "{The Seventh Data Release of the Sloan Digital Sky Survey}",
      journal = {\apjs},
         year = 2009,
        month = jun,
       volume = {182},
       number = {2},
        pages = {543-558},
          doi = {10.1088/0067-0049/182/2/543},
archivePrefix = {arXiv},
       eprint = {0812.0649},
 primaryClass = {astro-ph},
       adsurl = {https://ui.adsabs.harvard.edu/abs/2009ApJS..182..543A}
}

@ARTICLE{Holden+23a,
       author = {{Holden}, Luke R. and {Tadhunter}, Clive N. and {Morganti}, Raffaella and {Oosterloo}, Tom},
        title = "{Precise physical conditions for the warm gas outflows in the nearby active galaxy IC 5063}",
      journal = {\mnras},
         year = 2023,
        month = apr,
       volume = {520},
       number = {2},
        pages = {1848-1871},
          doi = {10.1093/mnras/stad123},
archivePrefix = {arXiv},
       eprint = {2301.03999},
 primaryClass = {astro-ph.GA},
       adsurl = {https://ui.adsabs.harvard.edu/abs/2023MNRAS.520.1848H}
}

@ARTICLE{Holden+23b,
       author = {{Holden}, Luke R. and {Tadhunter}, Clive N.},
        title = "{Outflow densities and ionization mechanisms in the NLRs of the prototypical Seyfert galaxies NGC 1068 and NGC 4151}",
      journal = {\mnras},
         year = 2023,
        month = sep,
       volume = {524},
       number = {1},
        pages = {886-905},
          doi = {10.1093/mnras/stad1677},
archivePrefix = {arXiv},
       eprint = {2306.03920},
 primaryClass = {astro-ph.GA},
       adsurl = {https://ui.adsabs.harvard.edu/abs/2023MNRAS.524..886H}
}

@ARTICLE{Davies+20,
       author = {{Davies}, R. and {Baron}, D. and {Shimizu}, T. and {Netzer}, H. and {Burtscher}, L. and {de Zeeuw}, P.~T. and {Genzel}, R. and {Hicks}, E.~K.~S. and {Koss}, M. and {Lin}, M. -Y. and {Lutz}, D. and {Maciejewski}, W. and {M{\"u}ller-S{\'a}nchez}, F. and {Orban de Xivry}, G. and {Ricci}, C. and {Riffel}, R. and {Riffel}, R.~A. and {Rosario}, D. and {Schartmann}, M. and {Schnorr-M{\"u}ller}, A. and {Shangguan}, J. and {Sternberg}, A. and {Sturm}, E. and {Storchi-Bergmann}, T. and {Tacconi}, L. and {Veilleux}, S.},
        title = "{Ionized outflows in local luminous AGN: what are the real densities and outflow rates?}",
      journal = {\mnras},
         year = 2020,
        month = nov,
       volume = {498},
       number = {3},
        pages = {4150-4177},
          doi = {10.1093/mnras/staa2413},
archivePrefix = {arXiv},
       eprint = {2003.06153},
 primaryClass = {astro-ph.GA},
       adsurl = {https://ui.adsabs.harvard.edu/abs/2020MNRAS.498.4150D}
}

@ARTICLE{RamosAlmeida+25,
       author = {{Ramos Almeida}, C. and {Garc{\'\i}a-Bernete}, I. and {Pereira-Santaella}, M. and {Speranza}, G. and {Maiolino}, R. and {Ji}, X. and {Audibert}, A. and {Cezar}, P.~H. and {Acosta-Pulido}, J.~A. and {Alonso-Herrero}, A. and {Garc{\'\i}a-Burillo}, S. and {Gonz{\'a}lez-Mart{\'\i}n}, O. and {Rigopoulou}, D. and {Tadhunter}, C.~N. and {Labiano}, A. and {Levenson}, N.~A. and {Donnan}, F.~R.},
        title = "{JWST MIRI reveals the diversity of nuclear mid-infrared spectra of nearby type 2 quasars}",
      journal = {\aap},
         year = 2025,
        month = jun,
       volume = {698},
          eid = {A194},
        pages = {A194},
          doi = {10.1051/0004-6361/202453549},
archivePrefix = {arXiv},
       eprint = {2504.01595},
 primaryClass = {astro-ph.GA},
       adsurl = {https://ui.adsabs.harvard.edu/abs/2025A&A...698A.194R}
}

@ARTICLE{Cresci+15,
       author = {{Cresci}, G. and {Mainieri}, V. and {Brusa}, M. and {Marconi}, A. and {Perna}, M. and {Mannucci}, F. and {Piconcelli}, E. and {Maiolino}, R. and {Feruglio}, C. and {Fiore}, F. and {Bongiorno}, A. and {Lanzuisi}, G. and {Merloni}, A. and {Schramm}, M. and {Silverman}, J.~D. and {Civano}, F.},
        title = "{Blowin' in the Wind: Both ``Negative'' and ``Positive'' Feedback in an Obscured High-z Quasar}",
      journal = {\apj},
         year = 2015,
        month = jan,
       volume = {799},
       number = {1},
          eid = {82},
        pages = {82},
          doi = {10.1088/0004-637X/799/1/82},
archivePrefix = {arXiv},
       eprint = {1411.4208},
 primaryClass = {astro-ph.GA},
       adsurl = {https://ui.adsabs.harvard.edu/abs/2015ApJ...799...82C}
}

@ARTICLE{Santoro+16,
       author = {{Santoro}, F. and {Oonk}, J.~B.~R. and {Morganti}, R. and {Oosterloo}, T.~A. and {Tadhunter}, C.},
        title = "{Embedded star formation in the extended narrow line region of Centaurus A: Extreme mixing observed by MUSE}",
      journal = {\aap},
         year = 2016,
        month = may,
       volume = {590},
          eid = {A37},
        pages = {A37},
          doi = {10.1051/0004-6361/201628353},
archivePrefix = {arXiv},
       eprint = {1604.03891},
 primaryClass = {astro-ph.GA},
       adsurl = {https://ui.adsabs.harvard.edu/abs/2016A&A...590A..37S}
}

@ARTICLE{Cresci+15b,
       author = {{Cresci}, G. and {Marconi}, A. and {Zibetti}, S. and {Risaliti}, G. and {Carniani}, S. and {Mannucci}, F. and {Gallazzi}, A. and {Maiolino}, R. and {Balmaverde}, B. and {Brusa}, M. and {Capetti}, A. and {Cicone}, C. and {Feruglio}, C. and {Bland-Hawthorn}, J. and {Nagao}, T. and {Oliva}, E. and {Salvato}, M. and {Sani}, E. and {Tozzi}, P. and {Urrutia}, T. and {Venturi}, G.},
        title = "{The MAGNUM survey: positive feedback in the nuclear region of NGC 5643 suggested by MUSE}",
      journal = {\aap},
         year = 2015,
        month = oct,
       volume = {582},
          eid = {A63},
        pages = {A63},
          doi = {10.1051/0004-6361/201526581},
archivePrefix = {arXiv},
       eprint = {1508.04464},
 primaryClass = {astro-ph.GA},
       adsurl = {https://ui.adsabs.harvard.edu/abs/2015A&A...582A..63C}
}

@ARTICLE{Maiolino+17,
       author = {{Maiolino}, R. and {Russell}, H.~R. and {Fabian}, A.~C. and {Carniani}, S. and {Gallagher}, R. and {Cazzoli}, S. and {Arribas}, S. and {Belfiore}, F. and {Bellocchi}, E. and {Colina}, L. and {Cresci}, G. and {Ishibashi}, W. and {Marconi}, A. and {Mannucci}, F. and {Oliva}, E. and {Sturm}, E.},
        title = "{Star formation inside a galactic outflow}",
      journal = {\nat},
         year = 2017,
        month = mar,
       volume = {544},
       number = {7649},
        pages = {202-206},
          doi = {10.1038/nature21677},
archivePrefix = {arXiv},
       eprint = {1703.08587},
 primaryClass = {astro-ph.GA},
       adsurl = {https://ui.adsabs.harvard.edu/abs/2017Natur.544..202M}
}

@ARTICLE{Gallagher+19,
       author = {{Gallagher}, R. and {Maiolino}, R. and {Belfiore}, F. and {Drory}, N. and {Riffel}, R. and {Riffel}, R.~A.},
        title = "{Widespread star formation inside galactic outflows}",
      journal = {\mnras},
         year = 2019,
        month = may,
       volume = {485},
       number = {3},
        pages = {3409-3429},
          doi = {10.1093/mnras/stz564},
archivePrefix = {arXiv},
       eprint = {1806.03311},
 primaryClass = {astro-ph.GA},
       adsurl = {https://ui.adsabs.harvard.edu/abs/2019MNRAS.485.3409G}
}

@ARTICLE{Audibert+25,
       author = {{Audibert}, A. and {Ramos Almeida}, C. and {Garc{\'\i}a-Burillo}, S. and {Speranza}, G. and {Lamperti}, I. and {Pereira-Santaella}, M. and {Panessa}, F.},
        title = "{Molecular gas excitation and outflow properties of obscured quasars at z {\ensuremath{\sim}} 0.1}",
      journal = {\aap},
         year = 2025,
        month = jul,
       volume = {699},
          eid = {A83},
        pages = {A83},
          doi = {10.1051/0004-6361/202453291},
archivePrefix = {arXiv},
       eprint = {2505.02759},
 primaryClass = {astro-ph.GA},
       adsurl = {https://ui.adsabs.harvard.edu/abs/2025A&A...699A..83A}
}

@ARTICLE{Audibert+23,
       author = {{Audibert}, A. and {Ramos Almeida}, C. and {Garc{\'\i}a-Burillo}, S. and {Combes}, F. and {Bischetti}, M. and {Meenakshi}, M. and {Mukherjee}, D. and {Bicknell}, G. and {Wagner}, A.~Y.},
        title = "{Jet-induced molecular gas excitation and turbulence in the Teacup}",
      journal = {\aap},
         year = 2023,
        month = mar,
       volume = {671},
          eid = {L12},
        pages = {L12},
          doi = {10.1051/0004-6361/202345964},
archivePrefix = {arXiv},
       eprint = {2302.13884},
 primaryClass = {astro-ph.GA},
       adsurl = {https://ui.adsabs.harvard.edu/abs/2023A&A...671L..12A}
}

@ARTICLE{Mercedes-Feliz+23,
       author = {{Mercedes-Feliz}, Jonathan and {Angl{\'e}s-Alc{\'a}zar}, Daniel and {Hayward}, Christopher C. and {Cochrane}, Rachel K. and {Terrazas}, Bryan A. and {Wellons}, Sarah and {Richings}, Alexander J. and {Faucher-Gigu{\`e}re}, Claude-Andr{\'e} and {Moreno}, Jorge and {Su}, Kung Yi and {Hopkins}, Philip F. and {Quataert}, Eliot and {Kere{\v{s}}}, Du{\v{s}}an},
        title = "{Local positive feedback in the overall negative: the impact of quasar winds on star formation in the FIRE cosmological simulations}",
      journal = {\mnras},
         year = 2023,
        month = sep,
       volume = {524},
       number = {3},
        pages = {3446-3463},
          doi = {10.1093/mnras/stad2079},
archivePrefix = {arXiv},
       eprint = {2301.01784},
 primaryClass = {astro-ph.GA},
       adsurl = {https://ui.adsabs.harvard.edu/abs/2023MNRAS.524.3446M}
}

@ARTICLE{Trindade+22,
       author = {{Trindade Falc{\~a}o}, Anna and {Kraemer}, S.~B. and {Crenshaw}, D.~M. and {Melendez}, M. and {Revalski}, M. and {Fischer}, T.~C. and {Schmitt}, H.~R. and {Turner}, T.~J.},
        title = "{Tracking X-ray outflows with optical/infrared footprint lines}",
      journal = {\mnras},
         year = 2022,
        month = feb,
       volume = {511},
       number = {1},
        pages = {1420-1430},
          doi = {10.1093/mnras/stac173},
archivePrefix = {arXiv},
       eprint = {2110.11436},
 primaryClass = {astro-ph.GA},
       adsurl = {https://ui.adsabs.harvard.edu/abs/2022MNRAS.511.1420T}
}

@ARTICLE{Fonseca+23,
       author = {{Fonseca-Faria}, M.~A. and {Rodr{\'\i}guez-Ardila}, A. and {Contini}, M. and {Dahmer-Hahn}, L.~G. and {Morganti}, R.},
        title = "{Physical conditions and extension of the coronal line region in IC 5063}",
      journal = {\mnras},
         year = 2023,
        month = sep,
       volume = {524},
       number = {1},
        pages = {143-160},
          doi = {10.1093/mnras/stad1871},
archivePrefix = {arXiv},
       eprint = {2306.09570},
 primaryClass = {astro-ph.GA},
       adsurl = {https://ui.adsabs.harvard.edu/abs/2023MNRAS.524..143F}
}

@ARTICLE{Delaney+25,
       author = {{Delaney}, Dan and {Berger}, Cassidy and {Hicks}, Erin and {Burtscher}, Leonard and {Rosario}, David and {M{\"u}ller-S{\'a}nchez}, Francisco and {Malkan}, Matthew},
        title = "{The LUNIS-AGN Catalog: Trends of Emission Line Velocity Dispersion and Surface Brightness within the Circumnuclear Regions of Seyfert Galaxies}",
      journal = {\apj},
         year = 2025,
        month = may,
       volume = {984},
       number = {2},
          eid = {163},
        pages = {163},
          doi = {10.3847/1538-4357/adc1cc},
archivePrefix = {arXiv},
       eprint = {2504.14136},
 primaryClass = {astro-ph.GA},
       adsurl = {https://ui.adsabs.harvard.edu/abs/2025ApJ...984..163D}
}

@ARTICLE{Woo+16,
       author = {{Woo}, Jong-Hak and {Bae}, Hyun-Jin and {Son}, Donghoon and {Karouzos}, Marios},
        title = "{The Prevalence of Gas Outflows in Type 2 AGNs}",
      journal = {\apj},
         year = 2016,
        month = feb,
       volume = {817},
       number = {2},
          eid = {108},
        pages = {108},
          doi = {10.3847/0004-637X/817/2/108},
archivePrefix = {arXiv},
       eprint = {1511.05142},
 primaryClass = {astro-ph.GA},
       adsurl = {https://ui.adsabs.harvard.edu/abs/2016ApJ...817..108W}
}

@ARTICLE{Cicone+18,
       author = {{Cicone}, Claudia and {Brusa}, Marcella and {Ramos Almeida}, Cristina and {Cresci}, Giovanni and {Husemann}, Bernd and {Mainieri}, Vincenzo},
        title = "{The largely unconstrained multiphase nature of outflows in AGN host galaxies}",
      journal = {Nature Astronomy},
         year = 2018,
        month = feb,
       volume = {2},
        pages = {176-178},
          doi = {10.1038/s41550-018-0406-3},
archivePrefix = {arXiv},
       eprint = {1802.10308},
 primaryClass = {astro-ph.GA},
       adsurl = {https://ui.adsabs.harvard.edu/abs/2018NatAs...2..176C}
}

@ARTICLE{Riffel+25,
       author = {{Riffel}, Rogemar A. and {Souza-Oliveira}, Gabriel L. and {Costa-Souza}, Jos{\'e} Henrique and {Zakamska}, Nadia L. and {Storchi-Bergmann}, Thaisa and {Riffel}, Rog{\'e}rio and {Bianchin}, Marina},
        title = "{Blowing Star Formation Away in AGN Hosts (BAH). II. Investigating the Origin of the H$_{2}$ Emission Excess in Nearby Galaxies with JWST MIRI}",
      journal = {\apj},
         year = 2025,
        month = apr,
       volume = {982},
       number = {2},
          eid = {69},
        pages = {69},
          doi = {10.3847/1538-4357/adb8dd},
archivePrefix = {arXiv},
       eprint = {2410.06960},
 primaryClass = {astro-ph.GA},
       adsurl = {https://ui.adsabs.harvard.edu/abs/2025ApJ...982...69R}
}

@ARTICLE{Bohn+24,
       author = {{Bohn}, Thomas and {Inami}, Hanae and {Togi}, Aditya and {Armus}, Lee and {Lai}, Thomas S. -Y. and {Barcos-Munoz}, Loreto and {Song}, Yiqing and {Linden}, S.~T. and {Surace}, Jason and {Bianchin}, Marina and {U}, Vivian and {Evans}, Aaron S. and {B{\"o}ker}, Torsten and {Malkan}, Matthew A. and {Larson}, Kirsten L. and {Stierwalt}, Sabrina and {Buiten}, Victorine A. and {Charmandaris}, Vassilis and {Diaz-Santos}, Tanio and {Howell}, Justin H. and {Privon}, George C. and {Ricci}, Claudio and {van der Werf}, Paul P. and {Aalto}, Susanne and {Hayward}, Christopher C. and {Kader}, Justin A. and {Mazzarella}, Joseph M. and {Muller-Sanchez}, Francisco and {Sanders}, David B.},
        title = "{GOALS-JWST: The Warm Molecular Outflows of the Merging Starburst Galaxy NGC 3256}",
      journal = {\apj},
         year = 2024,
        month = dec,
       volume = {977},
       number = {1},
          eid = {36},
        pages = {36},
          doi = {10.3847/1538-4357/ad87d3},
archivePrefix = {arXiv},
       eprint = {2403.14751},
 primaryClass = {astro-ph.GA},
       adsurl = {https://ui.adsabs.harvard.edu/abs/2024ApJ...977...36B}
}

@ARTICLE{Marconcini+25,
       author = {{Marconcini}, C. and {Feltre}, A. and {Lamperti}, I. and {Ceci}, M. and {Marconi}, A. and {Ulivi}, L. and {Mannucci}, F. and {Cresci}, G. and {Belfiore}, F. and {Bertola}, E. and {Carniani}, S. and {D'Amato}, Q. and {Fernandez-Ontiveros}, J.~A. and {Fritz}, J. and {Ginolfi}, M. and {Hatziminaoglou}, E. and {Hern{\'a}n-Caballero}, A. and {Hirschmann}, M. and {Mingozzi}, M. and {Rojas}, A.~F. and {Sabatini}, G. and {Salvestrini}, F. and {Scialpi}, M. and {Tozzi}, G. and {Venturi}, G. and {Vidal-Garc{\'\i}a}, A. and {Vignali}, C. and {Zanchettin}, M.~V. and {Amiri}, A.},
        title = "{MIRACLE: I. Unveiling the multi-phase, multi-scale physical properties of the active galaxy NGC 424 with MIRI, MUSE, and ALMA}",
      journal = {\aap},
         year = 2025,
        month = sep,
       volume = {701},
          eid = {A113},
        pages = {A113},
          doi = {10.1051/0004-6361/202554797},
archivePrefix = {arXiv},
       eprint = {2503.21921},
 primaryClass = {astro-ph.GA},
       adsurl = {https://ui.adsabs.harvard.edu/abs/2025A&A...701A.113M}
}

@ARTICLE{Feruglio+10,
       author = {{Feruglio}, C. and {Maiolino}, R. and {Piconcelli}, E. and {Menci}, N. and {Aussel}, H. and {Lamastra}, A. and {Fiore}, F.},
        title = "{Quasar feedback revealed by giant molecular outflows}",
      journal = {\aap},
         year = 2010,
        month = jul,
       volume = {518},
          eid = {L155},
        pages = {L155},
          doi = {10.1051/0004-6361/201015164},
archivePrefix = {arXiv},
       eprint = {1006.1655},
 primaryClass = {astro-ph.CO},
       adsurl = {https://ui.adsabs.harvard.edu/abs/2010A&A...518L.155F}
}

@ARTICLE{Cicone+14,
       author = {{Cicone}, C. and {Maiolino}, R. and {Sturm}, E. and {Graci{\'a}-Carpio}, J. and {Feruglio}, C. and {Neri}, R. and {Aalto}, S. and {Davies}, R. and {Fiore}, F. and {Fischer}, J. and {Garc{\'\i}a-Burillo}, S. and {Gonz{\'a}lez-Alfonso}, E. and {Hailey-Dunsheath}, S. and {Piconcelli}, E. and {Veilleux}, S.},
        title = "{Massive molecular outflows and evidence for AGN feedback from CO observations}",
      journal = {\aap},
         year = 2014,
        month = feb,
       volume = {562},
          eid = {A21},
        pages = {A21},
          doi = {10.1051/0004-6361/201322464},
archivePrefix = {arXiv},
       eprint = {1311.2595},
 primaryClass = {astro-ph.CO},
       adsurl = {https://ui.adsabs.harvard.edu/abs/2014A&A...562A..21C}
}

@ARTICLE{Fluetsch+19,
       author = {{Fluetsch}, A. and {Maiolino}, R. and {Carniani}, S. and {Marconi}, A. and {Cicone}, C. and {Bourne}, M.~A. and {Costa}, T. and {Fabian}, A.~C. and {Ishibashi}, W. and {Venturi}, G.},
        title = "{Cold molecular outflows in the local Universe and their feedback effect on galaxies}",
      journal = {\mnras},
         year = 2019,
        month = mar,
       volume = {483},
       number = {4},
        pages = {4586-4614},
          doi = {10.1093/mnras/sty3449},
archivePrefix = {arXiv},
       eprint = {1805.05352},
 primaryClass = {astro-ph.GA},
       adsurl = {https://ui.adsabs.harvard.edu/abs/2019MNRAS.483.4586F}
}

@ARTICLE{Lamperti+22,
       author = {{Lamperti}, I. and {Pereira-Santaella}, M. and {Perna}, M. and {Colina}, L. and {Arribas}, S. and {Garc{\'\i}a-Burillo}, S. and {Gonz{\'a}lez-Alfonso}, E. and {Aalto}, S. and {Alonso-Herrero}, A. and {Combes}, F. and {Labiano}, A. and {Piqueras-L{\'o}pez}, J. and {Rigopoulou}, D. and {van der Werf}, P.},
        title = "{Physics of ULIRGs with MUSE and ALMA: The PUMA project. IV. No tight relation between cold molecular outflow rates and AGN luminosities}",
      journal = {\aap},
         year = 2022,
        month = dec,
       volume = {668},
          eid = {A45},
        pages = {A45},
          doi = {10.1051/0004-6361/202244054},
archivePrefix = {arXiv},
       eprint = {2209.03380},
 primaryClass = {astro-ph.GA},
       adsurl = {https://ui.adsabs.harvard.edu/abs/2022A&A...668A..45L}
}

@ARTICLE{Pereira-Santaella+18,
       author = {{Pereira-Santaella}, M. and {Colina}, L. and {Garc{\'\i}a-Burillo}, S. and {Combes}, F. and {Emonts}, B. and {Aalto}, S. and {Alonso-Herrero}, A. and {Arribas}, S. and {Henkel}, C. and {Labiano}, A. and {Muller}, S. and {Piqueras L{\'o}pez}, J. and {Rigopoulou}, D. and {van der Werf}, P.},
        title = "{Spatially resolved cold molecular outflows in ULIRGs}",
      journal = {\aap},
         year = 2018,
        month = aug,
       volume = {616},
          eid = {A171},
        pages = {A171},
          doi = {10.1051/0004-6361/201833089},
archivePrefix = {arXiv},
       eprint = {1805.03667},
 primaryClass = {astro-ph.GA},
       adsurl = {https://ui.adsabs.harvard.edu/abs/2018A&A...616A.171P}
}

@ARTICLE{DallAgnol+23,
       author = {{Dall'Agnol de Oliveira}, Bruno and {Storchi-Bergmann}, Thaisa and {Morganti}, Raffaella and {Riffel}, Rogemar A. and {Ramakrishnan}, Venkatessh},
        title = "{Cold molecular gas outflow encasing the ionized one in the Seyfert galaxy NGC 3281}",
      journal = {\mnras},
         year = 2023,
        month = jul,
       volume = {522},
       number = {3},
        pages = {3753-3765},
          doi = {10.1093/mnras/stad1076},
archivePrefix = {arXiv},
       eprint = {2304.04004},
 primaryClass = {astro-ph.GA},
       adsurl = {https://ui.adsabs.harvard.edu/abs/2023MNRAS.522.3753D}
}

@ARTICLE{Vayner+21,
       author = {{Vayner}, Andrey and {Zakamska}, Nadia and {Wright}, Shelley A. and {Armus}, Lee and {Murray}, Norman and {Walth}, Gregory},
        title = "{Multiphase Outflows in High-redshift Quasar Host Galaxies}",
      journal = {\apj},
         year = 2021,
        month = dec,
       volume = {923},
       number = {1},
          eid = {59},
        pages = {59},
          doi = {10.3847/1538-4357/ac2b9e},
archivePrefix = {arXiv},
       eprint = {2110.00019},
 primaryClass = {astro-ph.GA},
       adsurl = {https://ui.adsabs.harvard.edu/abs/2021ApJ...923...59V}
}

@ARTICLE{Harrison+14,
       author = {{Harrison}, C.~M. and {Alexander}, D.~M. and {Mullaney}, J.~R. and {Swinbank}, A.~M.},
        title = "{Kiloparsec-scale outflows are prevalent among luminous AGN: outflows and feedback in the context of the overall AGN population}",
      journal = {\mnras},
         year = 2014,
        month = jul,
       volume = {441},
       number = {4},
        pages = {3306-3347},
          doi = {10.1093/mnras/stu515},
archivePrefix = {arXiv},
       eprint = {1403.3086},
 primaryClass = {astro-ph.GA},
       adsurl = {https://ui.adsabs.harvard.edu/abs/2014MNRAS.441.3306H}
}

@ARTICLE{VillarMartin+16,
       author = {{Villar-Mart{\'\i}n}, M. and {Arribas}, S. and {Emonts}, B. and {Humphrey}, A. and {Tadhunter}, C. and {Bessiere}, P. and {Cabrera Lavers}, A. and {Ramos Almeida}, C.},
        title = "{Ionized outflows in luminous type 2 AGNs at z < 0.6: no evidence for significant impact on the host galaxies}",
      journal = {\mnras},
         year = 2016,
        month = jul,
       volume = {460},
       number = {1},
        pages = {130-162},
          doi = {10.1093/mnras/stw901},
archivePrefix = {arXiv},
       eprint = {1604.04577},
 primaryClass = {astro-ph.GA},
       adsurl = {https://ui.adsabs.harvard.edu/abs/2016MNRAS.460..130V}
}

@ARTICLE{Riffel+23,
       author = {{Riffel}, R.~A. and {Storchi-Bergmann}, T. and {Riffel}, R. and {Bianchin}, M. and {Zakamska}, N.~L. and {Ruschel-Dutra}, D. and {Bentz}, M.~C. and {Burtscher}, L. and {Crenshaw}, D.~M. and {Dahmer-Hahn}, L.~G. and {Dametto}, N.~Z. and {Davies}, R.~I. and {Diniz}, M.~R. and {Fischer}, T.~C. and {Harrison}, C.~M. and {Mainieri}, V. and {Revalski}, M. and {Rodriguez-Ardila}, A. and {Rosario}, D.~J. and {Sch{\"o}nell}, A.~J.},
        title = "{The AGNIFS survey: spatially resolved observations of hot molecular and ionized outflows in nearby active galaxies}",
      journal = {\mnras},
         year = 2023,
        month = may,
       volume = {521},
       number = {2},
        pages = {1832-1848},
          doi = {10.1093/mnras/stad599},
archivePrefix = {arXiv},
       eprint = {2302.11324},
 primaryClass = {astro-ph.GA},
       adsurl = {https://ui.adsabs.harvard.edu/abs/2023MNRAS.521.1832R}
}

@ARTICLE{Kakkad+20,
       author = {{Kakkad}, D. and {Mainieri}, V. and {Vietri}, G. and {Carniani}, S. and {Harrison}, C.~M. and {Perna}, M. and {Scholtz}, J. and {Circosta}, C. and {Cresci}, G. and {Husemann}, B. and {Bischetti}, M. and {Feruglio}, C. and {Fiore}, F. and {Marconi}, A. and {Padovani}, P. and {Brusa}, M. and {Cicone}, C. and {Comastri}, A. and {Lanzuisi}, G. and {Mannucci}, F. and {Menci}, N. and {Netzer}, H. and {Piconcelli}, E. and {Puglisi}, A. and {Salvato}, M. and {Schramm}, M. and {Silverman}, J. and {Vignali}, C. and {Zamorani}, G. and {Zappacosta}, L.},
        title = "{SUPER. II. Spatially resolved ionised gas kinematics and scaling relations in z {\ensuremath{\sim}} 2 AGN host galaxies}",
      journal = {\aap},
         year = 2020,
        month = oct,
       volume = {642},
          eid = {A147},
        pages = {A147},
          doi = {10.1051/0004-6361/202038551},
archivePrefix = {arXiv},
       eprint = {2008.01728},
 primaryClass = {astro-ph.GA},
       adsurl = {https://ui.adsabs.harvard.edu/abs/2020A&A...642A.147K}
}

@ARTICLE{Rupke+17,
       author = {{Rupke}, David S.~N. and {G{\"u}ltekin}, Kayhan and {Veilleux}, Sylvain},
        title = "{Quasar-mode Feedback in Nearby Type 1 Quasars: Ubiquitous Kiloparsec-scale Outflows and Correlations with Black Hole Properties}",
      journal = {\apj},
         year = 2017,
        month = nov,
       volume = {850},
       number = {1},
          eid = {40},
        pages = {40},
          doi = {10.3847/1538-4357/aa94d1},
archivePrefix = {arXiv},
       eprint = {1708.05139},
 primaryClass = {astro-ph.GA},
       adsurl = {https://ui.adsabs.harvard.edu/abs/2017ApJ...850...40R}
}

@ARTICLE{Bertola+25,
       author = {{Bertola}, E. and {Cresci}, G. and {Venturi}, G. and {Perna}, M. and {Circosta}, C. and {Tozzi}, G. and {Lamperti}, I. and {Vignali}, C. and {Arribas}, S. and {Bunker}, A.~J. and {Charlot}, S. and {Carniani}, S. and {Maiolino}, R. and {Rodr{\'\i}guez Del Pino}, B. and {{\"U}bler}, H. and {Willott}, C.~J. and {B{\"o}ker}, T. and {Marshall}, M.~A. and {Parlanti}, E. and {Scholtz}, J.},
        title = "{GA-NIFS: Mapping z ≃ 3.5 AGN-driven ionized outflows in the COSMOS field}",
      journal = {\aap},
         year = 2025,
        month = jul,
       volume = {699},
          eid = {A220},
        pages = {A220},
          doi = {10.1051/0004-6361/202554281},
archivePrefix = {arXiv},
       eprint = {2505.08867},
 primaryClass = {astro-ph.GA},
       adsurl = {https://ui.adsabs.harvard.edu/abs/2025A&A...699A.220B}
}

@ARTICLE{Schawinski+15,
       author = {{Schawinski}, Kevin and {Koss}, Michael and {Berney}, Simon and {Sartori}, Lia F.},
        title = "{Active galactic nuclei flicker: an observational estimate of the duration of black hole growth phases of {\ensuremath{\sim}}{}10$^{5}$ yr}",
      journal = {\mnras},
         year = 2015,
        month = aug,
       volume = {451},
       number = {3},
        pages = {2517-2523},
          doi = {10.1093/mnras/stv1136},
archivePrefix = {arXiv},
       eprint = {1505.06733},
 primaryClass = {astro-ph.GA},
       adsurl = {https://ui.adsabs.harvard.edu/abs/2015MNRAS.451.2517S}
}

@ARTICLE{Hickox+14,
       author = {{Hickox}, Ryan C. and {Mullaney}, James R. and {Alexander}, David M. and {Chen}, Chien-Ting J. and {Civano}, Francesca M. and {Goulding}, Andy D. and {Hainline}, Kevin N.},
        title = "{Black Hole Variability and the Star Formation-Active Galactic Nucleus Connection: Do All Star-forming Galaxies Host an Active Galactic Nucleus?}",
      journal = {\apj},
         year = 2014,
        month = feb,
       volume = {782},
       number = {1},
          eid = {9},
        pages = {9},
          doi = {10.1088/0004-637X/782/1/9},
archivePrefix = {arXiv},
       eprint = {1306.3218},
 primaryClass = {astro-ph.CO},
       adsurl = {https://ui.adsabs.harvard.edu/abs/2014ApJ...782....9H}
}

@ARTICLE{Davies+24,
       author = {{Davies}, R. and {Shimizu}, T. and {Pereira-Santaella}, M. and {Alonso-Herrero}, A. and {Audibert}, A. and {Bellocchi}, E. and {Boorman}, P. and {Campbell}, S. and {Cao}, Y. and {Combes}, F. and {Delaney}, D. and {D{\'\i}az-Santos}, T. and {Eisenhauer}, F. and {Esparza Arredondo}, D. and {Feuchtgruber}, H. and {F{\"o}rster Schreiber}, N.~M. and {Fuller}, L. and {Gandhi}, P. and {Garc{\'\i}a-Bernete}, I. and {Garc{\'\i}a-Burillo}, S. and {Garc{\'\i}a-Lorenzo}, B. and {Genzel}, R. and {Gillessen}, S. and {Gonz{\'a}lez Mart{\'\i}n}, O. and {Haidar}, H. and {Hermosa Mu{\~n}oz}, L. and {Hicks}, E.~K.~S. and {H{\"o}nig}, S. and {Imanishi}, M. and {Izumi}, T. and {Labiano}, A. and {Leist}, M. and {Levenson}, N.~A. and {Lopez-Rodriguez}, E. and {Lutz}, D. and {Ott}, T. and {Packham}, C. and {Rabien}, S. and {Ramos Almeida}, C. and {Ricci}, C. and {Rigopoulou}, D. and {Rosario}, D. and {Rouan}, D. and {Santos}, D.~J.~D. and {Shangguan}, J. and {Stalevski}, M. and {Sternberg}, A. and {Sturm}, E. and {Tacconi}, L. and {Villar Mart{\'\i}n}, M. and {Ward}, M. and {Zhang}, L.},
        title = "{GATOS: missing molecular gas in the outflow of NGC 5728 revealed by JWST}",
      journal = {\aap},
         year = 2024,
        month = sep,
       volume = {689},
          eid = {A263},
        pages = {A263},
          doi = {10.1051/0004-6361/202449875},
archivePrefix = {arXiv},
       eprint = {2406.17072},
 primaryClass = {astro-ph.GA},
       adsurl = {https://ui.adsabs.harvard.edu/abs/2024A&A...689A.263D}
}

@ARTICLE{Bianchin+22,
       author = {{Bianchin}, M. and {Riffel}, R.~A. and {Storchi-Bergmann}, T. and {Riffel}, R. and {Ruschel-Dutra}, D. and {Harrison}, C.~M. and {Dahmer-Hahn}, L.~G. and {Mainieri}, V. and {Sch{\"o}nell}, A.~J. and {Dametto}, N.~Z.},
        title = "{Gemini NIFS survey of feeding and feedback in nearby active galaxies - V. Molecular and ionized gas kinematics}",
      journal = {\mnras},
         year = 2022,
        month = feb,
       volume = {510},
       number = {1},
        pages = {639-657},
          doi = {10.1093/mnras/stab3468},
archivePrefix = {arXiv},
       eprint = {2111.09130},
 primaryClass = {astro-ph.GA},
       adsurl = {https://ui.adsabs.harvard.edu/abs/2022MNRAS.510..639B}
}

@ARTICLE{Tadhunter+14,
       author = {{Tadhunter}, C. and {Morganti}, R. and {Rose}, M. and {Oonk}, J.~B.~R. and {Oosterloo}, T.},
        title = "{Jet acceleration of the fast molecular outflows in the Seyfert galaxy IC 5063}",
      journal = {\nat},
         year = 2014,
        month = jul,
       volume = {511},
       number = {7510},
        pages = {440-443},
          doi = {10.1038/nature13520},
archivePrefix = {arXiv},
       eprint = {1407.1332},
 primaryClass = {astro-ph.GA},
       adsurl = {https://ui.adsabs.harvard.edu/abs/2014Natur.511..440T}
}

@ARTICLE{Fluetsch+21,
       author = {{Fluetsch}, A. and {Maiolino}, R. and {Carniani}, S. and {Arribas}, S. and {Belfiore}, F. and {Bellocchi}, E. and {Cazzoli}, S. and {Cicone}, C. and {Cresci}, G. and {Fabian}, A.~C. and {Gallagher}, R. and {Ishibashi}, W. and {Mannucci}, F. and {Marconi}, A. and {Perna}, M. and {Sturm}, E. and {Venturi}, G.},
        title = "{Properties of the multiphase outflows in local (ultra)luminous infrared galaxies}",
      journal = {\mnras},
         year = 2021,
        month = aug,
       volume = {505},
       number = {4},
        pages = {5753-5783},
          doi = {10.1093/mnras/stab1666},
archivePrefix = {arXiv},
       eprint = {2006.13232},
 primaryClass = {astro-ph.GA},
       adsurl = {https://ui.adsabs.harvard.edu/abs/2021MNRAS.505.5753F}
}

@ARTICLE{Su+19,
       author = {{Su}, Kung-Yi and {Hopkins}, Philip F. and {Hayward}, Christopher C. and {Ma}, Xiangcheng and {Faucher-Gigu{\`e}re}, Claude-Andr{\'e} and {Kere{\v{s}}}, Du{\v{s}}an and {Orr}, Matthew E. and {Chan}, T.~K. and {Robles}, Victor H.},
        title = "{The failure of stellar feedback, magnetic fields, conduction, and morphological quenching in maintaining red galaxies}",
      journal = {\mnras},
         year = 2019,
        month = aug,
       volume = {487},
       number = {3},
        pages = {4393-4408},
          doi = {10.1093/mnras/stz1494},
archivePrefix = {arXiv},
       eprint = {1809.09120},
 primaryClass = {astro-ph.GA},
       adsurl = {https://ui.adsabs.harvard.edu/abs/2019MNRAS.487.4393S}
}

@ARTICLE{Ramos&Ricci+17,
       author = {{Ramos Almeida}, Cristina and {Ricci}, Claudio},
        title = "{Nuclear obscuration in active galactic nuclei}",
      journal = {Nature Astronomy},
         year = 2017,
        month = oct,
       volume = {1},
        pages = {679-689},
          doi = {10.1038/s41550-017-0232-z},
archivePrefix = {arXiv},
       eprint = {1709.00019},
 primaryClass = {astro-ph.GA},
       adsurl = {https://ui.adsabs.harvard.edu/abs/2017NatAs...1..679R}
}

@ARTICLE{Cardelli+89,
       author = {{Cardelli}, Jason A. and {Clayton}, Geoffrey C. and {Mathis}, John S.},
        title = "{The Relationship between Infrared, Optical, and Ultraviolet Extinction}",
      journal = {\apj},
         year = 1989,
        month = oct,
       volume = {345},
        pages = {245},
          doi = {10.1086/167900},
       adsurl = {https://ui.adsabs.harvard.edu/abs/1989ApJ...345..245C}
}

@ARTICLE{Hernandez+25,
       author = {{Hernandez}, Svea and {Smith}, Linda J. and {Jones}, Logan H. and {Togi}, Aditya and {Mel{\'e}ndez}, Marcio B. and {Abril-Melgarejo}, Valentina and {Adamo}, Angela and {Alonso Herrero}, Almudena and {D{\'\i}az-Santos}, Tanio and {Fischer}, Travis C. and {Garc{\'\i}a-Burillo}, Santiago and {Hirschauer}, Alec S. and {Hunt}, Leslie K. and {James}, Bethan and {Lebouteiller}, Vianney and {Long}, Knox S. and {Mingozzi}, Matilde and {Ramambason}, Lise and {Ramos Almeida}, Cristina},
        title = "{JWST/MIRI Detection of [Ne v] and [Ne VI] in M83: Evidence for the Long Sought-after Active Galactic Nucleus?}",
      journal = {\apj},
         year = 2025,
        month = apr,
       volume = {983},
       number = {2},
          eid = {154},
        pages = {154},
          doi = {10.3847/1538-4357/adba5d},
archivePrefix = {arXiv},
       eprint = {2502.17621},
 primaryClass = {astro-ph.GA},
       adsurl = {https://ui.adsabs.harvard.edu/abs/2025ApJ...983..154H}
}

@ARTICLE{Esparza-Arredondo+25,
       author = {{Esparza-Arredondo}, D. and {Ramos Almeida}, C. and {Audibert}, A. and {Pereira-Santaella}, M. and {Garc{\'\i}a-Bernete}, I. and {Garc{\'\i}a-Burillo}, S. and {Shimizu}, T. and {Davies}, R. and {Hermosa Mu{\~n}oz}, L. and {Alonso-Herrero}, A. and {Combes}, F. and {Speranza}, G. and {Zhang}, L. and {Campbell}, S. and {Bellocchi}, E. and {Bunker}, A.~J. and {D{\'\i}az-Santos}, T. and {Garc{\'\i}a-Lorenzo}, B. and {Gonz{\'a}lez-Mart{\'\i}n}, O. and {Hicks}, E.~K.~S. and {Labiano}, A. and {Levenson}, N.~A. and {Ricci}, C. and {Rosario}, D. and {Hoenig}, S. and {Packham}, C. and {Stalevski}, M. and {Fuller}, L. and {Izumi}, T. and {L{\'o}pez-Rodr{\'\i}guez}, E. and {Rigopoulou}, D. and {Rouan}, D. and {Ward}, M.},
        title = "{Molecular gas stratification and disturbed kinematics in the Seyfert galaxy MCG-05-23-16 revealed by JWST and ALMA}",
      journal = {\aap},
         year = 2025,
        month = jan,
       volume = {693},
          eid = {A174},
        pages = {A174},
          doi = {10.1051/0004-6361/202452488},
archivePrefix = {arXiv},
       eprint = {2411.12398},
 primaryClass = {astro-ph.GA},
       adsurl = {https://ui.adsabs.harvard.edu/abs/2025A&A...693A.174E}
}

@ARTICLE{Musiimenta+24,
       author = {{Musiimenta}, B. and {Speranza}, G. and {Urrutia}, T. and {Brusa}, M. and {Ramos Almeida}, C. and {Perna}, M. and {L{\'o}pez}, I.~E. and {Alexander}, D. and {Laloux}, B. and {Shankar}, F. and {Lapi}, A. and {Salvato}, M. and {Toba}, Y. and {Andonie}, C. and {Rodr{\'\i}guez}, I.~M.},
        title = "{Ionised AGN outflows in the Goldfish galaxy: The illuminating and interacting red quasar eFEDSJ091157.4+014327 at z {\ensuremath{\sim}} 0.6}",
      journal = {\aap},
         year = 2024,
        month = jul,
       volume = {687},
          eid = {A111},
        pages = {A111},
          doi = {10.1051/0004-6361/202449283},
archivePrefix = {arXiv},
       eprint = {2401.17299},
 primaryClass = {astro-ph.GA},
       adsurl = {https://ui.adsabs.harvard.edu/abs/2024A&A...687A.111M}
}

@ARTICLE{Girdhar+24,
       author = {{Girdhar}, A. and {Harrison}, C.~M. and {Mainieri}, V. and {Fern{\'a}ndez Aranda}, R. and {Alexander}, D.~M. and {Arrigoni Battaia}, F. and {Bianchin}, M. and {Calistro Rivera}, G. and {Circosta}, C. and {Costa}, T. and {Edge}, A.~C. and {Farina}, E.~P. and {Kakkad}, D. and {Kharb}, P. and {Molyneux}, S.~J. and {Mukherjee}, D. and {Njeri}, A. and {Silpa}, S. and {Venturi}, G. and {Ward}, S.~R.},
        title = "{Quasar feedback survey: molecular gas affected by central outflows and by  10-kpc radio lobes reveal dual feedback effects in 'radio quiet' quasars}",
      journal = {\mnras},
         year = 2024,
        month = jan,
       volume = {527},
       number = {3},
        pages = {9322-9342},
          doi = {10.1093/mnras/stad3453},
archivePrefix = {arXiv},
       eprint = {2311.03453},
 primaryClass = {astro-ph.GA},
       adsurl = {https://ui.adsabs.harvard.edu/abs/2024MNRAS.527.9322G}
}

@ARTICLE{Dan+25,
       author = {{Dan}, Kylie Yui and {Seebeck}, Jerome and {Veilleux}, Sylvain and {Rupke}, David and {Gonzalez-Alfonso}, Eduardo and {Garcia-Bernete}, Ismael and {Liu}, Weizhe and {Lutz}, Dieter and {Melendez}, Marcio and {Pereira Santaella}, Miguel and {Sturm}, Eckhard and {Tombesi}, Francesco},
        title = "{JWST Discovery of a Very Fast Biconical Outflow of Warm Molecular Gas in the Nearby Ultraluminous Infrared Galaxy F08572+3915 NW}",
      journal = {\apj},
         year = 2025,
        month = jan,
       volume = {979},
       number = {1},
          eid = {68},
        pages = {68},
          doi = {10.3847/1538-4357/ad9a50},
archivePrefix = {arXiv},
       eprint = {2412.05859},
 primaryClass = {astro-ph.GA},
       adsurl = {https://ui.adsabs.harvard.edu/abs/2025ApJ...979...68D}
}

@ARTICLE{Perna+25,
       author = {{Perna}, Michele and {Arribas}, Santiago and {Ji}, Xihan and {Marconcini}, Cosimo and {Lamperti}, Isabella and {Bertola}, Elena and {Circosta}, Chiara and {D'Eugenio}, Francesco and {{\"U}bler}, Hannah and {B{\"o}ker}, Torsten and {Maiolino}, Roberto and {Bunker}, Andrew J. and {Carniani}, Stefano and {Charlot}, St{\'e}phane and {Willott}, Chris J. and {Cresci}, Giovanni and {Marconi}, Alessandro and {Parlanti}, Eleonora and {Rodr{\'\i}guez Del Pino}, Bruno and {Scholtz}, Jan and {Venturi}, Giacomo},
        title = "{GA-NIFS: A galaxy-wide outflow in a Compton-thick mini-broad-absorption-line quasar at z = 3.5 probed in emission and absorption}",
      journal = {\aap},
         year = 2025,
        month = feb,
       volume = {694},
          eid = {A170},
        pages = {A170},
          doi = {10.1051/0004-6361/202453090},
archivePrefix = {arXiv},
       eprint = {2411.13698},
 primaryClass = {astro-ph.GA},
       adsurl = {https://ui.adsabs.harvard.edu/abs/2025A&A...694A.170P}
}

@ARTICLE{Zanchettin+21,
       author = {{Zanchettin}, M.~V. and {Feruglio}, C. and {Bischetti}, M. and {Malizia}, A. and {Molina}, M. and {Bongiorno}, A. and {Dadina}, M. and {Gruppioni}, C. and {Piconcelli}, E. and {Tombesi}, F. and {Travascio}, A. and {Fiore}, F.},
        title = "{The IBISCO survey. I. Multiphase discs and winds in the Seyfert galaxy Markarian 509}",
      journal = {\aap},
         year = 2021,
        month = nov,
       volume = {655},
          eid = {A25},
        pages = {A25},
          doi = {10.1051/0004-6361/202039773},
archivePrefix = {arXiv},
       eprint = {2107.06756},
 primaryClass = {astro-ph.GA},
       adsurl = {https://ui.adsabs.harvard.edu/abs/2021A&A...655A..25Z}
}

@ARTICLE{Zanchettin+23,
       author = {{Zanchettin}, M.~V. and {Feruglio}, C. and {Massardi}, M. and {Lapi}, A. and {Bischetti}, M. and {Cantalupo}, S. and {Fiore}, F. and {Bongiorno}, A. and {Malizia}, A. and {Marinucci}, A. and {Molina}, M. and {Piconcelli}, E. and {Tombesi}, F. and {Travascio}, A. and {Tozzi}, G. and {Tripodi}, R.},
        title = "{NGC 2992: Interplay between the multiphase disc, wind, and radio bubbles}",
      journal = {\aap},
         year = 2023,
        month = nov,
       volume = {679},
          eid = {A88},
        pages = {A88},
          doi = {10.1051/0004-6361/202245729},
archivePrefix = {arXiv},
       eprint = {2308.04108},
 primaryClass = {astro-ph.GA},
       adsurl = {https://ui.adsabs.harvard.edu/abs/2023A&A...679A..88Z}
}

@ARTICLE{Garcia-Bernete+21,
       author = {{Garc{\'\i}a-Bernete}, I. and {Alonso-Herrero}, A. and {Garc{\'\i}a-Burillo}, S. and {Pereira-Santaella}, M. and {Garc{\'\i}a-Lorenzo}, B. and {Carrera}, F.~J. and {Rigopoulou}, D. and {Ramos Almeida}, C. and {Villar Mart{\'\i}n}, M. and {Gonz{\'a}lez-Mart{\'\i}n}, O. and {Hicks}, E.~K.~S. and {Labiano}, A. and {Ricci}, C. and {Mateos}, S.},
        title = "{Multiphase feedback processes in the Sy2 galaxy NGC 5643}",
      journal = {\aap},
         year = 2021,
        month = jan,
       volume = {645},
          eid = {A21},
        pages = {A21},
          doi = {10.1051/0004-6361/202038256},
archivePrefix = {arXiv},
       eprint = {2009.12385},
 primaryClass = {astro-ph.GA},
       adsurl = {https://ui.adsabs.harvard.edu/abs/2021A&A...645A..21G}
}

@ARTICLE{Holden+24,
       author = {{Holden}, Luke R. and {Tadhunter}, Clive and {Audibert}, Anelise and {Oosterloo}, Tom and {Ramos Almeida}, Cristina and {Morganti}, Raffaella and {Pereira-Santaella}, Miguel and {Lamperti}, Isabella},
        title = "{ALMA reveals a compact and massive molecular outflow driven by the young AGN in a nearby ULIRG}",
      journal = {\mnras},
         year = 2024,
        month = may,
       volume = {530},
       number = {1},
        pages = {446-456},
          doi = {10.1093/mnras/stae810},
archivePrefix = {arXiv},
       eprint = {2403.08869},
 primaryClass = {astro-ph.GA},
       adsurl = {https://ui.adsabs.harvard.edu/abs/2024MNRAS.530..446H}
}

@ARTICLE{Dave+19,
       author = {{Dav{\'e}}, Romeel and {Angl{\'e}s-Alc{\'a}zar}, Daniel and {Narayanan}, Desika and {Li}, Qi and {Rafieferantsoa}, Mika H. and {Appleby}, Sarah},
        title = "{SIMBA: Cosmological simulations with black hole growth and feedback}",
      journal = {\mnras},
         year = 2019,
        month = jun,
       volume = {486},
       number = {2},
        pages = {2827-2849},
          doi = {10.1093/mnras/stz937},
archivePrefix = {arXiv},
       eprint = {1901.10203},
 primaryClass = {astro-ph.GA},
       adsurl = {https://ui.adsabs.harvard.edu/abs/2019MNRAS.486.2827D}
}

@ARTICLE{Zinger+20,
       author = {{Zinger}, Elad and {Pillepich}, Annalisa and {Nelson}, Dylan and {Weinberger}, Rainer and {Pakmor}, R{\"u}diger and {Springel}, Volker and {Hernquist}, Lars and {Marinacci}, Federico and {Vogelsberger}, Mark},
        title = "{Ejective and preventative: the IllustrisTNG black hole feedback and its effects on the thermodynamics of the gas within and around galaxies}",
      journal = {\mnras},
         year = 2020,
        month = nov,
       volume = {499},
       number = {1},
        pages = {768-792},
          doi = {10.1093/mnras/staa2607},
archivePrefix = {arXiv},
       eprint = {2004.06132},
 primaryClass = {astro-ph.GA},
       adsurl = {https://ui.adsabs.harvard.edu/abs/2020MNRAS.499..768Z}
}

@ARTICLE{DiMatteo+05,
       author = {{Di Matteo}, Tiziana and {Springel}, Volker and {Hernquist}, Lars},
        title = "{Energy input from quasars regulates the growth and activity of black holes and their host galaxies}",
      journal = {\nat},
         year = 2005,
        month = feb,
       volume = {433},
       number = {7026},
        pages = {604-607},
          doi = {10.1038/nature03335},
archivePrefix = {arXiv},
       eprint = {astro-ph/0502199},
 primaryClass = {astro-ph},
       adsurl = {https://ui.adsabs.harvard.edu/abs/2005Natur.433..604D}
}

@ARTICLE{Zubovas+23,
       author = {{Zubovas}, Kastytis and {Maskeli{\={u}}nas}, Gediminas},
        title = "{Life after AGN switch off: evolution and properties of fossil galactic outflows}",
      journal = {\mnras},
         year = 2023,
        month = oct,
       volume = {524},
       number = {4},
        pages = {4819-4840},
          doi = {10.1093/mnras/stad1661},
archivePrefix = {arXiv},
       eprint = {2306.00518},
 primaryClass = {astro-ph.GA},
       adsurl = {https://ui.adsabs.harvard.edu/abs/2023MNRAS.524.4819Z}
}

@ARTICLE{McKaig+24,
       author = {{McKaig}, Jeffrey D. and {Satyapal}, Shobita and {Laor}, Ari and {Abel}, Nicholas P. and {Doan}, Sara M. and {Ricci}, Claudio and {Cann}, Jenna M.},
        title = "{Why Are Optical Coronal Lines Faint in Active Galactic Nuclei?}",
      journal = {\apj},
         year = 2024,
        month = nov,
       volume = {976},
       number = {1},
          eid = {130},
        pages = {130},
          doi = {10.3847/1538-4357/ad7a79},
archivePrefix = {arXiv},
       eprint = {2408.15229},
 primaryClass = {astro-ph.GA},
       adsurl = {https://ui.adsabs.harvard.edu/abs/2024ApJ...976..130M}
}

@ARTICLE{Doan+25,
       author = {{Doan}, Sara and {Satyapal}, Shobita and {Reefe}, Michael and {Sexton}, Remington O. and {Matzko}, William and {McKaig}, Jeffrey D. and {Secrest}, Nathan J. and {Cann}, Jenna M. and {Laor}, Ari and {Canalizo}, Gabriela},
        title = "{The CLASS Quasar Catalog: Coronal Line Activity in Type 1 SDSS Quasars}",
      journal = {\apjs},
         year = 2025,
        month = oct,
       volume = {280},
       number = {2},
          eid = {57},
        pages = {57},
          doi = {10.3847/1538-4365/ade304},
archivePrefix = {arXiv},
       eprint = {2501.17067},
 primaryClass = {astro-ph.GA},
       adsurl = {https://ui.adsabs.harvard.edu/abs/2025ApJS..280...57D}
}

@ARTICLE{Baron+19,
       author = {{Baron}, Dalya and {Netzer}, Hagai},
        title = "{Discovering AGN-driven winds through their infrared emission - II. Mass outflow rate and energetics}",
      journal = {\mnras},
         year = 2019,
        month = jul,
       volume = {486},
       number = {3},
        pages = {4290-4303},
          doi = {10.1093/mnras/stz1070},
archivePrefix = {arXiv},
       eprint = {1903.11076},
 primaryClass = {astro-ph.GA},
       adsurl = {https://ui.adsabs.harvard.edu/abs/2019MNRAS.486.4290B}
}

@ARTICLE{Mukherjee+18,
       author = {{Mukherjee}, Dipanjan and {Bicknell}, Geoffrey V. and {Wagner}, Alexander Y. and {Sutherland}, Ralph S. and {Silk}, Joseph},
        title = "{Relativistic jet feedback - III. Feedback on gas discs}",
      journal = {\mnras},
         year = 2018,
        month = oct,
       volume = {479},
       number = {4},
        pages = {5544-5566},
          doi = {10.1093/mnras/sty1776},
archivePrefix = {arXiv},
       eprint = {1803.08305},
 primaryClass = {astro-ph.HE},
       adsurl = {https://ui.adsabs.harvard.edu/abs/2018MNRAS.479.5544M}
}

@ARTICLE{Kakkad+25,
       author = {{Kakkad}, D. and {Mainieri}, V. and {Tanaka}, Takumi S. and {Silverman}, John D. and {Law}, D. and {Riffel}, Rogemar A. and {Circosta}, C. and {Bertola}, E. and {Bianchin}, M. and {Bischetti}, M. and {Rivera}, G. Calistro and {Carniani}, S. and {Cicone}, C. and {Cresci}, G. and {Costa}, T. and {Harrison}, C.~M. and {Lamperti}, I. and {Kalita}, B. and {Koekemoer}, Anton M. and {Marconi}, A. and {Perna}, M. and {Piconcelli}, E. and {Puglisi}, A. and {Ilha}, Gabriele S. and {Tozzi}, G. and {Vietri}, G. and {Vignali}, C. and {Ward}, S. and {Zamorani}, G. and {Zappacosta}, L.},
        title = "{JWST MIRI/MRS observations of hot molecular gas in an AGN host galaxy at Cosmic Noon}",
      journal = {\mnras},
         year = 2025,
        month = jul,
       volume = {541},
       number = {4},
        pages = {3534-3548},
          doi = {10.1093/mnras/staf1125},
archivePrefix = {arXiv},
       eprint = {2507.05354},
 primaryClass = {astro-ph.GA},
       adsurl = {https://ui.adsabs.harvard.edu/abs/2025MNRAS.541.3534K}
}

@ARTICLE{Holden+25,
       author = {{Holden}, Luke R. and {Smith}, Daniel J.~B. and {Arnaudova}, Marina I. and {Tadhunter}, Clive N. and {Ramos Almeida}, Cristina and {Shenoy}, Shravya and {Cezar}, Pedro H. and {Das}, Soumyadeep and {Binu}, Akshara},
        title = "{Electron densities from [S II] lines significantly overestimate the impact of ionised AGN outflows}",
      journal = {\mnras},
         year = 2025,
        month = nov,
          doi = {10.1093/mnras/staf2075},
archivePrefix = {arXiv},
       eprint = {2511.15791},
 primaryClass = {astro-ph.GA},
       adsurl = {https://ui.adsabs.harvard.edu/abs/2025MNRAS.tmp.1961H}
}
